\input amstex
\loadbold
\documentstyle{amsppt}
\pagewidth{32pc}
\pageheight{45pc}
\mag=1200
\baselineskip=15 pt
\NoBlackBoxes
\TagsOnRight

\def\gap{\vskip 0.1in\noindent}
\def\ref#1#2#3#4#5#6{#1, {\it #2,} #3 {\bf #4} (#5), #6.}

\def\akh {1}        
\def\ag {2}         
\def\as {3}         
\def\bgm {4}        
\def\ds {5}         
\def\gs {6}         
\def\gdl {7}        
\def\ham {8}        
\def\im {9}         
\def\ir {10}        
\def\jl {11}        
\def\kato {12}      
\def\katz {13}      
\def\kei {14}       
\def\krch {15}      
\def\krsti {16}     
\def\krstu {17}     
\def\krkol {18}     
\def\land {19}      
\def\lang {20}      
\def\liv {21}       
\def\loe {22}       
\def\mmc {23}       
\def\naI {24}      
\def\naII {25}      
\def\naIII {26}      
\def\naIV {27}      
\def\nel {28}       
\def\nev {29}       
\def\nusI {30}     
\def\nusII {31}     
\def\pea {32}       
\def\rsI {33}       
\def\rsII {34}      
\def\st {35}        
\def\simap {36}     
\def\simjfa {37}    
\def\sti {38}       
\def\sto {39}       

\topmatter
\title The Classical Moment Problem as a Self-Adjoint 
Finite Difference Operator
\endtitle
\rightheadtext{The Classical Moment Problem}
\author  Barry Simon$^*$
\endauthor
\leftheadtext{B.~Simon}
\affil Division of Physics, Mathematics, and Astronomy \\
California Institute of Technology \\
Pasadena, CA 91125
\endaffil
\date November 14, 1997
\enddate
\thanks$^*$ This material is based upon work supported by the 
National Science Foundation under Grant No.~DMS-9401491. The 
Government has certain rights in this material.
\endthanks
\thanks To appear in {\it{Advances in Mathematics}}
\endthanks
\abstract This is a comprehensive exposition of the classical 
moment problem using methods from the theory of finite difference 
operators. Among the advantages of this approach is that the 
Nevanlinna functions appear as elements of a transfer matrix 
and convergence of Pad\'e approximants appears as the strong 
resolvent convergence of finite matrix approximations to a 
Jacobi matrix. As a bonus of this, we obtain new results on the 
convergence of certain Pad\'e approximants for series of Hamburger.
\endabstract
\endtopmatter

\document

\vskip 0.1in
\flushpar{\bf \S 1. Introduction}
\vskip 0.1in

The classical moment problem was central to the development of 
analysis in the period from 1894 (when Stieltjes wrote his 
famous memoir [\sti]) until the 1950's (when Krein completed his 
series on the subject [\krch, \krsti, \krstu]). The notion of 
measure (Stieltjes integrals), Pad\'e approximants, orthogonal 
polynomials, extensions of positive linear functionals 
(Riesz-Markov theorem), boundary values of analytic functions, 
and the Herglotz-Nevanlinna-Riesz representation theorem all 
have their roots firmly in the study of the moment problem.

This expository note attempts to present the main results of 
the theory with two main ideas in mind. It is known from early 
on (see below) that a moment problem is associated to a certain 
semi-infinite Jacobi matrix, $A$. The first idea is that the basic 
theorems can be viewed as results on the self-adjoint extensions 
of $A$. The second idea is that techniques from the theory of 
second-order difference and differential equations should be 
useful in the theory.

Of course, neither of these ideas is new. Both appear, for 
example, in Stone's treatment [\sto], in Dunford-Schwartz [\ds], 
and in Akhiezer's brilliant book on the subject [\akh]. But, 
normally, (and in [\akh], in particular), these ideas are presented 
as applications of the basic theory rather than as the central 
tenets. By placing them front and center, many parts of the 
theory become more transparent --- in particular, our realization 
of the Nevanlinna matrix as a transfer matrix makes its 
properties (as an entire function of order 1 and minimal type) 
evident (see Theorem~4.8 below).

Two basic moment problems will concern us here:

\example{Hamburger Moment Problem} Given a sequence 
$\gamma_0, \gamma_1,\gamma_2, \dots$ of reals, when is there a 
measure, $d\rho$, on $(-\infty, \infty)$ so that
$$
\gamma_n = \int_{-\infty}^\infty x^n\, d\rho(x) \tag 1.1
$$
and if such a $\rho$ exists, is it unique? We let $\Cal M_H (\gamma)$ 
denote the set of solutions of (1.1).
\endexample

\example{Stieltjes Moment Problem} Given a sequence $\gamma_0, 
\gamma_1, \gamma_2, \dots$ of reals, when is there a measure, 
$d\rho$ on $[0,\infty)$ so that
$$
\gamma_n = \int_0^\infty x^n\, d\rho(x) \tag 1.2
$$
and if such a $\rho$ exists, is it unique? We let $\Cal M_S (\gamma)$ 
denote the set of solutions of (1.2).
\endexample

\vskip 0.1in

We will not attempt a comprehensive historical attribution of 
classical results; see Akhiezer [\akh] and Shohat-Tamarkin 
[\st] for that.

We will always normalize the $\gamma$'s so that $\gamma_0 =1$. 
By replacing $\gamma_n$ by $\gamma_n/\gamma_0$, we can reduce 
general $\gamma$'s to this normalized case. We will also demand 
that $d\rho$ have infinite support, that is, that $\rho$ not be a 
pure point measure supported on a finite set. This eliminates 
certain degenerate cases.

One immediately defines two sesquilinear forms, $H_N$ and $S_N$, 
on $\Bbb C^N$ for each $N$ by
$$\align
H_N (\alpha,\beta) &= \sum \Sb n=0,1,\dots, N-1 \\ 
m=0,1,\dots, N-1 \endSb \bar\alpha_n \beta_m \gamma_{n+m} 
\tag 1.3 \\
S_N(\alpha,\beta) &= \sum \Sb n=0,1,\dots, N-1 \\ 
m=0,1,\dots, N-1 \endSb \bar\alpha_n \beta_m \gamma_{n+m+1}
\tag 1.4
\endalign
$$
and corresponding matrices $\Cal H_N$ and $\Cal S_N$ so that 
$H_N (\alpha,\beta)=\langle \alpha, \Cal H_N \beta\rangle$ 
and $S_N(\alpha,\beta)=\langle \alpha, \Cal S_N \beta\rangle$ 
in the usual Euclidean inner product. Our inner products are 
linear in the second factor and anti-linear in the first.

A standard piece of linear algebra says:

\proclaim{Lemma 1.1} An $N\times N$ Hermitean matrix $A$ is 
strictly positive definite if and only if each submatrix 
$A^{[1,J]} = (a_{ij})_{1\leq i, j\leq J}$ has $\det(A^{[1,J]})
>0$ for $J=1,2,\dots, N$.
\endproclaim

\demo{Proof} If $A$ is strictly positive definite, so is each 
$A^{[1,J]}$, so their eigenvalues are all strictly positive and 
so their determinants are all strictly positive.

For the converse, suppose that $A^{[1,N-1]}$ is positive 
definite. By the min-max principle, the $(N-1)$ eigenvalues of 
$A^{[1,N-1]}$ interlace the $N$ eigenvalues of $A^{[1,N]} 
\equiv A$. If all the eigenvalues of $A^{[1,N-1]}$ are positive, 
so are the $N-1$ largest eigenvalues of $A$. If the 
$N^{\text{\rom{th}}}$ eigenvalue were non-positive, $\det(A) 
\leq 0$. Thus, $A^{[1,N-1]}$ strictly positive definite and 
$\det(A^{[N,N]}) >0$ imply $A^{[1,N]}$ is strictly positive 
definite. An obvious induction completes the proof. \qed
\enddemo

This immediately implies

\proclaim{Proposition 1.2} $\{H_N\}^\infty_{N=1}$ are strictly 
positive definite forms if and only if $\det(\Cal H_N)>0$ for 
$N=1,2,\dots$. Similarly, $\{S_N\}^\infty_{N=1}$ are strictly 
positive definite forms if and only if $\det(\Cal S_N)>0$ for 
$N=1,2,\dots$.
\endproclaim

Suppose that the $\gamma_n$ obey (1.1). Then by an elementary 
calculation, 
$$\align
\int \biggl| \sum_{n=0}^{N-1} \alpha_n x^n\biggr|^2\, d\rho(x) 
&= H_N (\alpha,\alpha) \tag 1.5 \\
\int x \biggl| \sum_{n=0}^{N-1} \alpha_n x^n\biggr|^2 \, 
d\rho(x) &= S_N (\alpha, \alpha). \tag 1.6
\endalign
$$

Taking into account that if $\int |P(x)|^2\, d\rho(x)=0$, $\rho$ 
must be supported on the zeros of $P$, we have:

\proclaim{Proposition 1.3} A necessary condition that 
{\rom{(1.1)}} holds for some measure $d\rho$ on \linebreak 
$(-\infty, \infty)$ with infinite support is that each 
sesquilinear form $H_N$ is strictly positive definite. A 
necessary condition that there be a $d\rho$ supported on 
$[0,\infty)$ is that each $H_N$ and each $S_N$ be strictly 
positive definite.
\endproclaim

Suppose now that each $H_N$ is strictly positive definite. Let 
$\Bbb C[X]$ be the family of complex polynomials. Given $P(X) =
\sum_{n=0}^{N-1} \alpha_n X^n$, $Q(X)=\sum_{n=0}^{N-1} \beta_n 
X^n$ (we suppose the upper limits in the sums are equal by 
using some zero $\alpha$'s or $\beta$'s if need be), define
$$
\langle P,Q\rangle = H_N (\alpha,\beta). \tag 1.7
$$
This defines a positive definite inner product on $\Bbb C[X]$, 
and, in the usual way, we can complete $\Bbb C[X]$ to a Hilbert 
space $\Cal H^{(\gamma)}$ in which $\Bbb C[X]$ is dense. 
$\Bbb C[X]$ can be thought of as abstract polynomials or as  
infinite sequences $(\alpha_0, \alpha_1,\dots, \alpha_{N-1}, 0,
\dots)$ which are eventually $0$ via $\alpha\sim \sum_{j=0}^{N-1} 
\alpha_j X^j$.

We will start using some basic facts about symmetric and 
self-adjoint operators on a Hilbert space --- specifically, the 
spectral theorem and the von Neumann theory of self-adjoint 
extensions. For an exposition of these ideas, see Chapters~VII, 
VIII, and X of Reed-Simon, volumes~I and II [\rsI, \rsII] and 
see our brief sketch at the start of Section~2.

Define a densely defined operator $A$ on $\Cal H^{(\gamma)}$ with 
domain $D(A)=\Bbb C[X]$ by
$$
A[P(X)]=[XP(X)]. \tag 1.8
$$ 
In the sequence way of looking at things, $A$ is the right shift, 
that is, $A(\alpha_0, \alpha_1, \dots, \alpha_N, 0, \dots) 
\mathbreak = (0, \alpha_0, \alpha_1, \dots, \alpha_N, 0, \dots)$. 
This means that
$$
\langle P,A[Q]\rangle = S_N (\alpha,\beta) \tag 1.9
$$
and, in particular,
$$
\langle 1, A^n 1\rangle = \gamma_n. \tag 1.10
$$

Since $\Cal H_N$ and $\Cal S_N$ are real-symmetric matrices, $A$ 
is a symmetric operator, that is, $\langle A[P],Q\rangle = 
\langle P,A[Q]\rangle$. Moreover, if we define a complex 
conjugation $C$ on $\Bbb C[X]$ by $C(\sum_{n=0}^{N-1} \alpha_n 
X^n) = \sum_{n=0}^{N-1} \bar\alpha_n X^n$, then $CA = AC$. 
It follows by a theorem of von Neumann (see Corollary~2.4 in 
Section~2) that $A$ has self-adjoint extensions.

If each $S_N$ is positive definite, then $\langle P,A[P]\rangle 
\geq 0$ for all $P$, and it follows that $A$ has a non-negative 
self-adjoint extension $A_F$, the Friedrichs extension (discussed 
further in Section~3). We thus see:

\proclaim{Proposition 1.4} If all $H_N$ are positive definite, 
then $A$ has self-adjoint extensions. If all $S_N$ are positive 
definite, then $A$ has non-negative self-adjoint extensions.
\endproclaim

Let $\tilde A$ be a self-adjoint extension of $A$. By the spectral  
theorem, there is a spectral measure $d\tilde\mu$ for $\tilde A$ 
with vector $[1]\in\Cal H^{(\gamma)}$, that is, so that for 
any bounded function of $\tilde A$,
$$
\langle 1, f(\tilde A) 1\rangle = \int f(x)\, 
d\tilde\mu(x). \tag 1.11
$$
Since $1\in D(A^N)\subset D(\tilde A^N)$, (1.11) extends to 
polynomially bounded functions and we have by (1.10) that,
$$
\gamma_N = \int x^N \, d\tilde\mu(x).
$$
We see therefore that a self-adjoint extension of $A$ yields a 
solution of the Hamburger moment problem. Moreover, a non-negative 
self-adjoint extension has $\text{supp}(d\tilde\mu) \subset 
[0,\infty)$ and so yields a solution of the Stieltjes moment 
problem. Combining this with Propositions~1.2, 1.3, and 1.4, we 
have the first major result in the theory of moments.

\proclaim{Theorem 1 (Existence)} A necessary and sufficient 
condition for there to exist a measure $d\rho$ with infinite 
support obeying {\rom{(1.1)}} is that $\det(\Cal H_N)>0$ for 
$N=1,2, \dots$. A necessary and sufficient condition that also 
$d\rho$ be supported on $[0,\infty)$ is that both $\det(\Cal H_N)
>0$ and $\det(\Cal S_N)>0$ for $N=1,2,\dots$.
\endproclaim

Historically, existence was a major theme because the now 
standard tools on existence of measures were invented in the 
context of moment problems. We have settled it quickly, and the 
bulk of this paper is devoted to uniqueness, especially the 
study of cases of non-uniqueness. Non-uniqueness only occurs 
in somewhat pathological situations, but the theory is so elegant 
and beautiful that it has captivated analysts for a century.

Henceforth, we will call a set of moments 
$\{\gamma_n\}^\infty_{n=0}$ with $\det(\Cal H_N)>0$ for all 
$N$ a set of {\it{Hamburger moments}}. If both $\det(\Cal H_N)>0$ 
and $\det(\Cal S_N)>0$ for all $N$, we call them a set of 
{\it{Stieltjes moments}}.

We will call a solution of the moment problem which comes from 
a self-adjoint extension of $A$ a {\it von~Neumann solution}. 
The name is in honor of the use below of the von~Neumann 
theory of self-adjoint extensions of densely-defined symmetric 
operators. This name, like our use of Friedrichs solution and 
Krein solution later, is not standard, but it is natural from 
the point of view of self-adjoint operators. As far as I know, 
neither von~Neumann nor Friedrichs worked on the moment problem 
per se. While Krein did, his work on the moment problem was not 
in the context of the Krein extension we use to construct what 
we will call the Krein solution. What we call von~Neumann solutions, 
Akhiezer calls $N$-extremal and Shohat-Tamarkin call extremal. 
This last name is unfortunate since we will see there exist many 
solutions which are extreme points in the sense of convex set 
theory, but which are not von~Neumann solutions (and so, not 
extremal in the Shohat-Tamarkin sense).

Given the connection with self-adjoint extensions, the following 
result is reasonable (and true!):

\proclaim{Theorem 2 (Uniqueness)} A necessary and sufficient 
condition that the measure $d\rho$ in {\rom{(1.1)}} be unique 
is that the operator $A$ of {\rom{(1.8)}} is essentially 
self-adjoint \rom(i.e., has a unique self-adjoint extension\rom). 
A necessary and sufficient condition that there be a unique measure 
$d\rho$ in {\rom{(1.1)}} supported in $[0,\infty)$ is that $A$ 
have a unique non-negative self-adjoint extension.
\endproclaim

This result is surprisingly subtle. First of all, it is not 
obvious (but true, as we will see) that distinct self-adjoint 
extensions have distinct spectral measures $d\tilde\mu$, so 
there is something to be proven before multiple self-adjoint 
extensions imply multiple solutions of the moment problem. The 
other direction is even less clear cut, for not only is it not 
obvious, it is false that every solution of the moment problem 
is a von~Neumann solution (Reed-Simon [\rsII] has an incorrect 
proof of uniqueness that implicitly assumes every solution 
comes from a self-adjoint extension). As we will see, once 
there are multiple solutions, there are many, many more 
solutions than those that come from self-adjoint extensions in 
the von~Neumann sense of looking for extensions in 
$\overline{D(A)}$. But, as we will see in Section~6, there is 
a sense in which solutions are associated to self-adjoint 
operators in a larger space.

We also note we will see cases where the Stieltjes problem has 
a unique solution but the associated Hamburger problem does not.

The Hamburger part of Theorem~2 will be proven in Section~2 
(Theorems~2.10 and 2.12); the Stieltjes part will be proven in 
Sections~2 and 3 (Theorems~2.12 and 3.2). If there is a unique 
solution to (1.1), the moment problem is called {\it{determinate}}; 
if there are multiple solutions, it is called {\it{indeterminate}}. 
It is ironic that the English language literature uses these 
awkward terms, rather than determined and undetermined. Stieltjes 
was Dutch, but his fundamental paper was in French, and the names 
have stuck. Much of the interesting theory involves analyzing 
the indeterminate case, so we may as well give some examples that 
illuminate non-uniqueness.

\example{Example 1.1} Let $f$ be a non-zero $C^\infty$ function 
on $\Bbb R$ supported on $[0,1]$. Let $g(x) = \hat f(x)$, with 
$\hat f$ the Fourier transform of $f$. Then
$$
\int_{-\infty}^\infty x^n g(x)\, dx = \sqrt{2\pi} \, 
(-i)^n\, \frac{d^n f}{dx^n}\, (0)=0. 
$$
Let $d\rho_1 (x) = (\text{Re}\, g)_+ (x)\, dx$, the positive part 
of $\text{Re}\, g$, and let $d\rho_2 = (\text{Re}\, g)_- (x)\, dx$. 
By the above,
$$
\int_{-\infty}^\infty x^n d\rho_1 (x) = 
\int_{-\infty}^\infty x^n\, d\rho_2 (x)
$$
for all $n$. Since $d\rho_1$ and $d\rho_2$ have disjoint essential 
supports, they are unequal and we have non-unique solutions of the 
moment problem. (We will see eventually that neither is a 
von~Neumann solution.) The moments from $\rho_1, \rho_2$ may not 
be normalized. But we clearly have non-uniqueness after 
normalization.
\endexample

This non-uniqueness is associated to non-analyticity in a Fourier 
transform and suggests that if one can guarantee analyticity, 
one has uniqueness. Indeed, 

\proclaim{Proposition 1.5} Suppose that 
$\{\gamma_n\}^\infty_{n=0}$ is a set of Hamburger moments and 
that for some $C,R >0$,
$$
|\gamma_n |\leq CR^n n! \tag 1.12a
$$
Then the Hamburger moment problem is determinate.

If $\{\gamma_n\}^\infty_{n=0}$ is a set of Stieltjes moments 
and 
$$
|\gamma_n|\leq CR^n (2n)! \tag 1.12b
$$ 
then the Stieltjes moment problem is determinate.
\endproclaim

\demo{Proof} Let $d\rho$ obey (1.1). Then $x^{2n}\in L^1 
(\Bbb R, d\rho)$, and by the monotone convergence theorem,
$$\align
\int_{-\infty}^\infty \text{cosh} \biggl(\frac{x}{2R}\biggr)\, 
d\rho(x) &= \lim_{N\to\infty} \int_{-\infty}^\infty \sum_{n=0}^N 
\biggl(\frac{x}{2R}\biggr)^{2n} 
\frac{1}{(2n)!}\, d\rho(x)\\
&\leq C\, \lim_{N\to\infty} \sum_{n=0}^N \biggl(\frac12\biggr)^{2n} 
= \frac43 C < \infty. 
\endalign
$$
Thus, $e^{\alpha x} \in L^1 (\Bbb R, d\rho(x))$ for $|\alpha| <
\frac1{2R}$. It follows that $F_\rho (z) \equiv \int e^{izx} 
d\rho(x)$ has an analytic continuation to $\{ z\mid\, |\text{Im}\, 
z| < \frac1{2R}\}$. If $\mu$ is a second solution of (1.1), 
$F_\mu (z)$ is also analytic there. But $-i^n \frac{d^n F_\rho}{dz} 
(0)= \gamma_n$, so the Taylor series of $F_\rho$ and $F_\mu$ at $z=0$ 
agree, so $F_\rho = F_\mu$ in the entire strip by analyticity.

This implies that $\mu = \rho$ by a variety of means. For example, 
in the topology of bounded uniformly local convergence, (i.e., 
$f_n\to f$ means $\sup \|f_n\|_\infty <0$ and $f_n(x)\to f(x)$ 
uniformly for $x$ in any $[-\kappa, \kappa]$), linear combinations 
of the $\{ e^{iyx}\mid y\in\Bbb R\}$ are dense in all bounded 
continuous functions. Or alternatively, $G_\mu (z) = \int 
\frac{d\mu(x)}{x-z} = i\int_{-\infty}^0 e^{-iyz}F_\mu (y)\, dy$ 
for $\text{Im}\, z>0$, and $\mu$ can be recovered as a boundary 
value of $G_\mu$. 

The $(2n)!$ Stieltjes result follows from the $n!$ result and 
Proposition~1.6 below. \qed
\enddemo

We will generalize Proposition~1.5 later (see Corollary~4.5).

\proclaim{Proposition 1.6} Let $\{\gamma_n\}^\infty_{n=0}$ be a 
set of Stieltjes moments. Let 
$$\alignat2
\Gamma_{2m} &= \gamma_m, \qquad && m=0,1,\dots \\
\Gamma_{2m+1} &= 0, \qquad && m=0,1,\dots .
\endalignat
$$
If $\{\Gamma_j\}^\infty_{j=0}$ is a determinate Hamburger problem, 
then $\{\gamma_n\}^\infty_{n=0}$ is a determinate Stieljes 
problem.
\endproclaim

\demo{Proof} Let $d\rho$ solve the Stieltjes problem. Let
$$
d\mu(x) = \tfrac12 [\chi_{[0,\infty)} (x)\, d\rho (x^2) 
+\chi_{(-\infty, 0]}(x)\, d\rho (x^2)].
$$
Then the moments of $\mu$ are $\Gamma$. Thus uniqueness for the 
$\Gamma$ problem on $(-\infty, \infty)$ implies uniqueness for 
the $\gamma$ problem on $[0,\infty)$. \qed
\enddemo

\remark{Remark} We will see later (Theorem~2.13) that the 
converse of the last assertion in Proposition~1.6 is true. This 
is a more subtle result. 
\endremark

\example{Example 1.2} This example is due to Stieltjes [\sti]. 
It is interesting for us here because it is totally explicit and 
because it provides an example of non-uniqueness for the Stieltjes 
moment problem. Note first that
$$
\int_0^\infty u^k u^{-\ln u} \sin(2\pi \ln u)\, du = 0.
$$
This follows by the change of variables $v=-\frac{k+1}{2} + 
\ln u$, the periodicity of $\sin(\,\cdot\,)$, and the fact that 
sin is an odd function. Thus for any $\theta\in [-1,1]$,
$$
\frac{1}{\sqrt\pi} \int_0^\infty u^k u^{-\ln u} [1+\theta
\sin(2\pi \ln u)]\, du = e^{\frac14(k+1)^2}
$$
(by the same change of variables) so $\gamma_k =
e^{\frac14 (k+1)^2}$ is an indeterminate set of Stieltjes 
moments. Notice for $\theta\in (-1,1)$, if $d\rho_\theta (u)$ 
is the measure with these moments, then $\sin(2\pi \ln (u))/1+
\theta \sin(2\pi \ln (u))$ is in $L^2 (d\rho_\theta)$ 
and orthogonal to all polynomials, so $d\rho_\theta (u)$ is 
a measure with all moments finite but where the polynomials are 
not $L^2$ dense. As we shall see, this is typical of solutions of 
indeterminate moment problems which are not von~Neumann solutions. 
\endexample

Looking at the rapid growth of the moments in Example~1.2, you 
might hope that just as a condition of not too great growth 
(like Proposition~1.5) implies determinacy, there might be a 
condition of rapid growth that implies indeterminacy. But that 
is false! There are moments of essentially arbitrary rates of 
growth which lead to determinate problems (see the remark 
after Corollary~4.21 and Theorem~6.2).

\example{Example 1.3} There is a criterion of Krein [\krkol] 
for indeterminacy that again shows the indeterminacy of 
Example~1.2, also of the example of the moments of $\exp 
(-|x|^\alpha)\, dx$ (on $(-\infty, \infty)$) with $\alpha <1$ and 
also an example of Hamburger [\ham], the moments of
$$
\chi_{[0,\infty)}(x) \exp \biggl( - \frac{\pi\sqrt{x}}
{\ln^2 x + \pi^2}\biggr)\, dx.
$$
\endexample

\proclaim{Proposition 1.7 (Krein [\krkol])} Suppose that 
$d\rho(x) = F(x)\, dx$ where $0\leq F(x) \leq 1$ and either

\rom{(i)} $\text{\rom{supp}}(F)= (-\infty, \infty)$ and
$$
\int_{-\infty}^\infty - \frac{\ln (F(x))}{1+x^2}\, dx < \infty 
\tag 1.13a
$$
or

\rom{(ii)} $\text{\rom{supp}}(F) = [0,\infty)$ and
$$
\int_0^\infty -\frac{\ln (F(x))}{(1+x)}\, \frac{dx}
{\sqrt x} < \infty. \tag 1.13b
$$

Suppose also that for all $n$:
$$
\int_{-\infty}^\infty |x|^n F(x)\, dx < \infty. 
\tag 1.13c
$$
Then the moment problem \rom(Hamburger in case {\rom{(i)}}, 
Stieltjes in case {\rom{(ii))}} with moments
$$
\gamma_n = \frac{\int x^n F(x)\, dx}{\int F(x)\, dx}
$$
is indeterminate.
\endproclaim

\remark{Remarks} 1. Hamburger's example is close to borderline 
for (1.13b) to hold.

2. Since $\int_{-\infty}^\infty x^{2n} \exp(-|x|^\alpha)\, dx 
= 2\alpha^{-1} \Gamma (\frac{2n+1}{\alpha})\sim (\frac{2n}
{\alpha})!$ and (1.13a) holds for $F(x) = \exp(-|x|^\alpha)$ 
if $\alpha < 1$, we see that there are examples of Hamburger 
indeterminate moment problems with growth just slightly faster 
than the $n!$ growth, which would imply by Proposition~1.5 that 
the problem is determinate. Similarly, since $\int_0^\infty 
x^n \exp(-|x|^\alpha)\, dx = \alpha^{-1} \Gamma (\frac{n+1}
{\alpha})\sim (\frac{n}{\alpha})!$ and (1.13b) holds for $F(x) 
=\exp (-|x|^\alpha)$ if $\alpha < \frac12$, we see there are 
examples of Stieltjes indeterminate problems with growth just 
slightly faster than the $(2n)!$ of Proposition~1.5.

3. Since $F(x) = \frac12 e^{-|x|}$ has moments $\gamma_{2n} = 
(2n)!$ covered by Proposition~1.5, it is a determinate problem. The 
integral in (1.13a) is only barely divergent in this case. Similarly, 
$F(x) = \chi_{[0,\infty)}(x) e^{-\sqrt x}$ is a Stieltjes 
determinate moment problem by Proposition~1.5 and the integral 
in (1.13b) is barely divergent. This says that Krein's conditions 
are close to optimal.

4. Krein actually proved a stronger result by essentially the 
same method. $F$ need not be bounded (by a limiting argument 
from the bounded case) and the measure defining $\gamma_n$ can 
have an arbitrary singular part. Moreover, Krein proves (1.13a) 
is necessary and sufficient for $\{e^{i\alpha x}
\}_{0\leq\alpha<\infty}$ to not be dense in $L^2 
(\Bbb R, F(x)\, dx)$.

5. Krein's construction, as we shall see, involves finding a 
bounded analytic function $G$ in $\Bbb C_+ \equiv \{z\in\Bbb C 
\mid\text{Im}\, z>0\}$ so that $|G(x+i0)| \leq F(x)$. From this 
point of view, the fact that $F(x)$ cannot decay faster than 
$e^{-|x|}$ is connected to the Phragm\'en-Lindel\"of principle.

6. The analog of Krein's result for the circle instead of 
the half plane is due to Szego.
\endremark

\demo{Proof} Suppose we can find $G$ with $\text{Re}\, G
\not\equiv 0$ with $|G(x)|\leq F(x)$ so that
$$
\int x^n G(x)\, dx = 0.
$$
Then both $\frac{F(x)\, dx}{\int F(x)\, dx}$ and 
$\frac{[F(x)+\text{Re}\, G(x)]\, dx}{\int F(x)\, dx}$ solve the 
$\gamma_n$ moment problem, showing indeterminacy.

In case (i), define for $z\in\Bbb C_+$ (see (1.19) and the 
definition of Herglotz function below):
$$
Q(z) = \frac1{\pi} \int \biggl( \frac1{x-z} - \frac{x}{1+x^2} 
\biggr) [-\ln (F(x))]\, dx,
$$
which is a convergent integral by (1.13a). Then, $\text{Im}\, 
Q(z) \geq 0$ and by the theory of boundary values at such 
analytic functions [\katz], $\lim_{\varepsilon\downarrow 0} 
Q(x+i\varepsilon) \equiv Q(x+i0)$ exists for a.e.~$x\in\Bbb R$ 
and $\text{Im}\, Q(x+i0)=-\ln F(x)$.

Let $G(z) = \exp (iQ(z))$ for $z\in\Bbb C_+$ and $G(x) = \exp 
(iQ(x+i0))$. Then the properties of $Q$ imply $|G(z)| \leq 1$, 
$\lim_{\varepsilon\downarrow 0} G(x+i0)= G(x)$ for a.e.~$x$ 
in $\Bbb R$, and $|G(x)| = F(x)$. A standard contour integral 
argument then shows that for any $\varepsilon >0$ and $n>0$,
$$
\int x^n (1-i\varepsilon x)^{-n-2} G(x)\, dx =0.
$$
Since (1.13c) holds and $|G(x)|\leq F(x)$, we can take 
$\varepsilon\downarrow 0$ and obtain $\int x^n G(x)\, dx =0$.

In case (ii), define
$$
Q(z) = \frac{2z}{\pi} \int_0^\infty - \frac{\ln F(x^2)}
{x^2 - z^2}\, dx.
$$
Again, the integral converges by (1.13b) and $\text{Im}\, 
Q(z) >0$. Moreover, $Q(-\bar z) = -\overline{Q(z)}$. We have for 
a.e.~$x\in\Bbb R$, $\text{Im}\, Q(x+i0) = -\ln F(x^2)$. Thus, 
if $H(z) = \exp (iQ(z))$ for $z\in\Bbb C_+$ and $H(x) = \exp 
(iQ(x+i0))$ for $x\in\Bbb R$ and $G(x) = \text{Im}\, H(\sqrt x)$ 
for $x\in [0,\infty)$, then, as above,
$$
\int_{-\infty}^\infty x^{2n+1} H(x)\, dx =0,
$$
so since $H(-x) = \overline{H(x)}$, $\int_0^\infty 
x^{2n+1} G(x^2)\, dx = 0$ or $\int_0^\infty y^n 
G(y)\, dy =0$. Since $|H(x)| = F(x^2)$, we have that $|G(x)| 
\leq F(x)$. \qed
\enddemo

Additional examples of indeterminate moment problems (specified 
by their orthogonal polynomials) can be found in [\im, \ir].

One point to note for now is that in all these examples, we have 
non-uniqueness with measures $d\rho$ of the form $d\rho (x) = 
G(x)\, dx$ (i.e., absolutely continuous). This will have 
significance after we discuss Theorem~5 below.

To discuss the theory further, we must look more closely at the 
operator $A$ given by (1.8). Consider the set $\{ 1,X,X^2, 
\dots\}$ in $\Cal H^{(\gamma)}$. By the strict positivity of 
$H_N$, these elements in $\Cal H^{(\gamma)}$ are linearly 
independent and they span $\Cal H^{(\gamma)}$ by construction. 
Thus by a Gram-Schmidt procedure, we can obtain an orthogonal 
basis, $\{P_n(X)\}^\infty_{n=0}$, for $\Cal H^{(\gamma)}$. By 
construction,
$$\alignat2
P_n (X) &= c_{nn}X^n + \text{ lower order}, \qquad && 
\text{with }c_{nn} > 0 \tag 1.14a \\
\langle P_n, P_m\rangle &= 0, \qquad && m=0,1,2,\dots, n-1 
\tag 1.14b \\
\langle P_n, P_n\rangle &= 1. \tag 1.14c
\endalignat
$$
These are, of course, the well-known orthogonal polynomials for 
$d\rho$ determined by the moments $\{\gamma_n\}^\infty_{n=0}$. 
Note that often the normalization condition (1.14c) is 
replaced by $c_{nn}\equiv 1$, yielding a distinct set of 
``orthogonal" polynomials. There are explicit formulae for the 
$P_n(X)$ in terms of determinants and the $\gamma$'s. We 
discuss them in Appendix~A.

By construction, $\{P_j\}^n_{j=0}$ is an orthonormal basis for 
the polynomials of degree $n$. The realization of elements of 
$\Cal H^{(\gamma)}$ as $\sum_{n=0}^\infty \lambda_n P_n(X)$ 
with $\sum_{n=0}^\infty |\lambda_n|^2 <\infty$ gives a 
different realization of $\Cal H^{(\gamma)}$ as a set of 
sequences $\lambda = (\lambda_0, \dots)$ with the usual 
$\ell^2 (\{0,1,2,\dots \})$ inner product. $\Bbb C[X]$ 
corresponds to these $\lambda$'s with $\lambda_n=0$ for $n$ 
sufficiently large. But we need to bear in mind this change 
of realization as sequences.

Note that $\text{span}[1,\dots, X^n]=\text{span}[P_0, \dots, 
P_n(X)]$. In particular, $XP_n(X)$ has an expansion in $P_0, 
P_1, \dots, P_{n+1}$. But $\langle XP_n, P_j\rangle = 
\langle P_n, XP_j\rangle =0$ if $j<n-1$ since then $XP_j$ is 
of degree at most $n-1$, and (1.14b) holds. Thus for suitable 
sequences, $\{a_n\}^\infty_{n=0}$, $\{b_n\}^\infty_{n=0}$, 
and $\{c_n\}^\infty_{n=0}$ (with $P_{-1}(X)\equiv 0$),
$$
XP_n (X) = c_n P_{n+1}(X) + b_nP_n(X) + a_{n-1} P_{n-1}(X) 
\tag 1.15
$$
for $n=0,1,2,\dots$. Notice that by (1.14a), $c_n >0$ and
$$
c_n = \langle P_{n+1}, XP_n\rangle = \langle P_n, XP_{n+1}
\rangle = a_n.
$$
(1.15) thus becomes
$$
XP_n(X) = a_n P_{n+1}(X) + b_n P_n(X) + a_{n-1} P_{n-1}(X).  
\tag 1.15$^\prime$
$$
Since $\{P_n(X)\}$ are an orthonormal basis for 
$\Cal H^{(\gamma)}$, this says that in this basis, $A$ is given 
by a tridiagonal matrix, and $D(A)$ is the set of sequences of 
finite support.

Thus, given a set $\{\gamma_n\}^\infty_{n=0}$ of Hamburger 
moments, we can find $b_0, b_1, \dots$ real and $a_0, a_1, \dots$ 
positive so that the moment problem is associated to self-adjoint 
extensions of the Jacobi matrix,
$$
A=\pmatrix
b_0 & a_0 & 0 & 0 & \dots \\
a_0 & b_1 & a_1 & 0 & \dots \\
0 & a_1 & b_2 & a_2 & \dots \\
0 & 0 & a_2 & b_3 & \dots \\
\dots & \dots & \dots & \dots & \dots 
\endpmatrix . \tag 1.16
$$

There are explicit formulae for the $b_n$'s and $a_n$'s in 
terms of the determinants of the $\gamma_n$'s. We will discuss 
them in Appendix A. 

Conversely, given a matrix $A$ of the form (1.16), we can find 
a set of moments for which it is the associated Jacobi matrix. 
Indeed, if $\delta_0$ is the vector $(1,0, \dots, 0, \dots )$ in 
$\ell_2$, then
$$
\gamma_n = (\delta_0, A^n \delta_0).
$$
Thus the theory of Hamburger moments is the theory of 
semi-infinite Jacobi matrices.

So far we have considered the $P_n (X)$ as abstract polynomials, 
one for each $n$. It is useful to turn this around. First, 
replace the abstract $X$ by an explicit complex number. For 
each such $z$, consider the semi-infinite sequence $\pi(z) = 
(P_0(z), P_1(z), P_2(z),\dots)$. (1.15$^\prime$) now becomes 
(with $P_{-1}(z)$ interpreted as $0$):
$$
a_n P_{n+1}(z) + (b_n -z) P_n(z) + a_{n-1} P_{n-1}(z) = 0, 
\qquad n\geq 0 \tag 1.17
$$
so $P_n(z)$ obeys a second-order difference equation. If, for 
example, $\pi(x_0)\in\ell^2$, then $x_0$ is formally an eigenvalue 
of the Jacobi matrix $A$. (Of course, $\pi$ is never a finite 
sequence because it obeys a second-order difference equation and 
$\pi_0(z)=1$. Thus, $\pi\notin D(A)$. It may or may not happen 
that $\pi\in D(\bar A)$, so the formal relation may or may not 
correspond to an actual eigenvalue. It will always be true, 
though, as we shall see, that if $\pi(x_0)\in\ell^2$, then 
$(A^*-x_0)\pi(x_0)=0$.)

For convenience, set $a_{-1}=1$. For any $z\in \Bbb C$, the 
solutions of the equation
$$
a_n u_{n+1} + (b_n-z) u_n + a_{n-1} u_{n-1} = 0,  
\qquad n \geq 0 \tag 1.18
$$
are two-dimensional, determined by the initial data $(u_{-1}, 
u_0)$.

$P_n (z)$ corresponds to taking
$$
u_{-1} =0, \qquad u_0 = 1.
$$
There is a second solution, $Q_n (z)$, taking initial 
conditions
$$
u_{-1} = -1, \qquad u_0 = 0.
$$
We will also define $\xi(z)=(Q_0(z), Q_1(z), \dots)$. It can be 
seen by induction that for $n\geq 1$, $Q_n(X)$ is a polynomial of 
degree $n-1$. We will see later (Proposition~5.16) that the $Q$'s 
are orthogonal polynomials for another moment problem.

As we will see, this normalization is such that if some 
combination $\eta (z) \equiv t\pi(z) + \xi(z)\in\ell^2$, then 
$(A^* -z) \eta = \delta_0$ and $\langle \delta_0, \eta\rangle 
=t$. Here $\delta_n$ is the Kronecker vector with $1$ in 
position $n$ and zeros elsewhere. We will have a lot more to say 
about $Q_n(z)$ in Section~4.

There is a fundamental result relating the solution vectors 
$\pi,\xi$ to determinacy of the moment problem. In Section~4 
(see Proposition~4.4 and Theorem~4.7), we will prove 

\proclaim{Theorem 3} Fix a set of moments and associated 
Jacobi matrix. Then the following are equivalent:
\roster
\item"\rom{(i)}" The Hamburger moment problem is indeterminate.
\item"\rom{(ii)}" For some $z_0\in\Bbb C$ with $\text{\rom{Im}} 
\, z_0 \neq 0$, $\pi(z_0)\in\ell_2$.
\item"\rom{(iii)}" For some $z_0\in\Bbb C$ with 
$\text{\rom{Im}}\,z_0 \neq 0$, $\xi(z_0)\in\ell_2$.
\item"\rom{(iv)}" For some $x_0\in\Bbb R$, both $\pi(x_0)$ 
and $\xi(x_0)$ lie in $\ell_2$.
\item"\rom{(v)}" For some $x_0\in\Bbb R$, both $\pi(x_0)$ and 
$\frac{\partial\pi}{\partial x}(x_0)$ lie in $\ell_2$.
\item"\rom{(vi)}" For some $x_0\in\Bbb R$, both $\xi(x_0)$ 
and $\frac{\partial\xi}{\partial x}(x_0)$ lie in $\ell_2$.
\item"\rom{(vii)}" For all $z\in\Bbb C$, both $\pi(z)$ and 
$\xi(z)$ lie in $\ell_2$.
\endroster
\endproclaim

\remark{Remarks} 1. Appendix A has explicit determinantal 
formulae for $\sum_{n=0}^N |P_n (0)|^2$ and $\sum_{n=0}^N 
|Q_n (0)|^2$ providing ``explicit" criteria for determinacy in 
terms of limits of determinants.

2. Theorem~3 can be thought of as a discrete analog of Weyl's 
limit point/limit circle theory for self-adjoint extensions of 
differential operators; see [\rsII].
\endremark

This implies that if all solutions of (1.18) lie in $\ell^2$ 
for one $z\in\Bbb C$, then all solutions lie in $\ell^2$ for all 
$z\in\Bbb C$.

A high point of the theory, discussed in Section~4, is an 
explicit description of all solutions of the moment problem in 
the indeterminate case in terms of a certain $2\times 2$ 
matrix valued entire analytic function. It will suffice to 
only vaguely describe the full result in this introduction.

\definition{Definition} A {\it{Herglotz function}} is a function 
$\Phi(z)$ defined in $\Bbb C_+ \equiv\{z\in\Bbb C\mid\text{Im}\,
z>0\}$ and analytic there with $\text{Im}\,\Phi(z) >0$ there. 
\enddefinition

These are also sometimes called Nevanlinna functions. It is a 
fundamental result (see, e.g., [\ag]) that given such a $\Phi$, 
there exists $c\geq 0$, $d$ real, and a measure $d\mu$ on $\Bbb R$ 
with $\int\frac{d\mu(x)}{1+x^2} <\infty$ so that either $c\neq 
0$ or $d\mu \neq 0$ or both, and
$$
\Phi(z) = cz + d + \int \biggl[ \frac{1}{x-z} 
-\frac{x}{1+x^2}\biggr] \, d\mu (x). \tag 1.19
$$

\proclaim{Theorem 4} The solutions of the Hamburger moment 
problem in the indeterminate case are naturally parametrized by 
Herglotz functions together with the functions $\Phi(z)= t\in 
\Bbb R \cup\{\infty\}$. This later set of solutions are the 
von~Neumann solutions.

If the Stieltjes problem is also indeterminate, there is 
$t_0 >0$ so that the solutions of the Stieltjes problem are 
naturally parametrized by Herglotz functions, $\Phi(z)$, which 
obey $\Phi(x+i0)\in(t_0,\infty)$ for $x\in(-\infty,0)$ together 
with $[t_0,\infty) \cup\{\infty\}$. The later set of solutions 
are the von~Neumann solutions.
\endproclaim

We will prove Theorem~4 in Section~4 (see Theorems~4.14 and 
4.18). From the explicit Nevanlinna form of the solutions, one 
can prove (and we will in Section~4; see Theorems~4.11 and 4.17) 

\proclaim{Theorem 5} In the indeterminate case, the von~Neumann 
solutions are all pure point measures. Moreover, for any $t\in
\Bbb R$, there exists exactly one von~Neumann solution, 
$\mu^{(t)}$, with $\mu^{(t)}(\{t\}) >0$. Moreover, for any other 
solutions, $\rho$, of the moment problem, $\rho(\{t\}) <\mu^{(t)}
(\{t\})$.
\endproclaim

\remark{Remark} The parametrization $\mu^{(t)}$ of this theorem 
is inconvenient since $t\mapsto \mu^{(t)}$ is many to one (in 
fact, infinity to one since each $\mu^{(t)}$ has infinite support). 
We will instead use a parametrization $t\mapsto \mu_t$ given by 
$$
\int \frac{d\mu_t (x)}{x}=t,
$$
which we will see is a one-one map of $\Bbb R\cup \{\infty\}$ 
to the von~Neumann solutions.
\endremark

\example{Examples 1.1 and 1.2 revisited} As noted, the 
explicit non-unique measures we listed were absolutely 
continuous measures. But there are, of necessity, many pure 
point measures that lead to the same moments. Indeed, the 
measures $\mu_t$ associated to the von~Neumann solutions are each 
pure point. As we will see, there are many other pure point 
measures with the given moments. If $\nu$ is a Cantor measure, 
then it can be proven that $\int \mu_t\, d\nu(t)$ will be a 
singular continuous measure with the given moments. In the 
indeterminate case, the class of measures solving (1.1) is always 
extremely rich.

Given a set of moments $\{\gamma_n\}^\infty_{n=0}$ and $c\in
\Bbb R$, one can define a new set of moments 
$$
\gamma_n (c) = \sum_{j=0}^n \binom{n}{j} c^j \gamma_{n-j}. 
\tag 1.20
$$
For the Hamburger problem, the solutions of the 
$\{\gamma_n\}^\infty_{n=0}$ and each 
$\{\gamma_n(c)\}^\infty_{n=0}$ problem are in one-one 
correspondence. If $\mu$ solves the $\{\gamma_n\}^\infty_{n=0}$ 
problem, then $d\rho(x-c)=d\rho_c (x)$ solves the $\{\gamma_n 
(c)\}^\infty_{n=0}$ problem, and vice-versa. But translation 
does not preserve the condition $\text{supp}(\rho)\subset[0,
\infty)$, so that as $c$ decreases, the set of solutions of the 
Stieltjes problem shrinks. We will show that for any indeterminate 
Stieltjes moment problem, there is always a $c_0$, so that for 
$c>c_0$, $\{\gamma_n(c)\}^\infty_{n=0}$ is an indeterminate 
Stieltjes problem. For $c<c_0$, $\{\gamma_n(c)\}^\infty_{n=0}$ 
has no solution $d\rho$ with $\text{supp}(d\rho)\subset 
[0,\infty)$, and $\{\gamma_n (c_0)\}$ is a determinate Stieltjes 
problem (but, of course, an indeterminate Hamburger problem). This 
means there are lots of examples of moments which are determinate 
Stieltjes but indeterminate Hamburger. The existence of $c_0$ 
is intimately connected to Theorem~5. Among all the 
von~Neumann solutions, there is a distinguished one, $d\mu_F$ 
(the Friedrichs solution), with $f_0 =\inf(\text{supp}(\mu_t))$ 
maximal. One just takes $c_0 = -f_0$. That $f_0$ is a pure point 
of $d\mu_F$ is important in the analysis.
\endexample

Another topic in the moment problem concerns the theory of 
Pad\'e approximants, or what is esssentially the same thing ---  
continued fractions. Typical is Theorem~6 below, which we will 
prove in Section~5.

Consider a sequence of numbers $\{\kappa_n\}^\infty_{n=0}$,  
$$
\kappa_n = (-1)^n \int_0^\infty x^n \, d\rho(x) 
$$
for some $\rho$. We are interested in ``summing" the formal 
series (called a {\it{series of Stieltjes}})
$$
\sum_{n=0}^\infty \kappa_n z^n, \tag 1.21
$$
which is formally
$$
\int_0^\infty \frac{d\rho(x)}{1+xz}\, . \tag 1.22
$$
If the series (1.21) converges, then (1.22) is analytic in 
a circle, $|z|< R$, which implies that $\rho$ is supported 
in $[0,\frac1{R}]$. Thus, if $d\rho$ does not have compact 
support, then the series (1.21) will not converge for any 
$z\neq 0$.

\definition{Definition} The $[N,M]$ Pad\'e approximant to 
the series (1.21) is the unique rational function of the form 
$$
f^{[N,M]}(z)\equiv \frac{A^{[N,M]}(z)}{B^{[N,M]}(z)}\, ,
\tag 1.23
$$ 
where $A$ is a polynomial of degree $N$ and $B$ is a polynomial 
of degree $M$, and (as $z\to 0$)
$$
f^{[N,M]}(z) - \sum_{n=0}^{N+M} \kappa_n z^n = O(z^{N+M+1}).
\tag 1.24
$$

Note that $A/B$ has $(N+1)+(M+1)-1=N+M+1$ free parameters and 
$\{\kappa_n\}^{N+M}_{n=0}$ is $N+M+1$ free numbers.  There is an 
explicit solution for $A,B$ in terms of determinants; see Baker 
and Graves-Morris [\bgm]. We will say more about the definition 
of Pad\'e approximants in Section~5.
\enddefinition

\proclaim{Theorem 6} Let $\sum\kappa_n z^n$ be a series of 
Stieltjes.  Then for any $z\in\Bbb C\backslash(-\infty, 0)$, 
\linebreak $\lim_{N\to\infty} f^{[N-1,N]}(z) \equiv f_-(z)$ exists 
and $\lim_{N\to\infty} f^{[N,N]}(z)= f_+(z)$ exists. The 
convergence is uniform on compact subsets of $\Bbb C\backslash 
(-\infty,0]$. Moreover, $f_+ = f_-$ if and only if the 
associated Stieltjes problem is determinate. If the problem is 
indeterminate and $\rho$ is any solution of the associated 
Stieltjes problem, then
$$
f_- (x) \leq \int_0^\infty \frac{d\rho(y)}{1+xy} \leq f_+ (x) 
\tag 1.25
$$
for any $x\in [0,\infty)$. In any event, for $x>0$, $f^{[N,N]} 
(x)$ is monotone decreasing in $N$ and $f^{[N-1,N]}(x)$ is 
monotone increasing.
\endproclaim

In terms of the language of Section~3, $f_-$ is associated to 
the Friedrichs solution and $f_+$ to the Krein solution. For an 
interesting application of Theorem~6, see [\loe, \simap]. In 
Section~5, we will also discuss series of Hamburger and prove 
the new result that the $f^{[N,N]}(z)$ Pad\'e approximants always 
converge in that case.

As a by-product of the proof of Theorem~6, we will find 

\proclaim{Theorem 7} Let $\{\gamma_n\}^\infty_{n=0}$ be a set 
of Stieltjes moments. Then the Stieltjes problem is indeterminate 
if and only if
$$
\sum_{n=0}^\infty |P_n (0)|^2 < \infty \tag 1.26
$$
and
$$
\sup_n \biggl| \frac{Q_n(0)}{P_n(0)}\biggr| < \infty . \tag 1.27
$$
\endproclaim

\proclaim{Theorem 8} Let $\{\gamma_n\}_{n=0}^\infty$ be a set 
of Stieltjes moments. Then the Stieltjes problem is determinate 
while the Hamburger problem is indeterminate if and only if 
$$
\sum_{n=0}^\infty |Q_n (0)|^2 < \infty \tag 1.28 
$$
and
$$
\lim_{n\to\infty} \biggl| \frac{Q_n(0)}{P_n(0)}\biggr| = \infty . 
\tag 1.29
$$
\endproclaim

We will see that (1.26)/(1.27) are equivalent to a criterion of 
Stieltjes.

Theorem~6 is proven in Section~5 as parts of Theorem~5.2, 
Proposition~5.6, Proposition~5.8, and Proposition~5.11. Theorem~7 
is Theorem~5.21 and Theorem~8 is Proposition~5.22. 

In Section~6, which synthesizes and extends results in 
Akhiezer [\akh], we will study when the closure of the 
polynomials has finite codimension, and among our results proven 
as part of Theorem~6.4 is 

\proclaim{Theorem 9} Let $\rho$ lie in $\Cal M^H (\gamma)$ and 
let $\Cal H_0$ be the closure of the set of polynomials in 
$\Cal H_\rho \equiv L^2 (\Bbb R, d\rho)$. Then $\Cal H^\bot_0$ 
has finite dimension if and only if the Herglotz function, $\Phi$, 
in the Nevanlinna parametrization of Theorem~4 is a real rational 
function \rom(i.e., a ratio of real poynomials\rom). Equivalently, 
if and only if the measure $\mu$ of {\rom{(1.19)}} has finite support.
\endproclaim

In fact, we will see that $\dim(\Cal H^\bot_0)$ is the degree of 
the rational function $\Phi$.

One consequence of this will be the following, proven in 
Appendix~B (see Theorems~B.1 and B.4).

\proclaim{Theorem 10} $\Cal M^H(\gamma)$ is a compact convex 
set \rom(in the weak topology as a subset of the dual of $C(\Bbb R
\cup\{\infty\})$\rom). Its extreme points are dense in the set. 
\endproclaim

\remark{Remark} Do not confuse this density with the fact that 
the Krein-Millman theorem says that the convex combinations of 
the extreme points are dense. Here we have density without  
convex combinations, to be compared with the fact that in many 
other cases (e.g., probability measures on a compact Hausdorff 
space), the extreme points are closed.
\endremark

\vskip 0.1in

Here is a sketch of the contents of the rest of this paper. In 
Section~2, we review von Neumann's theory of self-adjoint 
extensions and use it to prove the uniqueness result (Theorem~2) 
in the Hamburger case. In Section~3, we review the 
Birman-Krein-Vishik theory of semi-bounded self-adjoint 
extensions and use it to prove the uniqueness result (Theorem~3) 
in the Stieltjes case. In Section~4, we turn to a detailed 
analysis of the polynomials $P$ and $Q$ in terms of transfer 
matrices for the difference equation associated to the Jacobi 
matrix and prove Theorems~3, 4, and 5. In Section~5, we discuss 
Pad\'e approximations and prove Theorems~6, 7, and 8. In Section~6, 
we discuss solutions, $\rho$, where the closure of the polynomials 
has finite codimension in $L^2 (\Bbb R, d\rho)$. Appendix~A is 
devoted to a variety of explicit formulae in terms of 
determinants of moments, and Appendix~B to the structure of 
$\Cal M^H(\gamma)$ as a compact convex set. Appendix~C summarizes 
notation and some constructions.

\vskip 0.3in
\flushpar{\bf \S 2. The Hamburger Moment Problem as a 
Self-Adjointness Problem}
\vskip 0.1in

Let us begin this section with a brief review of the von 
Neumann theory of self-adjoint extensions. For further details, 
see [\ag, \ds, \rsI, \rsII]. We start out with a densely defined 
operator $A$ on a Hilbert space $\Cal H$, that is, $D(A)$ is a 
dense subset of $\Cal H$ and $A:D(A)\to\Cal H$ a linear map. We 
will often consider its graph $\Gamma(A)\subset\Cal H \times 
\Cal H$ given by $\Gamma(A)=\{(\varphi, A\varphi)\mid\varphi\in 
D(A)\}$. Given operators $A,B$, we write $A\subset B$ and say $B$ 
is an {\it{extension}} of $A$ if and only if $\Gamma(A)\subset
\Gamma(B)$. 

One defines a new operator $A^*$ as follows: $\eta\in D(A^*)$ 
if and only if there is a $\psi\in\Cal H$ so that for all $\varphi 
\in D(A)$, $\langle \psi,\varphi\rangle = \langle\eta, A\varphi
\rangle$. We set $A^*\eta =\psi$. In essence, $A^*$, called 
the {\it{adjoint}} of $A$, is the maximal object obeying
$$
\langle A^*\eta, \varphi\rangle = \langle \eta, A\varphi\rangle 
\tag 2.1
$$
for all $\eta\in D(A^*)$, $\varphi\in D(A)$.

An operator is called {\it{closed}} if and only if $\Gamma(A)$ is 
a closed subset of $\Cal H\times\Cal H$ and it is called 
{\it{closable}} if and only if $\overline{\Gamma(A)}$ is the 
graph of an operator, in which case we define $\bar A$, the 
{\it{closure}} of $A$, by $\Gamma(\bar A)=\overline{\Gamma(A)}$. 
Thus, $D(\bar A)=\{\varphi\in\Cal H\mid\exists \varphi_n\in D(A),  
\text{ so that } \varphi_n\to\varphi \text{ and } A\varphi_n 
\text{ is Cauchy}\}$ and one sets $A\varphi = \lim A\varphi_n$.

Adjoints are easily seen to be related to these notions: $A^*$ 
is always closed; indeed,
$$
\Gamma (A^*)=\{(A\varphi, -\varphi)\in \Cal H \times \Cal H 
\mid \varphi\in D(A)\}^\bot . \tag 2.2
$$
$A^*$ is densely defined if and only if $A$ is closable, in 
which case (by (2.2)), $\bar A = (A^*)^*$.

An operator is called {\it{symmetric}} if $A\subset A^*$ 
(equivalently, if $\langle\varphi, A\psi\rangle = \langle A
\varphi, \psi\rangle$ for all $\varphi,\psi\in D(A)$), 
{\it{self-adjoint}} if $A=A^*$, and {\it{essentially self-adjoint}}  
if $\bar A$ is self-adjoint (equivalently, if $A^*$ is self-adjoint). 
Notice that symmetric operators are always closable since $D(A^*)
\supset D(A)$ is dense. Note also that if $A$ is symmetric, then 
$A^*$ is symmetric if and only if $A$ is essentially self-adjoint.

Von Neumann's theory solves the following fundamental question: 
Given a symmetric operator $A$, when does it have self-adjoint 
extensions, are they unique, and how can they be described? If 
$B$ is a self-adjoint extension of $A$, then $B$ is closed, so 
$\bar A \subset B$. Thus, looking at self-adjoint extensions 
of $A$ is the same as looking at self-adjoint extensions of 
$\bar A$.  So for now, we will suppose $A$ is a closed symmetric 
operator. 

Define $\Cal K_\pm = \ker(A^* \mp i)$, that is, $\Cal K_+ = 
\{\varphi\in\Cal H \mid A^*\varphi = i\varphi\}$. They are called 
the {\it deficiency subspaces}. Using $\|(A\pm i)\varphi\|^2 = 
\|A\varphi\|^2 + \|\varphi\|^2$, it is easy to see that if $A$ 
is closed, then $\text{Ran}(A\pm i)$ are closed, and since 
$\ker(A^*\mp i)=\text{Ran}(A\pm i)^\bot$, we have
$$
\text{Ran}(A\pm i) = \Cal K^\bot_\pm . \tag 2.3
$$
Let $d_\pm = \dim(\Cal K_\pm)$, the {\it{deficiency indices}}  
of $A$.

Place the graph norm on $D(A^*)$, that is, 
$\|\varphi\|^2_{A^*} = \langle \varphi, \varphi\rangle + 
\langle A^*\varphi, A^*\varphi\rangle$. This norm comes from an 
inner product, $\langle \varphi,\psi\rangle_{A^*} = \langle 
\varphi,\psi\rangle + \langle A^*\varphi, A^*\psi\rangle$.

\proclaim{Proposition 2.1} Let $A$ be a closed symmetric operator. 
Then
$$
D(A^*) = D(A) \oplus \Cal K_+ \oplus \Cal K_- , \tag 2.4
$$
where $\oplus$ means orthogonal direct sum in the $\langle \, 
\cdot\, , \,\cdot\,\rangle_{A^*}$ inner product.
\endproclaim

\demo{Proof} If $\varphi\in\Cal K_+$ and $\psi\in\Cal K_-$, 
then $\langle\varphi, \psi\rangle_{A^*} = \langle\varphi,\psi 
\rangle + \langle i\varphi, -i\psi\rangle =0$ so $\Cal K_+ 
\bot_{A^*} \Cal K_-$. If $\varphi\in D(A)$ and $\psi\in 
\Cal K_\pm$, then $\langle A^*\varphi, A^*\psi\rangle = 
\langle A\varphi, \pm i\psi\rangle = \langle \varphi, \pm i 
A^* \psi\rangle = -\langle \varphi, \psi\rangle$, so $D(A) 
\bot_{A^*} \Cal K_+ \oplus \Cal K_-$.

Let $\eta\in D(A^*)$. By (2.3), $\text{Ran}(A+i)+ \Cal K_+ = 
\Cal H$ so we can find $\varphi\in D(A)$ and $\psi\in \Cal K_+$, 
so $(A^* +i)\eta = (A+i)\varphi + 2i\psi$. But then $(A^* + i) 
[\eta - \varphi -\psi] =0$, so $\eta - \varphi - \psi\in\Cal K_-$, 
that is, $\eta\in D(A) + \Cal K_+ + \Cal K_-$. \qed
\enddemo

\proclaim{Corollary 2.2} \rom{(i)} Let $A$ be a closed symmetric 
operator. Then $A$ is self-adjoint if and only if $d_+ = d_- =0$. 

\rom{(ii)} Let $A$ be a symmetric operator. Then $A$ is essentially 
self-adjoint if and only if $d_+ = d_- =0$.
\endproclaim

If $A\subset B$, then $B^* \subset A^*$, so if $B$ is symmetric, 
then $A\subset B\subset B^* \subset A^*$. Thus, to look for 
symmetric extensions of $A$, we need only look for operators $B$ 
with $A\subset B\subset A^*$, that is, for restriction of $A^*$ to 
$D(B)$'s with $D(A)\subset D(B)$. By Proposition~2.1, every 
such $D(B)$ has the form $D(A)+S$ with $S\subset\Cal K_+ + 
\Cal K_-$. On $D(A^*)\times D(A^*)$, define the sesquilinear form 
(sesquilinear means linear in the second factor and anti-linear 
in the first)
$$
Q(\varphi,\psi) = \langle\varphi, A^*\psi\rangle_{\Cal H} - 
\langle A^* \varphi,\psi\rangle_{\Cal H}.
$$ 

\proclaim{Proposition 2.3} Let $A$ be a closed symmetric 
operator. Then
\roster
\item"\rom{(i)}" The operators $B$ with $A\subset B\subset A^*$ 
are in one-one correspondence with subspaces $S$ of $\Cal K_+ 
+\Cal K_-$ under $D(B)=D(A)+S$.
\item"\rom{(ii)}" $B$ is symmetric if and only if $Q\restriction 
D(B)\times D(B)=0$.
\item"\rom{(iii)}" $B$ is symmetric if and only if $Q\restriction 
S\times S=0$.
\item"\rom{(iv)}" $B$ is closed if and only if $S$ is closed in 
$\Cal K_+ \oplus \Cal K_-$ in $D(A^*)$ norm.
\item"\rom{(v)}" $\varphi\in D(A^*)$ lies in $D(B^*)$ if and 
only if $Q(\varphi,\psi)=0$ for all $\psi\in D(B)$.
\item"\rom{(vi)}" Let $J:\Cal K_+ \oplus \Cal K_- \to \Cal K_+ 
\oplus \Cal K_-$ by $J(\varphi, \psi) = (\varphi, -\psi)$. 
If $D(B) = D(A)+S$, then $D(B^*)=D(A)+J[S]^\bot$, where $^\bot$ 
is in $\Cal K_+ \oplus \Cal K_-$ in the $D(A^*)$ norm.
\item"\rom{(vii)}" $\Cal K_+(B) = \Cal K_+ \cap S^\bot$, 
$\Cal K_-(B) = \Cal K_- \cap S^\bot$ \rom(with $^\bot$ in the 
$\langle \, , \, \rangle_{A^*}$ inner product\rom).
\endroster
\endproclaim

\demo{Proof} We have already seen (i) holds and (ii) is obvious. 
(iii) holds since if $\varphi\in D(A)$ and $\psi\in D(A^*)$, then 
$Q(\varphi, \psi)=0$ by definition of $A^*$ and symmetry of $A$. 
Thus, if $\varphi_1, \varphi_2 \in D(A)$ and $\psi_1, \psi_2 \in 
S$, then $Q(\varphi_1 + \psi_1, \varphi_2 + \psi_2) = Q(\psi_1, 
\psi_2)$. (iv) is immediate if one notes that $\Gamma (B) = 
\Gamma(A)\oplus \{(\varphi, A^*\varphi)\mid \varphi\in S\}$ 
with $\oplus$ in $\Cal H\times\Cal H$ norm and that the 
$\Cal H\times\Cal H$ norm on $\{(\varphi, A^*\varphi)\mid 
\varphi\in S\}$ is just the $D(A^*)$ norm. (v) follows from the 
definition of adjoint.

To prove (vi), let $\eta = (\eta_1, \eta_2)$, $\varphi =
(\varphi_1, \varphi_2)\in \Cal K_+ \oplus\Cal K_-$. Then, direct 
calculations show that
$$\align
Q(\eta, \varphi) &= 2i [\langle\eta_1, \varphi_1\rangle_{\Cal H} 
- \langle \eta_2, \varphi_2\rangle_{\Cal H}] \tag 2.5 \\
\langle \eta, \varphi\rangle_{A^*} &= 2[\langle \eta_1, 
\varphi_1\rangle_{\Cal H} + \langle\eta_2, \varphi_2 
\rangle_{\Cal H}]. \tag 2.6
\endalign
$$
Thus, $Q(\eta, \varphi)=0$ if and only if $\eta\bot J\varphi$. 
(v) thus implies (vi).

To prove (vii), note that by (vi), $\Cal K_\pm(B) = \Cal K_\pm 
\cap D(B^*)=\Cal K_\pm\cap J[S]^\bot =\Cal K_\pm \cap S^\bot$ 
since $\varphi\in\Cal K_\pm$ lies in $J[S]^\bot$ if and only 
if $J\varphi$ lies in $S^\bot$ if and only if $\varphi$ lies 
in $S^\bot$ by $J\varphi = \pm\varphi$ if $\varphi\in\Cal K_\pm$. 
\qed
\enddemo

With these preliminaries out of the way, we can prove 
von~Neumann's classification theorem:

\proclaim{Theorem 2.4} Let $A$ be a closed symmetric operator. 
The closed symmetric extensions $B$ of $A$ are in one-one 
correspondence with partial isometries of $\Cal K_+$ into 
$\Cal K_-$, that is, maps $U:\Cal K_+ \to \Cal K_-$ for which 
there are closed subspaces $\Cal U_+\subset\Cal K_+$ and $\Cal U_- 
\subset\Cal K_-$ so that $U$ is a unitary map of $\Cal U_+ \to 
\Cal U_-$ and $U\equiv 0$ on $\Cal U^\bot_+$. If $B$ 
corresponds to $U$, then $D(B) =D(A) + \{\varphi+U\varphi\mid 
\varphi\in \Cal U_+\}$. Moreover, $\Cal K_\pm(B) = \Cal K_\pm
\cap \Cal U^\bot_\pm$ \rom($\Cal H$ orthogonal complement\rom). 
In particular, $A$ has self-adjoint extensions if and only if 
$d_+ =d_-$ and then the self-adjoint extensions are in one-one 
correspondence with unitary maps $U$ from $\Cal K_+$ to $\Cal K_-$.
\endproclaim

\demo{Proof} Let $B$ be a symmetric extension of $A$. Let 
$\varphi\in D(B)\cap (\Cal K_+ \oplus\Cal K_-)$, say $\varphi 
= (\varphi_1 + \varphi_2)$. By (2.5), $Q(\varphi,\varphi) = 
2i(\|\varphi_1\|^2 - \|\varphi_2\|^2)$, so $B$ symmetric 
implies that $\|\varphi_1\| = \|\varphi_2\|$, so $D(B)\cap 
(\Cal K_+ \oplus\Cal K_-)=\{\text{some }(\varphi_1, \varphi_2)\}$, 
where $\|\varphi_1\| = \| \varphi_2\|$. Since $D(B)$ is a subspace, 
one cannot have $(\varphi_1, \varphi_2)$ and $(\varphi_1, 
\varphi'_2)$ with $\varphi_2 \neq \varphi'_2$ (for then 
($0, \varphi_2 - \varphi'_2)\in D(B)$ so $\| \varphi_2 
-\varphi'_2\|=0$). Thus, $D(B)\cap (\Cal K_+ \oplus\Cal K_-)$ is 
the graph of a partial isometry. On the other hand, if $U$ is a 
partial isometry, $Q(\varphi +U\varphi, \psi + U\psi) = 2i 
(\langle \varphi, \psi\rangle - \langle U\varphi, U\psi\rangle) 
=0$ so each $U$ does yield a symmetric extension.

That $\Cal K_\pm(B)=\Cal K_\pm \cap \Cal U^\bot_\pm$ follows 
from (vii) of the last proposition if we note that if $\varphi 
\in\Cal K_+$, then $\varphi\perp_{A^*}\{(\psi+U\psi) \mid 
\psi\in\Cal U_+\}$ if and only if $\varphi\perp_{A^*}\{\psi\mid 
\psi\in\Cal U_+\}$ if and only if $\varphi\perp_\Cal H \Cal U_+$ 
(by (2.6)).

Thus $B$ is self-adjoint if and only if $U$ is a unitary from 
$\Cal K_+$ to $\Cal K_-$, which completes the proof. \qed
\enddemo

Recall that a map $T$ is {\it anti-linear} if $T(a\varphi + 
b\psi)=\bar a T(\varphi) + \bar b T(\psi)$ for $a,b\in 
\Bbb C$, $\varphi,\psi\in\Cal H$; that $T$ is {\it anti-unitary} 
if it is anti-linear, a bijection, and norm-preserving; and that 
a {\it complex conjugation} is an anti-unitary map whose 
sequence is $1$.

\proclaim{Corollary 2.5} Let $A$ be a symmetric operator. Suppose 
there exists a complex conjugation $C:\Cal H\to\Cal H$ so that 
$C:D(A)\to D(A)$ and $CA\varphi = AC\varphi$ for all $\varphi 
\in D(A)$. Then $A$ has self-adjoint extensions. If $d_+ =1$, 
every self-adjoint extension $B$ is real, that is, obeys $C:
D(B) \to D(B)$ and $CB\varphi=BC\varphi$ for all $\varphi 
\in D(B)$. If $d_+ \geq 2$, there are non-real self-adjoint 
extensions.
\endproclaim

\demo{Proof} $C$ is an anti-unitary from $\Cal K_+$ to 
$\Cal K_-$ so $d_+ = d_-$.

We claim that if $B$ is self-adjoint and is associated to $U:
\Cal K_+\to\Cal K_-$, then $B$ is real if and only if $CUCU=1$ 
for it is easy to see that $C:D(A^*)\to D(A^*)$ and $CA^*\varphi 
= A^*C\varphi$, so $B$ is real if and only if $C$ maps $D(B)$ to 
itself. Thus, $C(\varphi +U\varphi)$ must be of the form $\psi +
U\psi$. Since $\psi$ must be $CU\varphi$, this happens if and only 
if $C\varphi =UCU\varphi$, that is, $CUCU\varphi=\varphi$. This 
proves our claim.

If $d_+=1$ and $\varphi\in\Cal K_+$, $CU\varphi=e^{i\theta}
\varphi$ for some $\theta$. Then $(CUCU)\varphi = Ce^{i\theta} 
U\varphi=e^{-i\theta}CU\varphi=\varphi$ showing that every 
self-adjoint extension is real.

If $d_+ \geq 2$, pick $\varphi,\psi\in\Cal K_+$ with $\varphi 
\perp\psi$ and let $U\varphi =C\psi$, $U\psi=iC\varphi$. 
Then $CUCU\varphi =CUCC\psi=CU\psi = CiC\psi=-i\varphi$, 
so $CUCU\not\equiv 1$. Thus, the $B$ associated to this $U$ 
will not be real. \qed
\enddemo

Next we analyze a special situation that will always apply to 
indeterminate Hamburger moment problems.

\proclaim{Theorem 2.6} Suppose that $A$ is a closed symmetric 
operator so that there exists a complex conjugation under which 
$A$ is real. Suppose that $d_+ =1$ and that $\ker(A)=\{0\}$, 
$\dim\ker(A^*)=1$. Pick $\varphi\in\ker(A^*)$, $C\varphi=
\varphi$, and $\eta\in D(A^*)$, not in $D(A)+\ker(A^*)$. Then
$\langle \varphi, A^*\eta\rangle\neq 0$ and $\psi = \{\eta - 
[\langle\eta, A^*\eta\rangle/\langle\varphi, A^*\eta\rangle]
\varphi\}/ \langle\varphi, A^*\eta\rangle$ are such that in 
$\varphi,\psi$ basis, $\langle \,\cdot\, , A^*\,\cdot\,\rangle$ 
has the form
$$
\langle\,\cdot\, , A^*\,\cdot\,\rangle = 
\pmatrix 0 & 1 \\ 0 & 0 \endpmatrix . \tag 2.7
$$
The self-adjoint extensions, $B_t$, can be labelled by a 
real number or $\infty$ where
$$\alignat2
D(B_t) &= D(A) + \{\alpha(t\varphi + \psi) \mid \alpha\in 
\Bbb C\} \qquad && t\in\Bbb R \\
&= D(A) + \{\alpha\varphi \mid \alpha\in\Bbb C \} \qquad && 
t=\infty.
\endalignat
$$
The operators $B_t$ are independent of which real $\psi$ in 
$D(A^*)\backslash D(A)$ is chosen so that {\rom{(2.7)}} holds.
\endproclaim

\demo{Proof} If $\langle\varphi, A^*\eta\rangle =0$, then the 
$\langle\,\cdot\, ,A^*\,\cdot\,\rangle$ matrix with basis 
$\varphi, \eta$ would have the form $\left( \smallmatrix a & 0 \\ 
0 & 0 \endsmallmatrix\right)$ with $a\in\Bbb R$, in which case 
$Q\equiv 0$ on the span of $\varphi$ and $\eta$, which is 
incompatible with the fact that $d_+=1$. Thus, $\langle\varphi, 
A^*\eta\rangle \neq 0$ and (2.7) holds by an elementary calculation.

Once (2.7) holds, it is easy to see that the subspaces $S$ of 
$D(A^*)$ with $D(A)\subset S$, $\dim (S/D(A))=1$, and $Q
\restriction S\times S=0$ are precisely the $D(B_t)$.

If $\tilde\psi$ is a second $\psi$ for which (2.7) holds, 
then $Q(\psi -\tilde\psi, \rho)=0$ for all $\rho\in D(A^*)$, 
so $\psi-\tilde\psi\in D(A)$ and the $D(B_t)$'s are the same 
for $\psi$ and $\tilde\psi$. \qed
\enddemo

\remark{Remarks} 1. It is easy to see that if $\ker(A)=\{0\}$, 
then $\dim\ker(A^*)\leq d_+$, so if $d_+=1$ and $\ker(A)=\{0\}$, 
$\dim\ker(A^*)$ is either $0$ or $1$. The example of $A = 
-\frac{d^2}{dx^2}$ on $L^2 (0,\infty)$ with $D(A)=C^\infty_0 
(0,\infty)$ shows that it can happen that $A$ is real, $d_+=1$,  
but $\dim\ker(A^*)=0$.

2. One example where this formalism is natural is if $A\geq 
\alpha >0$ (for $\alpha\in\Bbb R$), in which case $\dim\ker(A^*) 
=d_+$. One takes $\eta= A^{-1}_F \varphi$, where $A_F$ is the 
Friedrichs extension. We will discuss this further in Section~3. 
A second example, as we will see, is the indeterminate moment 
problem, in which case one can take $\varphi = \pi (0)$, 
$\psi = \xi (0)$.
\endremark

\vskip 0.1in

We can now turn to the analysis of the symmetric operator $A$ on 
$\Cal H^{(\gamma)}$ described in Section~1. In terms of the 
explicit basis $P_n (X)$ in $\Cal H^{(\gamma)}$, we know $A$ 
has the Jacobi matrix form (1.16). We will first explicitly 
describe $A^*$ and the form $Q$. Given any sequence $s=
(s_0, s_1, \dots)$, we define a new sequence $\Cal F(s)$ (think 
of $\Cal F$ for ``formal adjoint") by
$$
\Cal F(s)_n = \cases b_0 s_0 + a_0 s_1 & \text{if } n=0 \\
a_{n-1} s_{n-1} + b_ns_n + a_n s_{n+1} & \text{if } n\geq 1. 
\endcases
$$
Given any two sequences, $s_n$ and $t_n$, we define their 
Wronskian,
$$
W(s,t)(n) = a_n(s_{n+1}t_n - s_n t_{n+1}). \tag 2.8
$$
The Wronskian has the following standard property if
$$
a_n s_{n+1} + c_n s_n + a_{n-1} s_{n-1} = 0 \tag 2.9
$$
and
$$
a_n t_{n+1} + d_n t_n + a_{n-1} t_{n-1} =0. \tag 2.10
$$
Then multiplying (2.9) by $t_n$ and subtracting (2.10) 
multiplied by $s_n$, we see that
$$
W(s,t)(n) - W(s,t)(n-1) = (d_n -c_n)s_n t_n. \tag 2.11
$$
In particular, if $d=c$, then $W$ is constant.

\proclaim{Theorem 2.7} Let $A$ be the Jacobi matrix {\rom{(1.16)}} 
with $D(A) =\{s\mid s_n =0$ for $n$ sufficiently large$\}$. Then
$$
D(A^*) =\{s\in\ell_2\mid \Cal F(s)\in\ell_2\} \tag 2.12
$$
with
$$
A^* s=\Cal F(s) \tag 2.13
$$
for $s\in D(A^*)$. Moreover, if $s,t\in D(A^*)$, then
$$
\lim_{n\to\infty} W(\bar s,t)(n) = \langle A^* s,t\rangle 
-\langle s, A^* t\rangle. \tag 2.14
$$
\endproclaim

\demo{Proof} Since the matrix (1.16) is a real-symmetric matrix, 
it is easy to see that for any $t\in D(A)$, and any sequences 
$$
\langle s,At\rangle = \langle \Cal F(s),t\rangle. \tag 2.15
$$
Since $t$ and $At$ are finite sequences, the sum in $\langle 
\, \cdot\, , \, \cdot\,\rangle$ makes sense even if $s$ or 
$\Cal F(s)$ are not in $\ell_2$. Since $D(A)$ is dense in $\ell_2$, 
(2.15) says that $s\in\ell_2$ lies in $D(A^*)$ precisely if 
$\Cal F(s)\in \ell_2$ and then (2.13) holds. That proves the 
first part of the theorem.

For the second, we perform a calculation identical to that 
leading to (2.11):
$$\align
\sum_{n=0}^N \left[\, \overline{\Cal F s_n}\, t_n - 
\overline{s_n}\, (\Cal Ft_n)\right] &= W(\bar s,t)(0) + 
\sum_{n=1}^N W(\bar s,t)(n)-W(\bar s,t)(n-1) \\
&= W(\bar s,t)(N). \tag 2.16
\endalign
$$
If $s,t\in D(A^*)$, then the left side of (2.16) converges to 
$\langle A^* s,t\rangle - \langle s, A^* t\rangle$ as $N\to
\infty$, so (2.14) holds. \qed
\enddemo

\proclaim{Lemma 2.8} If $\varphi\in D(A^*)$, $(A^*-z)\varphi 
=0$ for some $z$ and $\varphi_0 =0$, then $\varphi\equiv 0$. 
\endproclaim

\demo{Proof} Essentially, this is so because $(A^*-z)s =0$ is a 
second-order difference equation with $s_{-1}\equiv 0$, so 
solutions are determined by $s_0$. Explicitly, suppose $A^* s=zs$ 
and $s_0 =0$. Then
$$\alignat2
s_{n+1} &= a^{-1}_n [(-b_n +z)s_n - a_{n-1} s_{n-1}], \qquad 
&& n\geq 1 \\
&= a^{-1}_0 [(-b_0 +z) s_0], \qquad && n=0. 
\endalignat
$$
By a simple induction, $s\equiv 0$. \qed
\enddemo

\proclaim{Corollary 2.9} The operator $A$ associated to a Hamburger 
moment problem always has either deficiency indices $(1,1)$ or 
deficiency indices $(0,0)$.
\endproclaim

\demo{Proof} We have already seen that $d_+ =d_-$ so it suffices 
to show that $\ker(A^* -i)$ has dimension at most $1$. By Lemma~2.8, 
the map from $\ker(A^* -i)$ to $\Bbb C$ that takes $s\mapsto s_0$ 
is one-one, so $\ker(A^* -i)$ is of dimension $0$ or $1$. \qed
\enddemo

We are now ready to prove the more subtle half of the Hamburger 
part of Theorem~2.

\proclaim{Theorem 2.10 (one quarter of Theorem~2)} Let $A$ be 
essentially self-adjoint. Then the Hamburger moment problem has 
a unique solution. 
\endproclaim

\demo{Proof} Pick $z$ with $\text{Im}\, z>0$ and a measure $\rho$ 
obeying (1.1). Since $A$ is essentially self-adjoint, $(A-i)
[D(A)]$ is dense in $\ell^2$, and by an identical argument, 
$(A-z)[D(A)]$ is dense in $\ell^2$. Thus, there exists a 
sequence $R^{(z)}_n (X)$ of polynomials so that
$$
\|(X-z)R^{(z)}_n (X)-1\| \to 0
$$
in $\Cal H^{(\gamma)}$, and thus 
$$
\int [(x-z) R^{(z)}_n (x) -1]^2\, d\rho(x) \to 0
$$
since $d\rho(x)$ realizes $\Cal H^{(\gamma)}$ on polynomials. 
Now $\frac1{x-z}$ is bounded for $x\in\Bbb R$ since $\text{Im} 
\, z>0$. Thus
$$
\int \biggl|R^{(z)}_n (x) - \frac{1}{x-z}\biggr|^2\, 
d\rho(x)\to 0.
$$
It follows that
$$
G_\rho (x) \equiv \int\frac{d\rho(x)}{x-z} = \lim_{n\to\infty} \int 
R^{(z)}_n(x)\, d\rho(x)
$$
is independent of $\rho$. Since $G_\rho(x)$ determines $\rho$ 
(because $\rho(\{a\})=\lim_{\varepsilon\downarrow 0} \varepsilon \, 
\text{Im}\, G_\rho (a+i\varepsilon)$ and $\rho ((a,b)) + \rho 
([a,b])=\frac2{\pi} \lim_{\varepsilon\downarrow 0} \int_a^b
\text{Im}\, G_\rho (y+i\varepsilon)\, dy$), all $\rho$'s must 
be the same. \qed
\enddemo

\proclaim{Theorem 2.11} Let $A$ be a Jacobi matrix with $D(A)$ 
the sequences of finite support. Suppose $A$ is not essentially 
self-adjoint and $B,F$ are distinct self-adjoint extensions of 
$A$. Then
$$
\biggl( \delta_0, \frac{1}{B-i}\, \delta_0\biggr) \neq 
\biggl( \delta_0, \frac{1}{F-i}\, \delta_0\biggr). \tag 2.17
$$
\endproclaim

\remark{Remarks} 1. The proof shows for any $z\in\Bbb C\backslash 
\Bbb R$, $(\delta_0, (B-z)^{-1}\delta_0) \neq (\delta_0 (F-z)^{-1} 
\delta_0)$. It also works for $z\in\Bbb R$ as long as $z$ is in 
the resolvent set for both $B$ and $F$.

2. In Section~4, we will have a lot more to say about the 
possible values of $(\delta_0, \frac1{B-z}\delta_0)$ as $B$ 
runs through the self-adjoint extensions of $A$.
\endremark

\demo{Proof} We first claim that $\delta_0 \notin\text{Ran}(A-i)$. 
For suppose on the contrary that there exists $\eta\in D(A)$ and 
$\delta_0 = (A-i)\eta$ and that $(A^* +i)\varphi =0$. Then 
$(\delta_0, \varphi) = ((A-i)\eta, \varphi) = (\eta, (A^* +i)
\varphi)=0$. By Lemma~2.8, $\varphi\equiv 0$. Thus, if $\delta_0 
\in\text{Ran}(A-i)$, then $\ker(A^* +i)=\{0\}$, so $A$ is 
essentially self-adjoint. By hypothesis, this is false, so 
$\delta_0 \notin \text{Ran}(A-i)$.

Thus, $(B-i)^{-1}\delta_0$ and $(F-i)^{-1}\delta_0$ are in 
$D(A^*)\backslash D(A)$. Since $\dim(D(B)/D(A)) =\dim(D(F)/D(A)) 
=1$, if these vectors were equal, $D(B)$ would equal $D(F)$, so 
$B=A^*\restriction D(B)$ would equal $F=A^*\restriction D(F)$. 
Thus, $(B-i)^{-1}\delta_0 \neq (F-i)^{-1}\delta_0$.

Let $\eta = (B-i)^{-1}\delta_0 - (F-i)^{-1}\delta_0$. Then
$$
(A^* -i)\eta = (A^* -i)(B-i)^{-1}\delta_0 - (A^* -i)
(F-i)^{-1}\delta_0 = \delta_0 - \delta_0 =0 ,
$$
so $\eta\neq 0$ implies $\eta_0 \neq 0$ by Lemma~2.8. Thus, 
$(\delta_0, \eta) = (\delta_0, (B-i)^{-1}\delta_0) - 
(\delta_0, (F-i)^{-1}\delta_0)\neq 0$. \qed
\enddemo

As a corollary of Theorem~2.11, we have

\proclaim{Theorem 2.12 (two quarters of Theorem~2)} A Hamburger 
moment problem for which $A$ is not essentially self-adjoint 
is indeterminate. A Stieltjes moment problem for which 
$A$ has multiple non-negative self-adjoint extensions is 
indeterminate.
\endproclaim

\demo{Proof} Theorem~2.11 implies that distinct self-adjoint 
extensions lead to distinct spectral measures since
$$
(\delta_0, (B-i)^{-1}\delta_0) = \int \frac{d\mu^B (x)}{x-i} \, , 
$$
where $\mu^B$ is the solution to (1.1) associated to $B$. 
Positive self-adjoint extensions yield solutions of the 
Stieltjes moment problem. \qed
\enddemo

With this machinery available, we can prove the converse of 
Proposition~1.6:

\proclaim{Theorem 2.13} Let $\{\gamma_n\}^\infty_{n=0}$ be a set 
of moments for which the Stieltjes problem has solutions. Let
$$\alignat2
\Gamma_{2m} &= \gamma_m, \qquad && m=0,1,\dots \tag 2.18a \\
\Gamma_{2m+1} &= 0, \qquad && m=0,1,\dots. \tag 2.18b
\endalignat
$$
Then $\{\Gamma_j\}^\infty_{j=0}$ is a determinate Hamburger 
problem if and only if $\{\gamma_n\}^\infty_{n=0}$ is a 
determinate Stieltjes problem.
\endproclaim

\demo{Proof} The proof of Proposition~1.6 shows there is a 
one-one correspondence between solutions of the Stieltjes problem 
for $\gamma$ and those solutions $d\rho$ of the Hamburger 
problem for $\Gamma$ with $d\rho(-x)=d\rho(x)$. We will call 
such solutions symmetric. Thus it suffices to show that if the 
Hamburger problem for $\Gamma$ is indeterminate, then there are 
multiple symmetric solutions.

Let $U$ map $\Bbb C(X)$ to itself by $U[P(X)]=P(-X)$. By 
$\Gamma_{2m+1}=0$, $U$ is unitary, and so extends to a unitary 
map from $\Cal H^{(\Gamma)}$ to itself. Clearly, $U$ maps $D(A) 
=\Bbb C(X)$ to itself and $U\!AU^{-1}=-A$.

Thus, $U$ maps $D(A^*)$ to itself and $U\!A^*U^{-1}=-A^*$. This 
means that $U$ maps $\Cal K_+$ to $\Cal K_-$.

Let $C$ be the complex conjugation under which $A$ is real. Then 
$UC=CU$ is an anti-unitary map of $\Cal K_+$ to itself. So if 
$\varphi$ is a unit vector in $\Cal K_+$, then $UC\varphi = 
e^{i\theta}\varphi$ for some real $\theta$. Thus, $UC
(e^{i\theta/2}\varphi) = e^{-i\theta/2}UC\varphi = e^{i\theta/2} 
\varphi$ so there exists $\psi\neq 0$ in $\Cal K_+$ with 
$UC\psi=\psi$.

In particular, since $UC=CU$, $U(\psi\pm C\psi) = \pm(\psi \pm 
C\psi)$ and therefore, if $B=A^*\restriction D(B)$ and $F=
A^*\restriction D(F)$, where $D(B)=D(A) + [\psi + C\psi]$ and 
$D(F)=D(A)+[\psi - C\psi]$, then $U$ leaves both $D(B)$ and 
$D(F)$ invariant, and so $B$ and $F$ are distinct extensions 
with $U\!BU^{-1}=-B$ and $U\!FU^{-1}=-F$. Their spectral measures 
thus yield distinct symmetric extensions. \qed
\enddemo

\remark{Remark} Since $U(\psi + e^{i\theta}C\psi)=e^{i\theta} 
(\psi + e^{-i\theta}C\psi)$, we see that if $\theta\neq 0,\pi$, 
then $U$ does not leave $D(B_\theta)$ invariant. Thus among 
von~Neumann solutions of the $\Gamma$ problem, exactly two are 
symmetric.
\endremark

As a consequence of our identification of $A^*$, we can prove 
one small part of Theorem~3. Recall that $\pi(x_0)$ is the 
sequence $\{P_n(x_0)\}^\infty_{n=0}$ and $\xi(x_0)$ is the 
sequence $\{Q_n(x_0)\}^\infty_{n=i}$.

\proclaim{Theorem 2.14} Let $\{\gamma_n\}^\infty_{n=0}$ be a set 
of Hamburger moments. If either of the following holds, then 
the Hamburger moment problem is indeterminate.
\roster
\item"\rom{(i)}" For some $x_0\in\Bbb R$, $\pi(x_0)$ and 
$\frac{\partial\pi}{\partial x} (x_0)$ lie in $\ell_2$.
\item"\rom{(ii)}" For some $x_0\in\Bbb R$, $\xi(x_0)$ and 
$\frac{\partial\xi}{\partial x}(x_0)$ lie in $\ell_2$. 
\endroster
\endproclaim

\remark{Remark} Since $P_n(z)$ is a polynomial, $\frac{\partial\pi}
{\partial x}$ makes sense as a sequence.
\endremark

\demo{Proof} In terms of the formal adjoint $\Cal F$, we have
$$\align
\Cal F(\pi(x_0)) &= x_0 \pi(x_0) \tag 2.19 \\
\Cal F \bigg(\frac{\partial\pi}{\partial x}\, (x_0) \biggr) 
&= \pi(x_0) + x_0 \, \frac{\partial \pi}{\partial x}\, 
(x_0) \tag 2.20 \\
\Cal F (\xi(x_0)) &= \delta_0 + x_0 \xi(x_0) \tag 2.21 \\
\Cal F \biggl( \frac{\partial\xi}{\partial x}\, (x_0)\biggr) 
&= \xi(x_0) + x_0 \, \frac{\partial\xi}{\partial x_0}\, . 
\tag 2.22
\endalign
$$
Thus, if (i) holds, we conclude $\pi$, $\frac{\partial\pi}
{\partial x}$ are in $D(A^*)$ and 
$$
(A^* -x_0) \pi(x_0)=0, \qquad \left. (A^* - x_0) \frac{\partial\pi}
{\partial x}\right|_{x=x_0} = \pi(x_0).
$$
If $A^*$ were self-adjoint, then
$$
\| \pi(x_0)\|^2 = \bigg\langle \left. \pi, (A^*-x_0)
\frac{\partial\pi}{\partial x}\right|_{x=x_0}\bigg\rangle = 
\biggl\langle \left. (A^* - x_0) \pi, \frac{\partial\pi}
{\partial x}\right|_{x=x_0} \bigg\rangle =0, 
$$
which is impossible since $P_0 (x_0) =1$. Thus, $A^*$ is not 
self-adjoint and the problem is indeterminate.

If (ii) holds, $\xi, \frac{\partial\xi}{\partial x}$ are in 
$D(A^*)$ and 
$$
(A^* - x_0) \xi(x_0) = \delta_0, \qquad \left. (A^* - x_0) 
\frac{\partial\xi}{\partial x}\right|_{x=x_0} = \xi(x_0).
$$
If $A^*$ were self-adjoint, then
$$\align
\|\xi(x_0)\|^2 &= \bigg\langle \left. \xi(x_0), (A^* - x_0) 
\frac{\partial\xi}{\partial x}\right|_{x=x_0} \bigg\rangle \\
&= \bigg\langle\left. (A^* - x_0) \xi(x_0), \frac{\partial \xi}
{\partial x}\right|_{x=x_0} \bigg\rangle \\
&= \left. \frac{\partial Q_0(x)}{\partial x} \right|_{x=x_0} 
=0
\endalign
$$
since $Q_0(x_0)\equiv 0$. This is impossible since $Q_1(x_0)\neq 0$ 
and so again, the problem is indeterminate. \qed
\enddemo

While we have used the theory of essential self-adjointness 
to study the moment problem, following Nussbaum [\nusI], one can 
turn the analysis around:

\definition{Definition} Let $B$ be a symmetric operator on a 
Hilbert space $\Cal H$ and let $\varphi\in C^\infty (B)= 
\cap_n D(B^n)$. $\varphi$ is called {\it{a vector of uniqueness 
for}} $B$ if and only if the Hamburger moment problem  for 
$$
\gamma_n = \frac{(\varphi, B^n \varphi)}{\|\varphi\|^2} \tag 2.23
$$
is determinate.
\enddefinition

\proclaim{Theorem 2.15 (Nussbaum [\nusI])} Let $B$ be a 
\rom(densely defined\rom) symmetric operator on a Hilbert space 
$\Cal H$. Suppose that $D(B)$ contains a dense subset of vectors 
of uniqueness. Then $B$ is essentially self-adjoint.
\endproclaim

\demo{Proof} Let $\Cal U$ be the vectors of uniqueness for $B$. 
Let $\varphi\in\Cal U$ and let $A_\varphi$ be the restriction 
of $B$ to the closure of the space spanned by 
$\{B^n\varphi\}^\infty_{n=0}$. $A_\varphi$ is unitarily 
equivalent to the Jacobi matrix for the moment problem (2.23). 
Thus, $A_\varphi$ is essentially self-adjoint on $\Cal H_\varphi$, 
the closure of the span of $\{B^n\varphi\}^\infty_{n=0}$. 
Therefore, $\overline{(A_\varphi +i)[D(A_\varphi)]} = 
\Cal H_\varphi$ and, in particular, $\varphi \in 
\overline{(A_\varphi +i)[D(A_\varphi)]} \subset 
\overline{(B+i)[D(B)]}$. It follows that $\Cal U\subset 
\overline{(B+i)[D(B)]}$ and thus, $\overline{\text{Ran}(B+i)} = 
\Cal H$. Similarly, $\overline{\text{Ran}(B-i)}=\Cal H$, so 
$B$ is essentially self-adjoint by Corollary~2.2. \qed
\enddemo

The motivation for Nussbaum was understanding a result of 
Nelson [\nel].

\definition{Definition} Let $B$ be a symmetric operator. 
$\varphi\in C^\infty$ is called
\roster
\item"{}" {\it{analytic}} if $\| B^n \varphi\| \leq CR^n n!$
\item"{}" {\it{semi-analytic}} if $\|B^n \varphi\| \leq 
CR^n (2n)!$ 
\item"{}" {\it{quasi-analytic}} if $\sum_{n=1}^\infty 
\|B^{2n}\varphi\|^{1/2n} <\infty$.
\item"{}" {\it{Stieltjes}} if $\sum_{n=1}^\infty 
\|B^n \varphi\|^{1/2n}<\infty$.
\endroster
\enddefinition

\proclaim{Corollary 2.16} If $B$ is a symmetric operator, then 
any analytic or quasi-analytic vector is a vector of uniqueness. 
If $B$ is also bounded from below, then any semi-analytic or 
Stieltjes vector is a vector of uniqueness. In particular, 
\roster
\item"\rom{(i)}" {\rom{(Nelson [\nel])}} If $D(B)$ contains a 
dense set of analytic vectors, then $B$ is essentially self-adjoint. 
\item"\rom{(ii)}" {\rom{(Nussbaum [\nusII], Masson-McClary 
[\mmc])}} If $D(B)$ contains a dense set of semi-analytic vectors 
and is bounded from below, then $B$ is essentially self-adjoint.
\item"\rom{(iii)}" {\rom{(Nussbaum [\nusI])}} If $D(B)$ contains 
a dense set of quasi-analytic vectors, then $B$ is essentially 
self-adjoint.
\item"\rom{(iv)}" {\rom{(Nussbaum [\nusII], Masson-McClary 
[\mmc])}} If $D(B)$ contains a dense set of Stieltjes vectors 
and is bounded from below, then $B$ is essentially self-adjoint.
\endroster
\endproclaim

\demo{Proof} The first two assertions follow Proposition~1.5, 
Corollary~3.4, and Corollary~4.5. Theorem~2.14 completes the 
proof. \qed
\enddemo

\vskip 0.3in
\flushpar{\bf \S 3. The Stieltjes Moment Problem as a 
Self-Adjointness Problem}
\vskip 0.1in

One goal in this section is to prove the remaining part of 
Theorem~2, namely, that if the operator $A$ has a unique 
non-negative self-adjoint extension, then the Stieltjes problem  
is determinate. In the indeterminate case, we will also introduce 
two distinguished non-negative von~Neumann solutions: the 
Friedrichs and Krein solutions.

The name Friedrichs and Krein for the associated self-adjoint 
extensions is standard. We will naturally carry over the names 
to the measures $d\mu_F$ and $d\mu_K$ that solve the Stieltjes 
problem. Those names are not standard. Krein did important work 
on the moment problem, but not in the context of what we will 
call the Krein solution because of his work on self-adjoint 
extensions.

We begin with a very brief summary of the Birman-Krein-Vishik 
theory of extensions of strictly positive operators (see [\as] 
for a complete exposition). Suppose $A$ is a closed symmetric 
operator and for some $\alpha >0$, $(\varphi, A\varphi) \geq\alpha 
\| \varphi\|^2$ for all $\varphi\in D(A)$. One closes the quadratic 
form analogously to closing the operator, that is, $\varphi\in 
Q(A_F)$ if and only if there is $\varphi_n\in D(A)$ so that 
$\varphi_n \to \varphi$ and $(\varphi_n - \varphi_m, A(\varphi_n 
-\varphi_m))\to 0$ as $n\to\infty$. One then sets $(\varphi, 
A_F \varphi)=\lim(\varphi_n, A\varphi_n)$. The quadratic form
$(\varphi, A_F \varphi)$ on $Q(A_F)$ is closed, that is, 
$Q(A_F)$ is complete in the norm $\sqrt{(\varphi, A_F \varphi)}$. 

Not every quadratic form abstractly defined is closable (e.g., 
on $L^2(\Bbb R, dx)$, the form with form domain $C^\infty_0 
(\Bbb R)$ given by $\varphi\mapsto |\varphi(0)|^2$ is not 
closable), but it is not hard to see that if a quadratic form 
comes from a non-negative symmetric operator, it is closable 
(see Theorem~X.23 in [\rsII] or [\kato], Chapter~VI).

While a closed symmetric operator need not be self-adjoint, it 
is a deep result (see [\kato, \rsII]) that a closed positive quadratic 
form is always the quadratic form of a self-adjoint operator. It 
follows that $A_F$ is the quadratic form of a self-adjoint 
operator, the {\it{Friedrichs extension}} of $A$. It follows that 
$\inf \,\text{spec}(A_F)=\alpha >0$.

Define $N=\ker(A^*)$ which is closed since $A^*$ is a closed 
operator. Then $D(A^*)=D(A)+N + A^{-1}_F [N]$, where $+$ means 
independent sum. In particular, $N\neq \{0\}$ if $A$ is not 
self-adjoint (which we suppose in the rest of this paragraph). 
The {\it{Krein extension}} is the one with $D(A_K)=D(A)+N$. Then 
$Q(A_K)=Q(A_F) + N$. Since $N\subset D(A_K)$, we have that 
$\inf\,\text{spec}(A_K)=0$ so $A_K \neq A_F$. The set of 
non-negative self-adjoint extensions of $A$ is precisely the 
set of self-adjoint operators $B$ with $A_K \leq B \leq A_F$ 
(where $0\leq C\leq D$ if and only if $Q(D) \subset Q(C)$ and 
$(\varphi, C\varphi)\leq (\varphi, D\varphi)$ for all $\varphi\in 
Q(D)$; equivalently, if and only if $(D+1)^{-1}\leq (C+1)^{-1}$ 
as bounded operators). These $B$'s can be completely described 
in terms of closed but not necessarily densely defined quadratic forms, 
$C$, on $N$. One can think of $C=\infty$ on $Q(C)^\bot$. $A_F$ 
corresponds to $C\equiv\infty$ (i.e., $Q(C)=\{0\}$) and $A_K$ to 
$C\equiv 0$. In general, $Q(A_C) = Q(A_F) + Q(C)$ and if $\varphi 
\in Q(A_F)$ and $\eta\in Q(C)$, then $(\varphi + \eta, A_C
(\varphi + \eta)) = (\varphi, A_F \varphi)+(\eta, C\eta)$.

Two aspects of Jacobi matrices (defined on sequences of finite 
support) make the theory special. First, $\dim(N)\leq 1$. 
Second, by Theorem~5 (which we will prove in the next section 
without using any results from the present section), if $A$ is 
not essentially self-adjoint, $A_F$ has a discrete spectrum with 
the property that if $t\in\text{spec}(A_F)$, then $\mu_F 
(\{t\}) > \mu (\{t\})$ for any other solution, $\mu$, of the 
Hamburger moment problem. Moreover, if $t\in\text{spec}(A_F)$, 
$t\notin\text{spec}(B)$ for any other self-adjoint extension, 
$B$.

\proclaim{Proposition 3.1} Let $A$ be a Jacobi matrix defined 
on the sequences of finite support, and suppose that $A$ has 
deficiency indices $(1, 1)$ and that $A$ is non-negative 
\rom(i.e., that the moment problem associated to $A$ has at 
least some solutions of the Stieltjes moment problem\rom). 
Then $A$ has a unique non-negative self-adjoint extension if 
and only if $0$ is an eigenvalue in the spectrum of $A_F$.
\endproclaim

\demo{Proof} Let $\alpha=\inf \,\text{spec}(A_F)$. By hypothesis, 
$\alpha \geq 0$, and by Theorem~5, $\alpha$ is a discrete, 
simple eigenvalue of $A_F$. We want to show that if $\alpha >0$, 
then $A$ has additional non-negative self-adjoint extensions; and 
contrary-wise, if $\alpha =0$, then it has no additional 
non-negative self-adjoint extensions.

If $\alpha >0$, $A\geq \alpha >0$, then $A$ has a Krein extension 
distinct from the $A_F$, so there are multiple non-negative 
self-adjoint extensions.

Suppose $\alpha =0$ and that $B$ is some non-negative self-adjoint 
extension. Since $A_F$ is the largest such extension, $0\leq B 
\leq A_F$, so
$$
(A_F +1)^{-1} \leq (B+1)^{-1} \leq 1 . \tag 3.1
$$
Since $\alpha =0$, $\|(A_F +1)^{-1}\| =1$, so (3.1) implies that 
$\|(B+1)^{-1}\|=1$. Since $B$ has discrete spectrum by Theorem~5, 
this means $0$ is an eigenvalue of $B$. But, by Theorem~5 again, 
$A_F$ is the unique self-adjoint spectrum with a zero eigenvalue. 
Thus, $B=A_F$, that is, there is a unique non-negative 
self-adjoint extension.   \qed
\enddemo

\proclaim{Theorem 3.2 (last quarter of Theorem~2)} Suppose that 
$A$ is a Jacobi matrix on the vectors of finite support and that 
$A$ is non-negative. If $A$ has a unique non-negative self-adjoint 
extension, then the associated Stieltjes moment problem is 
determinate.
\endproclaim

\demo{Proof} Clearly, if $A$ is essentially self-adjoint, then 
the Hamburger problem is determinate and so, a fortiori, the 
Stieltjes problem is determinate. Thus, we need only consider the 
case where $A$ has multiple self-adjoint extensions, but only one 
that is non-negative.

Then $A$ has deficiency indices $(1,1)$. By Proposition~3.1, 
$\alpha =\inf\,\text{spec}(A_F)=0$. Moreover, by Theorem~5, $0$ 
is a pure point of $\mu_F$. Let $\varphi = P_{\{0\}}(A_F)\delta_0$. 
By Lemma~2.8, $\varphi\neq 0$. Then there exists $\varphi_n\in D(A)$ 
so $\varphi_n\to\varphi$ with $\|\varphi_n\|=\|\varphi\|$ and 
$(\varphi_n, A_F \varphi_n)\to (A_F \varphi, \varphi)=0$. If 
$\mu_F (\{0\}) =\tau$, then $\|\varphi\|^2 = \langle \varphi, 
\delta_0\rangle = \tau$ so $\|\varphi_n\|=\sqrt\tau$ and 
$\langle \varphi_n, \delta_0\rangle \to\tau$.

Suppose $\rho$ is some solution of the Stieltjes moment problem. 
Since $\varphi_n\to\varphi$ in $L^2 (d\mu_F)$, $\varphi_n$ is 
Cauchy in $L^2(d\rho)$, and so $\varphi_n \to f$ in $L^2(d\rho)$ 
and $\|f\|=\sqrt\tau$. Since $\int x\varphi_n(x)^2\, d\rho
\to 0$, we conclude that $\int xf^2\, d\rho =0$. Since $\rho$ 
is a measure supported on $[0,\infty)$, we conclude $f(x)=0$ for 
$x\neq 0$, and thus $\rho$ has a pure point at zero also. Since 
$(\varphi_n, \delta_0)\to\tau$, we see that $\int f\, d\rho=\lim 
\int \varphi_n(x)\, d\rho(x)=\lim (\varphi_n, \delta_0)=\tau$. 
But $\int f^2 \, d\rho = \|f\|^2 =\tau$ and $\int f\,d\rho=\tau$. 
Thus $f(0)=1$ and so $\rho(\{0\})=\tau$. But Theorem~5 asserts 
that any solution of the Hamburger problem distinct from $\mu_F$ 
has $\mu_F (\{0\}) <\tau$. We conclude that $\rho=\mu_F$, that 
is, that the solution of the Stieltjes problem is unique. \qed
\enddemo

We will call the solution of the Stieltjes problem associated to 
the Friedrichs extension of $A$ the {\it Friedrichs solution}. 
If the Stieltjes problem is indeterminate, then $A\geq \alpha >0$ 
by Proposition~3.1 and Theorem~3.2. Moreover, since the Hamburger 
problem is indeterminate, $N=\ker(A)=\text{span}[\pi(0)]$ has 
dimension $1$, so the Krein extension exists and is distinct from 
the Friedrichs extension. We will call its spectral measure the 
{\it Krein solution} of the Stieltjes problem.

As discussed in the introduction, every indeterminate Stieltjes 
problem has associated to it an example of a determinate 
Stieltjes problem where the associated Hamburger problem is 
indeterminate. For by Proposition~3.1 and Theorem~3.2, the 
indeterminate Stieltjes problem is associated to an $A$ where 
the bottom of the spectrum of $A_F$ is some $f_0 >0$. The 
moment problem $\{\gamma_m (-f_0)\}$ with $\gamma_m (-f_0)= 
\sum_{j=0}^m \binom{m}{j} \gamma_j (-f_0)^{m-j}$ will be 
determinate Stieltjes (since the associated $A_F$ is the old 
$A_F - f_0$) but indeterminate Hamburger.

We summarize with

\proclaim{Theorem 3.3} Suppose that $\{\gamma_n\}^\infty_{n=0}$ 
is a set of Stieltjes moments. Then, if $\gamma$ is Hamburger 
indeterminate, there is a unique $c_0 \leq 0$ so that $\gamma 
(c_0)$ is Stieltjes determinate. Moreover, $\gamma(c)$ is 
Stieltjes indeterminate if $c>c_0$ and $\gamma(c)$ are not 
Stieltjes moments if $c<c_0$. In particular, if $\gamma(c)$ is 
Stieltjes determinate for two values of $c$, then $\gamma$ is 
Hamburger determinate.
\endproclaim

\proclaim{Corollary~3.4} If $\{\gamma_n\}^\infty_{n=0}$ is a 
set of Stieltjes moments and obeys
$$
|\gamma_n| \leq CR^n (2n)! \tag 3.2
$$
then $\{\gamma_n\}^\infty_{n=0}$ is Hamburger determinate. 
\endproclaim

\demo{Proof} By (3.2),
$$
|\gamma_n (c)| \leq CR^n (2n)! \sum_{j=0}^n \binom{n}{j} c^j 
= C[R(c+1)]^n (2n)!
$$
so by Proposition~1.5, $\{\gamma_n\}^\infty_{n=0}$ is Stieltjes 
determinate for all $c\geq 0$. By Theorem~3.3, it is 
Hamburger determinate.
\enddemo

\vskip 0.3in

\flushpar{\bf \S 4. Transfer Matrices and the Indeterminate 
Moment Problem}
\vskip 0.1in

In this section, we will analyze the indeterminate Hamburger 
problem in great detail, using the second-order difference 
equation associated to the Jacobi matrix $A$. In particular, 
we will prove Theorems~3, 4, and 5.

Throughout, we will fix a set of Hamburger moments 
$\{\gamma_m\}^\infty_{m=0}$. It will be convenient to define a 
linear functional $E(\,\cdot\,)$ on polynomials $P(X)$ by 
$E(X^m)\equiv \gamma_m$. Where we have polynomials $P$ of 
several variables $X,Y,\dots$, we will indicate by $E_X 
(\,\cdot\,)$ the obvious map to polynomials of $Y,\dots$ with 
$$
E_X (X^m Y^{n_1} \dots)=\gamma_m Y^{n_1} \dots .
$$

In (1.14), we defined the set of polynomials $P_n (X)$ by the 
three conditions, $\deg(P_n(X))=n$, $E(P_n (X)P_m(X))=\delta_{nm}$, 
and $P_n(X) = c_{nn}X^n +\cdots$ with $c_{nn} >0$. We noted that 
with $\{a_n\}^\infty_{n=0}$, $\{b_n\}^\infty_{n=0}$, and $a_{-1}
\equiv 1$, the elements of the Jacobi matrix associated to 
$\gamma$, the sequence $u_n = P_n(z)$; $n=0,1,2,\dots$ obeys
$$
a_n u_{n+1} + (b_n -z) u_n + a_{n-1} u_{n-1} =0 \tag 4.1
$$
with
$$
u_{-1} =0, \qquad u_0 =1  . \tag 4.2
$$

We also introduced the sequence of polynomials $Q_n (X)$ of 
degree $n-1$ by requiring that $u_n = Q_n(z)$ solve (4.1) with 
the initial conditions
$$
u_{-1} = -1, \qquad u_0 =0  . \tag 4.3
$$

Notice that the Wronskian $W(Q_\cdot (z), P_\cdot(z))(-1) = a_{-1} 
[Q_0(z) P_{-1}(z) - Q_{-1}(z) P_0(z)]=1$. Thus, by (2.11) we 
have:

\proclaim{Proposition 4.1} 
$$
a_{k-1} [Q_k(z) P_{k-1}(z) - 
Q_{k-1}(z) P_k(z)]\equiv 1
$$
\endproclaim

The following is a useful formula for the $Q$'s in terms of the 
$P$'s. Note that since $\frac{X^n - Y^n}{(X-Y)} = 
\sum_{j=0}^{n-1} X^j Y^{n-1-j}$ for any polynomial $P$, 
$\frac{P(X)-P(Y)}{X-Y}$ is a polynomial in $X$ and $Y$.

\proclaim{Theorem 4.2} For $n\geq 0$, 
$$
E_X \biggl(\frac{P_n(X) - P_n(Y)}{X-Y}\biggr) = Q_n (Y).
$$
\endproclaim

\demo{Proof} $\frac{[P_n(X)-P_n(Y)]}{(X-Y)}$ is a polynomial in 
$X$ and $Y$ of degree $n-1$, so we can define polynomials
$$
R_n(Y)\equiv E_X \biggl(\frac{P_n(X)-P_n(Y)}{X-Y}\biggr).
$$
Subtract the equations,
$$\align
a_n P_{n+1}(X) + b_n P_n(X) + a_{n-1} P_{n-1}(X) &= XP_n(X) \\
a_n P_{n+1}(Y) + b_n P_n (Y) + a_{n-1} P_{n-1}(Y) &= YP_n(Y)
\endalign
$$
and see that for $n=0,1,2,\dots$ with $R_{-1}(Y)\equiv 0$:
$$\align
a_n R_{n+1}(Y) + b_n R_n(Y) + a_{n-1} R_{n-1}(Y) &= 
E_X\biggl(\frac{XP_n(X)-YP_n(Y)}{X-Y}\biggr) \\ 
&= E_X \biggl(\frac{YP_n(X) - YP_n(Y)}{X-Y} + P_n (X)\biggr)\\
&= YR_n (Y) + \delta_{n0}.
\endalign
$$
Replace the abstract variable $Y$ with a complex variable $z$ and 
let $\tilde R_n (z) = R_n(z)$ if $n\geq 0$ but $\tilde R_{-1}(z) 
=-1$. Then
$$
a_n \tilde R_{n+1}(z) + b_n \tilde R_n (z) + a_{n-1} 
\tilde R_{n-1}(z) = z\tilde R_n (z)
$$
(since the $\delta_{n0}$ term becomes $-a_{n-1}\tilde R_{n-1}(z)$ 
for $n=0$). Moreover,
$$
\tilde R_{-1}(z)=-1, \qquad \tilde R_0 (z) = E_X \biggl( 
\frac{1-1}{X-Y}\biggr) =0
$$
so by uniqueness of solutions of (4.1) with prescribed initial 
conditions, $\tilde R_n (z) = Q_n (z)$, that is, for $n\geq 0$, 
$R_n =Q_n$, as claimed. \qed
\enddemo

One consequence of Theorem~4.2 is that $P_n(X)$ and $Q_n(X)$ 
have the same leading coefficient, that is, if $P_n(X) = c_{nn} 
X^n + \cdots$, then $Q_n(X) = c_{nn} X^{n-1} + \cdots$.

The following calculation is part of the standard theory [\akh] 
and plays a critical role in what follows:

\proclaim{Theorem 4.3} Let $\rho$ solve the moment problem and 
assume $z\in\Bbb C\backslash\Bbb R$. Set $\zeta = G_\rho (z) 
\equiv \int\frac{d\rho(x)}{x-z}$. Then
$$
\langle (x-z)^{-1}, P_n (x)\rangle_{L^2 (d\rho)} = 
Q_n (z) + \zeta P_n (z). \tag 4.4
$$
In particular,
$$
 \sum_{n=0}^\infty |Q_n(z) + 
\zeta P_n(z)|^2 \leq \frac{\text{\rom{Im}}\,\zeta}
{\text{\rom{Im}}\, z} \tag 4.5
$$
with equality if $\rho$ is a von~Neumann solution.
\endproclaim

\remark{Remark} We will eventually see (Proposition~4.15 and 
its proof) that the ``if" in the last sentence can be replaced 
by ``if and only if."
\endremark

\demo{Proof}
$$
\int \frac{P_n(x)}{x-z}\, d\rho(x) = \zeta P_n(z) + \int 
\frac{P_n(x) - P_n(z)}{x-z}\, d\rho(x) = \zeta P_n(z) + 
Q_n(z)
$$
by Theorem~4.2. This is just (4.4). (4.5) is just Parseval's 
inequality for the orthonormal set $\{P_n(x)\}^\infty_{n=0}$ in 
$L^2(d\rho)$ if we note that
$$
\int\frac{d\rho(x)}{|x-z|^2} = \frac{1}{z-\bar z} \int d\rho(x) 
\biggl[ \frac{1}{x-z} - \frac{1}{\bar x - \bar z}\biggr] 
=\frac{\text{Im}\, \zeta}{\text{Im}\,z}\,.
$$
If $\rho$ comes from a self-adjoint extension, by construction 
of $\Cal H^{(\gamma)}$, $\{P_n(x)\}^\infty_{n=0}$ is an 
orthonormal basis for $L^2(d\rho)$, so equality holds in (4.5). 
\qed
\enddemo

\remark{Remark} In $\Cal H^{(\gamma)}$, thought of as limits 
of polynomials,  the vector with components $Q_n (z) + \zeta 
P_n(z)$ is represented by
$$
\sum_n [Q_n(z) + \zeta P_n(z)] P_n(X).
$$
This product of $P$'s may seem strange until one realizes that 
$\sum_{n=1}^N P_n (y)P_n(x)$ is a reproducing kernel in $L^2 
(d\rho)$ for polynomials of degree $N$. This links the construction 
to ideas in Landau [\land].
\endremark

Given Theorem~2, the following proves the equivalence of parts 
(i), (ii), (iii), and (vii) for $z\in\Bbb C\backslash\Bbb R$ of 
Theorem~3.

\proclaim{Proposition 4.4} Suppose that $z_0\in\Bbb C
\backslash\Bbb R$. Then the following are equivalent:
\roster
\item"\rom{(i)}" The Jacobi matrix $A$ is not essentially 
self-adjoint.
\item"\rom{(ii)}" $\pi(z_0) = \{P_n(z_0)\}^\infty_{n=0}$ is 
in $\ell^2$.
\item"\rom{(iii)}" $\xi(z_0) = \{Q_n(z_0)\}^\infty_{n=0}$ 
is in $\ell^2$.
\item"\rom{(iv)}" Both $\pi(z_0)$ and $\xi(z_0)$ are in 
$\ell^2$.
\endroster

Moreover, when any of these conditions holds, there is a closed 
disk of positive radius, $\Cal D(z_0)$ , in the same half-plane 
as $z_0$ so that for any solution $\rho$ of the moment problem, 
$\zeta=G_\rho (z_0)\in\Cal D(z_0)$. The values of $\zeta$ 
when $\rho$ is a von~Neumann solution lie on $\partial
\Cal D(z_0)$ and fill this entire boundary. Every point in 
$\Cal D(z_0)$ is the value of $G_\rho(z_0)$ for some 
$\rho$ solving the moment problem.

In addition, if these conditions hold, $\pi(z_0)$ and $\xi(z_0)$ 
lie in $D(A^*)\backslash D(A)$.
\endproclaim

\remark{Remarks} 1. One can show (using (2.11)) that the center 
of the disk is
$$
\lim_{n\to\infty} - \frac{Q_n(z_0)\, 
\overline{P_{n-1}(z_0)} - Q_{n-1}(z_0)\, 
\overline{P_n(z_0)}}{P_n(z_0) \overline{P_{n-1}(z_0)} - 
P_{n-1}(z_0)\, \overline{P_n(z_0)}}
$$
and the radius is
$$
\frac{1}{2 |\text{Im}\, z_0|} \,\, \frac{1}{\|\pi(z_0)\|^2} 
\tag 4.6
$$
but these explicit formulae will play no role in our discussions.

2. We will see later (Theorem~4.14) that if $\rho$ is not a 
von~Neumann solution, then $\zeta\in\Cal D(z_0)^{\text{\rom{int}}}$.
\endremark

\demo{Proof} Since $A$ is real, $A$ fails to be essentially 
self-adjoint if and only if there is a non-zero solution of 
$(A^*-z_0)u=0$. By Theorem~2.6 and the unique solubility of 
second-order difference operators given $u_{-1}=0$ and $u_0$, 
every such solution has $u_n = u_0 P_n(z_0)$ so (i) is equivalent 
to (ii). Let $\zeta = G_\rho(z_0)$ for some von~Neumann solution. 
Then $\xi(z_0) + \zeta \pi(z_0)\in\ell^2$, so (ii) is equivalent 
to (iii) or (iv).

If (i)--(iv) hold, then (4.5) has the form
$$
a|\zeta|^2 + b\zeta + \bar b\bar\zeta + c \leq 0, 
$$
where $a=\|\pi(z_0)\|^2$, $c=\|\xi(z_0)\|^2$, and $b=2 
\langle\xi(z_0), \pi(z_0)\rangle - \frac{i}{2\, \text{Im}\, z_0}$. 
The set where this inequality holds is always a disk, $\Cal D(z_0)$, 
although a priori, it could be empty depending on the values of 
$a,b,c$. However, by Theorem~2.11, we know that there are multiple 
$\zeta$'s obeying (4.5) so the disk must have strictly positive 
radius. By Theorem~4.3, $\zeta$'s for von~Neumann solutions obey 
equality in (4.5) and so lie in $\partial\Cal D(z_0)$. 

The self-adjoint extensions are parametrized by a circle 
(unitaries from $\Cal K_+$ to $\Cal K_-$) in such a way that 
the map from the parametrization to values of $(\delta_0, 
(B_t-z_0)^{-1}\delta_0)$ is continuous. By Theorem~2.11, this map 
is one-one. Every one-one continuous map of the circle to itself 
is surjective, so all of $\partial\Cal D(z_0)$ occurs.

Given a point $\zeta$ in $\Cal D(z_0)^{\text{\rom{int}}}$, find 
$\zeta_0$ and $\zeta_1$ in $\partial\Cal D(z_0)$ and $\theta\in 
(0,1)$ so $\zeta = \theta\zeta_0 + (1-\theta)\zeta_1$. If 
$\mu_0, \mu_1$ are the von~Neumann solutions that have 
$G_{\mu_i} (z_0) = \zeta_i$ ($i=0,1)$, then $\rho =\theta\mu_0 + 
(1-\theta)\mu_1$ is a measure solving the moment problem with 
$G_\rho (z_0)=\zeta$. 

If (iv) holds, then $\pi(\bar z_0), \xi(z_0)\in D(A^*)$. 
By (2.14) and Proposition~1.4, $\langle A^*\pi(\bar z_0), 
\xi(z_0)\rangle \mathbreak - \langle \pi(\bar z_0), A^*
\xi(z_0)\rangle =-1$, so neither $\pi(\bar z_0)$ nor $\xi(z_0)$ 
lies in $D(A)$. \qed
\enddemo

\remark{Remark} The Cayley transform way of looking at 
self-adjoint extensions says that for any $z_0 \in \Bbb C
\backslash\Bbb R$,
$$
(B_t - \bar z_0)(B_t - z_0)^{-1} = V(z_0) + 
e^{i\theta(z_0, t)} X(z_0),
$$
where $V(z_0)$ is the $t$-independent map $(\bar A - \bar z_0) 
(\bar A -z_0)^{-1}$ from $\overline{\text{Ran}(A-z_0)}$ to 
$\overline{\text{Ran}(A-\bar z_0)}$ and $X(z_0)$ is any 
isometry from $\ker(A^* - \bar z_0)$ to $\ker(A^* -z_0)$ 
extended to $\Cal H$ as a partial isometry. $\theta$ is a 
parametrization depending on $z_0, t$ (and the choice of 
$X(z_0)$), but the theory guarantees that as $t$ varies through 
all possible values, $e^{i\theta}$ runs through the full circle 
in an injective manner. Now $(B_t - \bar z_0)(B_t - z_0)^{-1} 
=1 + 2i(\text{Im}\, z_0)(B_t - z_0)^{-1}$, so
$$
(\delta_0, (B_t - z_0)^{-1}\delta_0) = 
[2i(\text{Im}\, z_0)]^{-1} [-1 + (\delta_0, V(z_0)\delta_0) 
+e^{i\theta(z_0, t)} (\delta_0, X(z_0)\delta_0)]
$$
is seen directly to be a circle. Since one can take $X(z_0)$ to be
$$
(\delta_n, X(z_0)\delta_m) = \frac{P_m (\bar z_0) P_n(z_0)}
{\sum_{j=0}^\infty |P_j (z_0)|^2}
$$
and $P_0(z)\equiv 1$, we even see that the radius of the circle 
is given by (4.6). 
\endremark

\proclaim{Corollary 4.5} If a Hamburger problem is indeterminate, 
then $\sum_{n=0}^\infty |a_n|^{-1} <\infty$. In particular, if 
$\sum_{n=0}^\infty |a_n|^{-1} =\infty$, \rom(e.g., $a_n \equiv 
1$\rom), then the Hamburger problem is determinate. If
$$
\sum_{n=1}^\infty \gamma^{-1/2 n}_{2n} = \infty,
$$
then the Hamburger problem is determinate. If
$$
\sum_{n=1}^\infty \gamma^{-1/2 n}_n = \infty 
$$
for a set of Stieltjes moments, that problem is both Stieltjes 
and Hamburger determinate. 
\endproclaim

\remark{Remarks} 1. The last pair of assertions is called 
Carleman's criterion. It generalizes Proposition~1.5 and proves 
uniqueness in some cases where growth doesn't imply analyticity of 
the Fourier transform.

2. See Corollary~5.24 for the Stieltjes analog of the first 
assertion.
\endremark

\demo{Proof} By Proposition~4.1 and the Schwarz inequality,
$$
\sum_{n=0}^N |a_n|^{-1} \leq 2 \biggl( \sum_{n=0}^{N+1} 
|P_n (z)|^2\biggr)^{1/2} \bigg( \sum_{n=0}^{N+1} |Q_n (z)|^2 
\biggr)^{1/2}. 
$$
Thus, divergence of $\sum_{n=0}^\infty |a_n|^{-1}$ implies 
that either $\pi$ or $\xi$ (or both) fail to be $\ell_2$, and so 
determinacy.

Consider Carleman's criterion in the Hamburger case. By induction 
in $n$ and (4.1) starting with $P_0(x)=1$, we see that
$$
P_n(x) = (a_1 \dots a_n)^{-1} x^n + \text{lower order},
$$
so $\langle(a_1 \dots a_n)^{-1}x^n, P_n(x)\rangle =1$ and 
thus, by the Schwartz inequality,
$$
1 \leq \gamma_{2n} (a_1 \dots a_n)^{-2}
$$
hence
$$
\gamma^{-1/2n}_{2n} \leq (a_1 \dots a_n)^{-1/n}.
$$
Therefore our result follows from the divergence criteria proven at 
the start of the proof and the inequality
$$
\sum_{j=1}^n (a_1 \dots a_j)^{-1/j} \leq 2e \sum_{j=1}^n a^{-1}_j
$$
(which is actually known to hold with $e$ in place of $2e$).

To prove this, note first that $1+x \leq e^x$ so $(1+n^{-1})^n 
\leq e$ so using induction, $n^n \leq e^n n!$, so by the 
geometric-arithmetic mean inequality,
$$\align
(a_1 \dots a_j)^{-1/j} &= (a^{-1}_1 2a^{-1}_2 \dots 
ja^{-1}_j)^{1/j} (j!)^{-1/j} \\
&\leq \frac{e}{j^2} \sum_{k=1}^j ka^{-1}_k.
\endalign
$$
Thus,
$$
\sum_{j=1}^n (a_1 \dots a_j)^{-1/j} \leq e \sum_{k=1}^n 
a^{-1}_k \biggl( \sum_{j=k}^n \frac{k}{j^2}\biggr) \leq 
2e \sum_{j=1}^n a^{-1}_j
$$
since
$$
\sum_{j=k}^\infty \frac{k}{j^2} \leq \frac{1}{k} + k 
\int_k^\infty \frac{dy}{y^2} \leq 2. 
$$

By Proposition~1.6, Carleman's criterion in the Hamburger case 
means that if a set of Stieltjes moments obeys $\sum_{n=1}^\infty 
(\gamma_n)^{-1/2n}=\infty$, then it is Stieltjes determinate. Thus 
by Theorem~3.3, it suffices to show that
$$
\sum_{n=1}^\infty (\gamma_n (c))^{-1/2n} \geq D(c) 
\sum_{n=1}^\infty \gamma_n^{-1/2n}
$$
for $c>0$ and $D(c) < \infty$ to conclude Hamburger determinacy. 

Note first that $\frac{\gamma_{j+1}}{\gamma_j} \geq 
\frac{\gamma_j}{\gamma_{j-1}}$ by the Schwarz inequality. Thus, 
if $\alpha = \frac{\gamma_0}{\gamma_1}$, we have that 
$$
\gamma_{n-j} \leq \gamma_n \alpha^j.
$$
Therefore,
$$
\gamma_n (c) \leq \gamma_n \sum_{j=0}^n \binom{n}{j} c^j 
\alpha^j = (1 + c\alpha)^n \gamma_n
$$
and so
$$
\sum_{n=1}^\infty (\gamma_n (c))^{-1/2n} \geq 
(1 + c\alpha)^{-1/2} \sum_{n=1}^\infty \gamma_n^{-1/2n}. \qed
$$
\enddemo

To complete the proof of Theorem~3, we need to show that if every 
solution of $[(A^* -z_0)u]_n =0$ ($n\geq 1$) is $\ell^2$ for a 
fixed $z=z_0$, the same is true for all $z\in\Bbb C$ and that 
$\frac{\partial\pi}{\partial x}$, $\frac{\partial\xi}{\partial x}$ 
are in $\ell^2$. We will do this by the standard method of 
variation of parameters. So we write a general solution of 
$[(A^*-z)u]_n =0$ ($n\geq 1$) in terms of $P_n(z_0)$ and 
$Q_n(z_0)$ using
$$
\binom{u_{n+1}}{u_n} \equiv \alpha_n 
\binom{P_{n+1}(z_0)}{P_n (z_0)} + \beta_n 
\binom{Q_{n+1}(z_0)}{Q_n (z_0)} . \tag 4.7
$$
Since $W(P_\cdot(z_0), Q_\cdot(z_0)) = -1$, the two vectors on 
the right of (4.7) are linearly independent, and so they span 
$\Bbb C^2$ and $(\alpha_n, \beta_n)$ exist and are unique. 
Indeed, by the Wronskian relation:
$$\align
\alpha_n &= W(Q_\cdot(z_0), u)(n) \tag 4.8a \\
\beta_n &= -W (P_\cdot(z_0), u)(n). \tag 4.8b
\endalign
$$
A straightforward calculation using the fact that $P,Q$ obey 
(4.1) with $z=z_0$, (2.11), and (4.8) shows that (4.1) at $n$ 
is equivalent to
$$
\binom{\alpha_n}{\beta_n} = [1 + (z-z_0) S(n, z_0)] 
\binom{\alpha_{n-1}}{\beta_{n-1}}, \tag 4.9
$$
where
$$
S(n,z_0) = \pmatrix -Q_n (z_0) P_n(z_0) & -Q_n (z_0) Q_n (z_0) \\
P_n (z_0) P_n (z_0) & P_n (z_0) Q_n (z_0) \endpmatrix .
\tag 4.10
$$
For example, using (2.11) and (4.7) for $n\to n-1$: 
$$\align
\alpha_n &= W(Q_\cdot (z_0), u)(n) \\
&= W(Q_\cdot (z_0), u)(n-1) + (z_0-z) Q_n (z_0)u_n \\
&= \alpha_{n-1} - (z-z_0) Q_n (z_0) \{\alpha_{n-1} P_n (z_0) 
+ \beta_{n-1} Q_n (z_0)\}.
\endalign
$$

Notice that
$$
S^2 = \det(S) = \text{Tr}(S)=0. \tag 4.11
$$
The following is obvious and explains why $\ell^2$ solutions are 
so natural: 

If $\pi(z_0) = \{P_n (z_0)\}^\infty_{n=0}$ and $\xi(z_0) = 
\{Q_n (z_0)\}^\infty_{n=0}$ are {\it both} in $\ell^2$, then 
$$
\sum_{n=0}^\infty \|S(n,z_0)\| <\infty. \tag 4.12
$$

\proclaim{Lemma 4.6} Let $A_n$ be a sequence of matrices with 
$\sum_{n=0}^\infty \|A_n\|<\infty$. Let $D_N(z) = (1+zA_N)
(1+zA_{N-1})\dots (1+zA_0)$. Then $D_\infty (z) = \lim D_N (z)$ 
exists for each $z\in\Bbb C$ and defines an entire function of $z$ 
obeying
$$
\|D_\infty (z)\| \leq c_\varepsilon \exp(\varepsilon |z|)
$$
for each $\varepsilon >0$.
\endproclaim

\demo{Proof} Notice first that
$$
\| (1+B_N)\dots (1+B_0)\| \leq \prod_{j=0}^N (1+\|B_j\|) 
\leq \exp \biggl(\sum_{j=0}^N \|B_j\|\biggr) \tag 4.13
$$
and that
$$
\| (1+B_N)\dots (1+B_0) - 1 \| \leq \prod_{j=0}^N 
(1+\|B_j\|) - 1 \leq \exp \biggl( \sum_{j=0}^N \|B_j\| 
\biggr) - 1. \tag 4.14
$$
From (4.14), we see that
$$\align
\| D_{N+j}(z) - D_N(z)\| &\leq \biggl[ \exp \biggl( |z| 
\sum_{N+1}^{N+j} \| A_j \|\biggr) -1 \biggr] \| D_N(z)\| \\
&\leq \biggl[ \exp \biggl( |z| \sum_{N+1}^\infty \|A_j\| 
\biggr) -1 \biggr] \exp\biggl( |z| \sum_{j=0}^N \|A_j\| 
\biggr),
\endalign
$$
from which it follows that $D_N (z)$ is Cauchy uniformly for 
$z$ in balls of $\Bbb C$. Thus, $D_\infty$ exists and is entire 
in $z$. By (4.13), 
$$
\|D_\infty (z)\| \leq \prod_{j=0}^N (1 + |z| \|A_j\|) 
\exp \biggl( |z| \sum_{j=N+1}^\infty \|A_j\|\biggr),
$$
so given $\varepsilon$, choose $N$ so $\sum_{j=N+1}^\infty 
\|A_j\|\leq\frac{\varepsilon}{2}$ and use the fact that the 
polymonial $\prod_{j=0}^N (1+|z| \|A_j\|)$ can be bounded by 
$c_\varepsilon \exp(\frac12\varepsilon |z|)$. \qed
\enddemo

Given Proposition~4.4 and Theorem~2.14, the following completes 
the proof of Theorem~3:

\proclaim{Theorem 4.7} Let $A$ be a Jacobi matrix and consider 
solutions of {\rom{(4.1)}} for $z\in\Bbb C$. If for some $z_0 
\in\Bbb C$, all solutions are in $\ell^2$, then that is true for 
each $z\in\Bbb C$. Moreover, $\pi(z)$ and $\xi(z)$ are analytic 
$\ell_2$-valued functions so, in particular, $\frac{\partial\pi}
{\partial z}, \frac{\partial\xi}{\partial z}\in\ell_2$ for all 
$z$. The disk $\Cal D(z_0)$ varies continuously as $z$ runs through 
$\Bbb C_+$.
\endproclaim

\demo{Proof} Define
$$
T(n,-1; z, z_0) = (1+(z-z_0) S(n,z_0)) \dots 
(1+(z-z_0)S(0,z_0)). 
$$
By (4.12) and Lemma~4.6,
$$
\sup_n \|T(n, -1,; z, z_0) \| \leq \exp \biggl( |z-z_0| 
\sum_{j=0}^\infty \|S(j,z)\|\biggr) < \infty.
$$
Thus for any initial $\binom{\alpha_{-1}}{\beta_{-1}}$, if 
$u$ has the form (4.7), then $\sup_n \|\binom{\alpha_n}
{\beta_n}\|\leq  \mathbreak \sup_n \| T(n,-1; z, z_0)\|  
\| \binom{\alpha_{-1}}{\beta_{-1}}\| \equiv C$. Thus, $u_n 
= \alpha_n P_n(z_0) + \beta_n Q_n (z_0)$ has $|u_n|^2 \leq 
C^2 [P_n(z_0)^2 + Q_n (z_0)^2]$ which is in $\ell_1$. Since 
$\pi(z)$ is associated to $T(n, -1; z, z_0)\binom{1}{0}$ and 
$\xi(z)$ is associated to $T(n,-1; z,z_0)\binom{0}{1}$, the 
claimed analyticity holds because $\sup_n \|T(n, -1; z, z_0)\|$ 
is bounded as $z$ varies in bounded sets. 

Continuity of $\Cal D(z_0)$ follows from the formula for $\partial
\Cal D(z_0)$, viz.~$w\in\partial\Cal D(z_0)$ if and only if
$$
\| \pi(z_0)\|^2 |w|^2 + 2\, \text{Re}\biggl[\biggl( 
2\langle \xi(z_0), \pi(z_0)\rangle - \frac{i}
{2\,\text{Im}\, z_0} \biggr) w \biggr] + \|\xi(z_0)\|^2 =0. 
\qed
$$
\enddemo

By Theorem~3 and Lemma~4.6, if the Hamburger problem is 
indeterminate, 
$$
T(\infty, -1; z, z_0) \equiv \lim_{n\to\infty} T(n,-1; 
z, z_0) \tag 4.15
$$
exists. We define four functions $A(z)$, $B(z)$, $C(z)$, $D(z)$ 
by
$$
T(\infty, -1; z,z_0=0) = \pmatrix -B(z) & -A(z) \\ 
D(z) & C(z) \endpmatrix . \tag 4.16
$$
The {\it Nevanlinna matrix} is defined by
$$
N(z) = \pmatrix A(z) & C(z) \\
B(z) & D(z) \endpmatrix . \tag 4.17
$$

\proclaim{Theorem 4.8} 
\roster
\item"\rom{(i)}" Each of the functions $A$, $B$, $C$, $D$ is 
an entire function obeying
$$
|f(z)| \leq c_\varepsilon \exp (\varepsilon |z|)
$$
for each $\varepsilon >0$.
\item"\rom{(ii)}" For $z$ small,
$$
B(z) = -1 + O(z) \tag 4.18a
$$
and
$$
D(z) = \alpha z + O(z^2) \tag 4.18b
$$
with $\alpha >0$.
\item"\rom{(iii)}" $AD-BC\equiv 1$.
\endroster
\endproclaim

\demo{Proof} (i) follows from Lemma~4.6. (ii) follows if we note 
that
$$
T(\infty, -1; z,z_0=0) = 1 + z\sum_{n=0}^\infty S(n, z_0=0) + 
O(z^2).
$$
(4.18a) is immediate and (4.18b) follows since (4.10) says that 
$\alpha = \sum_{n=0}^\infty P_n (0)^2 >0$. (iii) holds since 
by (4.11), $\det T =1$. \qed
\enddemo

For our purposes, the formula for $A,B,C,D$ as matrix elements of 
an infinite product is sufficient, but there is an infinite sum 
formula connected to standard equations for perturbed solutions 
(see, e.g., [\pea, \jl]) that lets us make contact with the more 
usual definitions [\akh].

\proclaim{Theorem 4.9} If the Hamburger problem is indeterminate, 
then,
$$\align
A(z) &= z\sum_{n=0}^\infty Q_n(0) Q_n(z) \\
B(z) &= -1 + z\sum_{n=0}^\infty Q_n(0) P_n(z) \\
C(z) &= 1 + z\sum_{n=0}^\infty P_n(0) Q_n(z) \\
D(z) &= z \sum_{n=0}^\infty P_n(0) P_n(z). 
\endalign
$$
\endproclaim

\demo{Proof} 
$$\align
T(n, -1; z,z_0 =0) &= (1+z S(n, z_0=0)) \dots (1+z S(0,z_0=0)) \\
&= 1 + \sum_{j=0}^n zS(j, z_0=0) \prod_{k=0}^{j-1} (1+z S(k,z_0=0)) 
\endalign
$$
so it suffices to show that
$$\split
S(j,z_0=0) (1 +z S(j-1, z_0&=0)) \dots (1+zS(0,z_0=0)) \\ 
&\quad = 
\pmatrix -Q_j(0) P_j(z) & -Q_j (0) Q_j(z) \\
P_j(0) P_j(z) & P_j(0) Q_j(z) \endpmatrix  
\endsplit . \tag 4.19
$$
By definition,
$$
(1+zS(j-1, z_0=0)) \dots (1+zS(0,z_0=0)) = 
\pmatrix \alpha^{(1)}_{j-1} & \alpha^{(2)}_{j-1} \\
\beta^{(1)}_{j-1} & \beta^{(2)}_{j-1} \endpmatrix ,
$$
where
$$
\alpha^{(1)}_{j-1} \binom{P_j(0)}{P_{j-1}(0)} + 
\beta^{(1)}_{j-1} \binom{Q_j(0)}{Q_{j-1}(0)} = 
\binom{P_j(z)}{P_{j-1}(z)}
$$
and
$$
\alpha^{(2)}_{j-1} \binom{P_j(0)}{P_{j-1}(0)} + 
\beta^{(2)}_{j-1} \binom{Q_j(0)}{Q_{j-1}(0)} = 
\binom{Q_j(z)}{Q_{j-1}(z)}. 
$$
Thus by (4.10),
$$\align
\text{LHS of (4.19)} &= \pmatrix
-Q_j(0)[\alpha^{(1)}_{j-1} P_j(0) + \beta^{(1)}_{j-1} Q_j(0)] 
& -Q_j(0) [\alpha^{(2)}_{j-1} P_j(0) + \beta^{(2)}_{j-1} Q_j (0)] \\
P_j(0)[\alpha^{(1)}_{j-1} P_j(0) + \beta^{(1)}_{j-1} Q_j(0)] 
& P_j(0) [\alpha^{(2)}_{j-1} P_j(0) + \beta^{(2)}_{j-1} 
Q_j(0)] \endpmatrix \\
&= \text{RHS of (4.19)}.  \qed
\endalign 
$$ 
\enddemo

$\pi(0) = \{P_n(0)\}$ and $\xi(0)=\{Q_n(0)\}$ span $D(A^*)/D(A)$ 
by Proposition~4.4. Notice that $A^* \pi(0)=0$ while $A^*\xi(0)=
\delta_0$, so $\langle \pi(0), A^* \xi(0)\rangle =1$ and (2.7) 
holds. Thus, we can parametrize the self-adjoint extensions $B_t$ 
of $A$ by
$$
D(B_t) = D(A) + \{\alpha(t\pi(0)+\xi(0))\mid \alpha\in
\Bbb C\} \tag 4.20a
$$
if $t<\infty$ and
$$
D(B_\infty) = D(A) + \{\alpha \pi(0)\}. \tag 4.20b
$$
This is equivalent to defining $t$ by
$$
t=(\delta_0, B^{-1}_t \delta_0). \tag 4.20c
$$

\proclaim{Theorem 4.10} For each $t\in\Bbb R\cup \{\infty\}$ and 
$z\in\Bbb C\backslash\Bbb R$,
$$
(\delta_0, (B_t -z)^{-1}\delta_0) = -\frac{C(z)t+A(z)}
{D(z)t +B(z)}\, . \tag 4.21
$$
\endproclaim

\demo{Proof} Let us sketch the proof before giving details:
$$
T(-1, \infty; z, z_0=0) \equiv T(\infty, -1; z, z_0=0)^{-1}. 
\tag 4.22
$$
Then
$$
\binom{\alpha_{-1}}{\beta_{-1}} \equiv T(-1,\infty; z, z_0=0) 
\binom{t}{1} \tag 4.23
$$
is such that the $u_n$ associated to $\binom{\alpha_{-1}}
{\beta_{-1}}$ obeys (4.1) and is asymptotically $t\pi (0) + 
\xi(0)$ and so in $D(B_t)$. $(B_t-z)^{-1} \delta_0$ will 
thus be $-\frac{u_t}{u_t (-1)}$ and $(\delta_0, (B_t -z)^{-1} 
\delta_0)$ will be $-\frac{u_t(0)}{u_t(-1)}$. But $u_t(0) =
\alpha_{-1}$ and $u_t (-1) = -\beta_{-1}$ so $(\delta_0, 
B_t-z)^{-1}\delta_0)$ will be $\frac{\alpha_{-1}}{\beta_{-1}}$, 
which is given by (4.21).

Here are the details. If $\varphi\in D(A^*)$, then by (2.14), 
$\varphi\in D(B_t)$ if and only if
$$
\lim_{n\to\infty} W(\varphi, t\pi(0) + \xi(0))(n)=0. 
\tag 4.24
$$
Suppose $u$ solves (4.1) for a given $z$. Then $u\in D(A^*)$ 
and $u$ has the form (4.7), where by (4.15), $\lim_{n\to\infty} 
\binom{\alpha_n}{\beta_n} = \binom{\alpha_\infty}{\beta_\infty}$ 
exists. Clearly,
$$
W(u, t\pi(0) + \xi(0))(n) = -\alpha_n + \beta_n t
$$
so (4.24) holds if and only if 
$$
\alpha_\infty = t\beta_\infty. 
$$

Thus, if $T(-1, \infty; z, z_0=0)$ is given by (4.22), then  
$\binom{\alpha_{-1}}{\beta_{-1}}$ given by (4.23) is initial data  
for a solution $u$ of (4.1) that has $u\in D(B_t)$. A solution 
$u$ of (4.1) has
$$
(A^* -z)u = -u_{-1} \delta_0
$$
and thus, if $u$ is associated to the data in (4.23), 
$$
(B_t -z)u = -u_{-1}\delta_0
$$
and
$$
(\delta_0, (B_t -z)^{-1} \delta_0) = -\frac{u_0}{u_{-1}}\, .
$$
But
$$
\binom{u_0}{u_{-1}} = \alpha_{-1} 
\binom{P_0(0)}{P_{-1}(0)} + \beta_{-1} 
\binom{Q_0(0)}{Q_{-1}(0)} = \binom{\alpha_{-1}}{-\beta_{-1}}
$$
so
$$
(\delta_0, (B_t -z)^{-1}\delta_0) = \frac{\alpha_{-1}}
{\beta_{-1}}\, .
$$
Since $T(\infty, -1; z, z_0=0)$ has the form (4.16) and has 
determinant $1$, its inverse has the form
$$
\pmatrix C(z) & A(z) \\
-D(z) & -B(z) \endpmatrix
$$
and so $\alpha_{-1} = C(z)t + A(z)$, $\beta_{-1} = -D(z)t - 
B(z)$. Thus (4.21) is proven. \qed
\enddemo

\remark{Remark} Our convention for the parameter $t$ differs from 
that in Akhiezer [\akh], which is the traditional one! If $s=
-t^{-1}$, then
$$
-\frac{Ct+A}{Dt+B} = -\frac{As - C}{Bs-D}\, .
$$
The parameter $s$ is what he calls $t$. Our $\Phi(z)$ later and 
his $\Phi(z)$, which I will call $\Psi(z)$, are related by $\Phi(z) 
= -\Psi(z)^{-1}$. Since this transformation takes Herglotz functions 
to themselves, we both deal with Herglotz functions. See [\im] for 
an interesting alternate reparametrization.
\endremark

\vskip 0.1in

We turn next to half of Theorem~5. We will prove that each 
$B_t$ has point spectrum with any two disjoint. The condition 
that $\mu_t (\{t\}) >\rho (\{t\})$ for any other solution of the 
moment problem will wait until after we have proven Theorem~4.

\proclaim{Theorem 4.11 (half of Theorem 5)} Suppose the 
Hamburger problem is indeterminate. Each $B_t$ has pure point 
spectrum only. The eigenvalues of the different $B_t$'s are 
distinct and every $x\in\Bbb R$ is an eigenvalue of some $B_t$. 
If $x$ is an eigenvalue of $B_t$, then
$$
\mu_t (\{x\}) = \frac{1}{\sum_{n=0}^\infty |P_n (x)|^2}\, . 
\tag 4.25
$$
Moreover, if $\lambda_n (B_t)$ are the eigenvalues of $B_t$ 
\rom(ordered in some convenient way\rom), then for any $p>1$,
$$
\sum_n |\lambda_n (B_t)|^{-p} <\infty. \tag 4.26
$$
\endproclaim

\demo{Proof} Let $\mu_t$ be the spectral measure for $B_t$. By 
Theorem~4.10,
$$
\int\frac{d\mu_t (x)}{x-z} = -\frac{C(z)t+A(z)}{D(z)t+B(z)}
$$
is a meromorphic function of $z$, since $A,B,C,D$ are entire. 
That implies that $\mu_t$ is pure point only and these points 
are given by solutions of
$$
D(z)t + B(z) =0.
$$
Since $A,B,C,D$ are real when $z=x$ is real, for any $x$, 
$x\in\text{spec}(B_t)$ for $t=-\frac{B(x)}{D(x)}$. Since $AD-BC 
\equiv 1$, $B$ and $D$ have no simultaneous zeros, so the 
eigenvalues are distinct for different $t$'s. (4.26) is a 
standard result on the zeros of an entire function like 
$D(z)t+ B(z)$, which obeys (i) of Theorem~4.8. 

To prove (4.25), note that if $x$ is an eigenvalue of $B_t$, 
then the corresponding normalized eigenvector $\varphi$ 
obeys $A^*\varphi = x\varphi$. It follows that $\varphi_n 
= \varphi_0 P_n (x)$ and then by normalization, that 
$\varphi^2_0 = 1/\sum_{n=0}^\infty |P_n(x)|^2$. Thus, $\mu_t 
(\{x\}) = \varphi^2_0$ is given by (4.25). \qed 
\enddemo

Now define the map $F(z): \Bbb C\cup\{\infty\} \to \Bbb C 
\cup\{\infty\}$ by
$$
F(z)(w) = -\frac{C(z)w + A(z)}{D(z)w + B(z)}\, . \tag 4.27
$$
$F(z)$ as a fractional linear transformation is one-one and 
onto, and takes $\Bbb R\cup\{\infty\}$ onto a circle or straight 
line. By Proposition~4.4 and (4.21), $F(z) [\Bbb R\cup\{\infty\}]$ 
is precisely the circle $\partial\Cal D(z)$ when $\text{Im}\, z 
\neq 0$.

\proclaim{Proposition 4.12} If $\text{\rom{Im}}\, z>0$, $F(z)$ 
maps the upper half-plane $\Bbb C_+ = \{z\in\Bbb C\mid
\text{\rom{Im}}\, z>0\}$ onto the interior of the disk 
$\Cal D(z)^{\text{\rom{int}}}$.
\endproclaim

\demo{Proof} For each $z\in\Bbb C\backslash\Bbb R$, $F(z)$ must 
map $\Bbb C_+$ onto either $\Cal D(z)^{\text{\rom{int}}}$ or 
$\Bbb C\backslash\Cal D(z)$ since $\Bbb R$ is mapped to 
$\partial\Cal D(z)$ and a fractional linear transformation 
maps $\Bbb C\cup\{\infty\}$ to itself bijectively. Thus it 
suffices to show $F(z)[i]\in\Cal D(z)^{\text{\rom{int}}}$ for 
all $z$ with $\text{Im}\,z>0$. But $F(z)[i]$ moves analytically, 
and so continuously. It can only move from 
$\Cal D(z)^{\text{\rom{int}}}$ to $\Bbb C\backslash\Cal D(z)$ by 
lying on $\partial\Cal D(z)$, which is impossible if $\text{Im}\, 
z>0$ since then $\partial\Cal D(z)=F(z)[\Bbb R\cup\{\infty\}]$. 
Thus it suffices to show $F(z)[i]\in\Cal D(z)^{\text{\rom{int}}}$ 
for $z=ix$ with $x>0$ and small.

Now, $F(z)[i]\in\Cal D(z)^{\text{\rom{int}}}$ if and only if 
$F(z)$ takes the lower half-plane to $\Bbb C\cup\{\infty\} 
\backslash\Cal D(z)$, which holds if and only if $F(z)^{-1}
[\infty]$ is in the lower half-plane. Thus it suffices to show 
for $x$ small and positive that $F(ix)^{-1}[\infty]\in\Bbb C_-$.

$F(ix)^{-1}[\infty]$ is that $w$ with $D(ix)w+B(ix)=0$, that is, 
$w=-\frac{B(ix)}{D(ix)}$. But by (4.18), $B(ix) =-1 + O(x)$ and 
$D(ix) = i\alpha x + O(x^2)$ with $\alpha >0$, so $w=-i
\alpha^{-1} x^{-1} + O(1)$ lies in $\Bbb C_-$, which suffices 
to prove the proposition. \qed
\enddemo

We now follow the classical argument of Nevanlinna [\nev] to 
obtain Theorem~4. We need to relate solutions of (1.1) to 
asymptotics of the Stieltjes transform of $\rho$ following 
Hamburger [\ham] and Nevanlinna [\nev].

\proclaim{Proposition 4.13} Let $\Cal M^H(\gamma) = \{\rho \mid 
\rho\text{\rom{ obeys (1.1)}}\}$. Let $G_\rho(z) = \int
\frac{d\rho(x)}{x-z}$. Then for any $N$ as $y\to\infty$,
$$
y^{N+1} \biggl[G_\rho (iy) + \sum_{n=0}^N (-i)^{n+1} 
y^{-n-1} \gamma_n\biggr] \to 0 \tag 4.28
$$
uniformly for $\rho$ in $\Cal M^H(\gamma)$. Conversely, if $G$ 
is a function on $\Bbb C_+$ with $\text{\rom{Im}}\, G(z)>0$ there 
and so that {\rom{(4.28)}} holds for each $N$, then $G(z) = 
G_\rho (z)$ for some $\rho$ in $\Cal M^H(\gamma)$.
\endproclaim

\demo{Proof} By the geometric series with remainder, the left side 
of (4.28) is
$$
-(-i)^{N+1} y^{N+1} \int \frac{d\rho(x)}{x-iy} \,\, 
\frac{x^{N+1}}{y^{N+1}} \equiv R_N(\rho) \tag 4.29
$$
so, using $|x-iy|^{-1}\leq y^{-1}$,
$$
|R_N(\rho)| \leq y^{-1} \int |x|^{N+1} d\rho(x) 
\cases = \gamma_{N+1} y^{-1} & \text{if $N$ is odd} \\
\leq \frac12 [\gamma_N + \gamma_{N+2}]y^{-1} & \text{if $N$ is even}. 
\endcases
$$
This shows that (4.28) holds and the convergence is uniform for 
$\rho\in \Cal M^H(\gamma)$.

For the converse, first use the Herglotz representative theorem 
which says that if $G$ maps $\Bbb C_+$ to $\Bbb C_+$, then 
for some measure $d\rho$, some $c\geq 0$ and some real $d$:
$$
G(z) = cz + d + \int d\rho(x) \biggl[ \frac{1}{x-z} - 
\frac{x}{1+x^2}\biggr] , \tag 4.30
$$
where, a priori, $\rho$ only obeys
$$
\int\frac{d\rho(x)}{1+ x^2} < \infty.
$$
By (4.28),
$$
y[G(iy)] \to i\gamma_0. \tag 4.31
$$
If (4.30) holds, then $y^{-1}G(iy)\to ic$, so $c=0$. Since 
$c=0$, (4.30) says
$$
y\text{ Im}\, G(iy) = \int\frac{y^2}{x^2 + y^2}\, d\rho(x),
$$
so (4.31) and the monotone convergence theorem implies that 
$\int d\rho(x)=\gamma_0$. Once this is true, (4.30) implies that 
as $y\to\infty$, 
$$
\text{Re}\, G(iy)\to d - \int d\rho(x)\, \frac{x}{1+x^2} 
\tag 4.32
$$
so (4.31) implies the right side of (4.32) is zero, and thus 
(4.30) becomes
$$
G(z) = \int \frac{d\rho(x)}{x-z}\, . \tag 4.33
$$

We will now prove inductively that $\rho$ obeys (1.1), that is, 
$\rho\in \Cal M^H(\gamma)$. Suppose that we know that (1.1) 
holds for $n=0,1,\dots, 2M-2$. (4.28) for $N=2M$ then implies 
that
$$
\int\frac{(iy)^2 x^{2M-1}}{x-iy}\, d\rho(x) + iy\gamma_{2M-1} 
\to -\gamma_{2M}.
$$
Taking real and imaginary parts, we see that
$$\align
\gamma_{2M-1} &= \lim_{y\to\infty} \int \frac{y^2 x^{2M-1}}
{x^2 + y^2}\, d\rho(x) \tag 4.34 \\
\gamma_{2M} &= \lim_{y\to\infty} \int \frac{y^2 x^{2M}}
{x^2 + y^2}\, d\rho(x). \tag 4.35
\endalign
$$
(4.35) and monotone convergence implies that $\int x^{2M}\, 
d\rho(x)<\infty$, so that dominated convergence and (4.34), 
(4.35) imply that (1.1) holds for $n=0,1,2,\dots, 2M$. \qed
\enddemo

Let $\Cal F$ be the set of analytic maps from $\Bbb C_+$ to 
$\bar\Bbb C_+ \cup\{\infty\}$. By the open mapping theorem for 
analytic functions, $\Phi \in\Cal F$ either maps $\Bbb C_+$ to 
$\Bbb C_+$ or $\Phi$ has a constant value in $\Bbb R\cup
\{\infty\}$.

\proclaim{Theorem 4.14 (Nevanlinna [\nev])} Let $\gamma$ be a 
set of indeterminate Hamburger moments. Then there is a one-one 
correspondence between solutions, $\rho$, of the moment problem 
and functions $\Phi\in\Cal F$ given by
$$
G_\rho \equiv \int \frac{d\rho(x)}{x-z} = 
-\frac{C(z)\Phi(z) + A(z)}{D(z) \Phi(z) + B(z)}\, . \tag 4.36
$$
In particular, if $\rho$ is not a von~Neumann solution, 
$G_\rho(z) \in D(z)^{\text{\rom{int}}}$ for all $z\in\Bbb C_+$.

\endproclaim

\remark{Remarks} 1. By (4.21), the von~Neumann solutions $\mu_t$ 
correspond precisely to the case $\Phi(z)\equiv t\in\Bbb R \cup
\{\infty\}$.

2. We will call the Herglotz function associated to some 
$\rho\in\Cal M^H(\gamma)$ the {\it Nevanlinna function} of 
$\rho$ and denote it by $\Phi_\rho$. Conversely, given $\Phi$ 
(and $\gamma$), we denote by $\rho_\Phi$ the associated solution 
of the moment problem.

3. In Section~6, we will discuss the $\rho_\Phi$'s associated to 
the case where the measure $\mu$ in the Herglotz representation 
(1.19) is a measure with finite support.
\endremark

\demo{Proof} Let $\Phi\in\Cal F$ and let $G(z)$ denote the right 
side of (4.36). In terms of the map $F(z)$ of (4.27), $G(z) = 
F(z)(\Phi(z))$, so by Proposition~4.12, $G(z)\in\Cal D(z)$. By 
the uniformity in Proposition~4.13, the fact that the $\mu_t\in 
\Cal M^H(\gamma)$ and that $\{G_{\mu_t}(z)\mid t\in\Bbb R\cup
\{\infty\} \}=\partial\Cal D(z)$, this implies that $G(z)$ obeys 
(4.28). Thus by Proposition~4.13, $G(z)=G_\rho(z)$ for some $\rho
\in \Cal M^H(\gamma)$.

Conversely, let $\rho\in\Cal M^H(\gamma)$ and consider 
$G_\rho(z)$. By Theorem~4.3, Proposition~4.4 and Proposition~4.12, 
$\Phi(z) \equiv F(z)^{-1} (G_\rho (z))\in \bar\Bbb C_+ \cup 
\{\infty\}$,  and so $\Phi\in\Cal F$. Thus, $G_\rho(z) = F(z) 
(\Phi(z))$, that is, (4.36) holds. 

If $\rho$ is not a von~Neumann solution, then $\text{Im}\, 
\Phi_\rho (z) >0$, so $G_\rho(z) = F(z)(\Phi_\rho(z))$ is in 
the interior of $\Cal D(z)$. \qed
\enddemo

\proclaim{Proposition 4.15} Let $\rho\in\Cal M^H(\gamma)$. If 
$\rho$ is a von~Neumann solution, then the polynomials are 
dense in $L^2(\Bbb R, d\rho)$ and for each $z_0\in\Bbb C_+$, there 
is no other $\mu$ in $\Cal M^H(\gamma)$ with $G_\mu (z_0)=
G_\rho(z_0)$. If $\rho$ is not a von~Neumann solution, then the 
polynomials are not dense in $L^2 (\Bbb R, d\rho)$ and for each 
$z_0\in \Bbb C_+$, there are distinct $\mu$'s in $\Cal M^H(\gamma)$ 
with $G_\mu (z_0)=G_\rho(z_0)$.
\endproclaim

\demo{Proof} If $\rho$ is a von~Neumann solution, the polynomials 
are dense by construction and the uniqueness result holds since 
$G_\mu(z_0)=G_\rho(z_0)$ if and only if $\Phi_\mu (z_0) = 
\Phi_\rho(z_0)$ and $\Phi_\rho(z_0)$ is then real.

Conversely, suppose $\rho$ is not a von~Neumann solution. Then, 
$G_\rho(z_0)\in D(z)^{\text{\rom{int}}}$, so by the proof of 
Theorem~4.3, $(x-z_0)^{-1}$ is not in the $L^2$ closure of the 
polynomials. Thus, the polynomials are not dense. If $\Phi_\rho$ 
is not a constant, set $\Psi(z) = -|\Phi_\rho(z_0)|^2 \Phi_\rho 
(z)^{-1} +2\text{Re}\, \Phi_\rho (z_0)$. If $\Phi_\rho(z)$ is 
a constant, $q$ (of necessity with $\text{Im}\, q>0$), set 
$\Psi(z) = cz+d$ with $c=\frac{\text{Im}\, q}{\text{Im}\, z_0}$ 
($>0$) and $d=\text{Re}\, q -c\,\text{Re}\, z_0$. In either case, 
$\Psi \neq \Phi$, $\Psi$ is Herglotz, and $\Psi(z_0) = \Phi(z_0)$. 
Thus, if $\mu = \rho_\Psi$, we have $G_\mu (z_0) = F(z_0) 
(\Psi(z_0)) = F(z_0)(\Phi(z_0)) = G_\rho(z_0)$. \qed
\enddemo

To complete the proof of Theorem~5, we need the following fact 
about Herglotz functions:

\proclaim{Theorem 4.16} Let $\Phi(z)$ be a Herglotz function so
$$
\Phi(z) = cz+ d + \int d\mu(x) \biggl[\frac{1}{x-z} - 
\frac{x}{1+x^2}\biggr]
$$
with $d$ real, $c\geq 0$, and either $c>0$ or $d\mu\neq 0$. 
Suppose for some $x_0$, $\Phi(x_0 + i\varepsilon)\to t\in
\Bbb R$. Then either $|\frac{[\Phi(x_0 +i\varepsilon)-t]}
{i\varepsilon}|\to\infty$ or else $\int \frac{d\mu(x)}
{(x-x_0)^2}<\infty$ 
and
$$
\lim_{\varepsilon\downarrow 0} \frac{\Phi(x_0 + i\varepsilon)
-t}{i\varepsilon} = c+ \int\frac{d\mu(x)}{(x-x_0)^2}\, . 
\tag 4.37
$$
\endproclaim

\remark{Remark} Do not confuse the $\mu$ in the Herglotz 
representation for $\Phi$ and the measure $\rho$ of (4.36). 
In particular, $\mu$ is allowed to have finite support even 
though we are suppposing that $\rho$ does not.
\endremark

\demo{Proof} Note first that
$$
\frac{\text{Im}\, \Phi(x_o + i\varepsilon)}{\varepsilon} = 
c + \int \frac{d\mu(x)}{(x-x_0)^2 + \varepsilon^2}\, . 
\tag 4.38
$$
On the other hand,
$$
\text{Im}\, \frac{\partial\Phi}{\partial z}\, (x_0 + i\varepsilon) 
=\int\frac{d\mu(x)\,2\varepsilon (x-x_0)}
{[(x-x_0)^2 + \varepsilon^2]^2}\, . \tag 4.39
$$
If $\int\frac{d\mu(x)}{(x-x_0)^2}=\infty$, then (4.38) implies 
$\text{Im}\,[\frac{\Phi(x_0 +i\varepsilon)-t}{\varepsilon}] 
\to\infty$ , so if the limit is finite, then $\int
\frac{d\mu(x)}{(x-x_0)^2}<\infty$. (4.39) and the dominated 
convergence theorem implies that $\text{Im}\, \frac{\partial\Phi}
{\partial z} (x_0 + i\varepsilon)\to 0$ so that $\frac{[\text{Re}\, 
\Phi(x_0 + i\varepsilon)-t]}{i\varepsilon} = \frac{1}{\varepsilon} 
\int_0^\varepsilon \text{Im}\, \frac{\partial\Phi}{\partial z} 
(x_0 + iy)\, dy \to 0$. This and (4.38) implies (4.37). \qed
\enddemo

\proclaim{Theorem 4.17 (end of Theorem~5)} Let $\gamma$ be an 
indeterminate moment problem. Let $\rho\in \Cal M^H(\gamma)$ 
correspond to a $\Phi$ which is not a constant in $\Bbb R \cup
\{\infty\}$ \rom(so that $\rho$ is not a von~Neumann solution\rom). 
Suppose $\alpha\equiv\rho(\{x_0\}) >0$ for some point $x_0\in
\Bbb R$. Then there is a von~Neumann solution $\mu_t$ so that 
$\mu_t (\{x_0\})>\alpha$. 
\endproclaim

\demo{Proof} We will suppose that $D(x_0)\neq 0$. If $D(x_0) 
=0$, then $B(x_0)\neq 0$ and we can use
$$
-\frac{C(z)\Phi(z)+A(z)}{D(z)\Phi(z)+B(z)} = 
-\frac{A(z) (-\Phi(z))^{-1} - C(z)}{B(z)(-\Phi(z))^{-1} 
- D(z)}
$$
and $-\Phi(z)^{-1}$ in place of $\Phi(z)$ to give an identical 
argument. Define $t=-\frac{B(x_0)}{D(x_0)}\in\Bbb R$. Since 
$AD-BC=1$,
$$
C(x_0)t + A(x_0) = \frac{1}{D(x_0)} \tag 4.40
$$
is non-zero and has the same sign as $D$.

$$
(-i\varepsilon) G_\rho (x_0+i\varepsilon) = (i\varepsilon) 
\biggl[ \frac{C(x_0 + i\varepsilon) \Phi(x_0 + i\varepsilon) 
+ A(x_0 + i\varepsilon)}{D(x_0+i\varepsilon)\Phi(x_0+i\varepsilon) 
+B(x_0+i\varepsilon)} \biggr] \to \alpha.
$$
This is only possible if $G_\rho (x_0+i\varepsilon) \to\infty$, 
which requires that $\Phi(x_0+i\varepsilon)\to t$. Then
$$
C(x_0+i\varepsilon) \Phi(x_0+i\varepsilon) +A(x_0+i\varepsilon) 
\to C(x_0)t +A(x_0)
$$
while
$$
\frac{D(x_0+i\varepsilon) \Phi(x_0+i\varepsilon) +
B(x_0+i\varepsilon)}{i\varepsilon} \to D'(x_0) t + B'(x_0) + 
D(x_0) \lim_{\varepsilon\downarrow 0} \frac{\Phi(x_0+i\varepsilon) 
-t}{i\varepsilon}\, .
$$
Thus using (4.40),
$$
\alpha = \frac{1}{D(x_0)^2 \lim_{\varepsilon\downarrow 0} 
\frac{\Phi(x_0+i\varepsilon) -t}{i\varepsilon} + D(x_0) 
[D'(x_0)t + B'(x)]}\, .
$$
On the other hand, taking $\Phi(x)\equiv t$ to describe $\mu_t$,
$$
\mu_t (\{x_0\}) = \frac{1}{D(x_0) [D'(x_0)t + B'(x_0)]}
$$
must be positive. It follows from Lemma~4.16 that $\int 
\frac{d\mu(x)}{(x-x_0)^2}<\infty$ and that
$$
\alpha^{-1} > \mu_t (\{x_0\})^{-1}. \qed
$$
\enddemo

Finally, we turn to the Nevanlinna parametrization in the 
indeterminate Stieltjes case. As we have seen, among the $B_t$ 
there are two distingtuished ones, $B_F$ and $B_K$. The Krein 
extension has $\ker(A^*)\subset D(B_K)$ so $\pi(0)\in D(B_K)$,  
which means $t=\infty$ in the notation of (4.20). The Friedrichs 
extension is $B_F=B_{t_0}$ where $t_0 = (\delta_0, A^{-1}_F 
\delta_0)$.

\proclaim{Theorem 4.18 (Stieltjes case of Theorem 4)} If the 
Stieltjes problem for $\gamma$ is indeterminate, its solutions 
obey {\rom{(4.36)}} for precisely those $\Phi$ that either are 
a constant in $[t_0, \infty)\cup\{\infty\}$ or that obey $\Phi(z)$ 
is analytic in $\Bbb C\backslash [0,\infty)$ with $\Phi(x) \in 
[t_0,\infty)$ if $x\in (-\infty, 0)$. Here $t_0 = (\delta_0, 
A^{-1}_F \delta_0)$.
\endproclaim

\remark{Remark} Such $\Phi$'s are precisely those of the form
$$
\Phi(z) = d + \int_0^\infty \frac{d\mu(x)}{x-z}\, , \tag 4.41
$$
where $d\geq t_0$ and $\int_0^\infty \frac{d\mu(x)}{x+1}<\infty$.
\endremark

\demo{Proof} Clearly, the solutions of the Stieltjes problem are
precisely the solutions $\rho$ of the Hamburger problem with 
$\lim_{\varepsilon\downarrow 0} G_\rho (-y + i\varepsilon)$ real 
for each $y\in (0,\infty)$. Since the $B_t$'s obey $t=(\delta_0, 
B^{-1}_t \delta_0)$ and $B_K < B_t < B_F$ means $t_0 < (\delta_0, 
B^{-1}_t \delta_0)<\infty$, we know that for $y\in (-\infty,0)$, 
$F(y)$ maps $[t_0, \infty)$ into $(0,\infty)$ ($F(\,\cdot\,)$ is 
given by (4.27)). It follows that if $\Phi$ obeys $\Phi(x)\in 
[t_0, \infty)$ for $x\in (-\infty, 0)$, then $G_\rho (y+i
\varepsilon)$ has a limit in $(0,\infty)$ on $(-\infty, 0)$, 
that is, $\Phi$ defines a solution of the Stieltjes moment 
problem. The converse follows from this argument and the next 
theorem. \qed
\enddemo

\proclaim{Theorem 4.19} Let $d\mu_F$ and $d\mu_K$ be the 
Friedrichs and Krein solutions of an indeterminate Stieltjes 
moment problem and let $d\rho$ be another solution. Then for 
all $y\in (0,\infty)$, 
$$
\int_0^\infty \frac{d\mu_F (x)}{x+y} \leq 
\int_0^\infty \frac{d\rho(x)}{x+y} \leq 
\int_0^\infty \frac{d\mu_K (x)}{x+y}\, . \tag 4.42
$$
\endproclaim

This will be proven below as part of Theorem~5.2.

\proclaim{Corollary 4.20} If $\rho\in\Cal M^S(\gamma)$ and 
$d\rho\neq d\mu_F$, then for $y\in [0,\infty)$,
$$
\int_0^\infty \frac{d\mu_F(x)}{x+y} < \int_0^\infty 
\frac{d\rho(x)}{x+y}\,. \tag 4.43
$$
\rom(Note that the inequality is strict.\rom)
\endproclaim

\demo{Proof} For each $y\in (0,\infty)$, $F(-y)(\, \cdot\,)$ is 
a strictly monotone map of $[t_0, \infty)$ to $[G_{\mu_F}(-y), 
\mathbreak G_{\mu_K}(-y))$ by the proof of Theorem~4.18 and the 
fact that any fractional linear map of $\Bbb R$ to $\Bbb R$ is 
monotone in any region where it is finite. By a limiting argument, 
the same is true for $y=0$. Thus, (4.43) follows from the relation 
of Nevanlinna functions $\Phi_\rho(y) > \Phi_{\mu_F}(y) \equiv
t_0$. This is immediate from (4.41). \qed
\enddemo

As a corollary of this, we see

\proclaim{Corollary 4.21} Let $\gamma$ be a set of indeterminate 
Stieltjes moments and let $d\mu_F$ be its Friedrichs solution. Let 
$r=\int x^{-1} d\mu_F(x)$ and let
$$
\tilde\gamma_0 = 1; \qquad \tilde\gamma_n = r^{-1}\gamma_{n-1}, 
\qquad n=1,2,\dots .
$$
Then $\{\tilde\gamma_j\}^\infty_{j=0}$ is a determinate 
Hamburger moment problem.
\endproclaim

\remark{Remark} This lets us produce Stieltjes determinate 
moment problems with essentially arbitrary rates of growth for 
$\gamma_n$. By using Theorem~2.13, we obtain determinate 
Hamburger problems with arbitrary fast growth also.
\endremark

\demo{Proof} $r^{-1} x^{-1} d\mu_F\equiv d\lambda_0(x)$ is a 
Stieltjes measure whose moments are $\tilde\gamma$. Let $d\lambda(x)$ 
be another such Stieltjes measure and let $d\rho(x)=rx\, d\lambda(x)$. 
Then the moments of $d\rho$ are $\gamma$ and $\int x^{-1} d\rho(x) 
=r$. So, by Corollary~4.20, $\rho=\mu_F$. Thus, $\lambda = 
\lambda_0$. Thus, $\lambda_0$ is the unique solution of the Stieltjes 
problem. But $0\not\in\text{supp}(d\lambda_0)$. Thus by 
Proposition~3.1, the Hamburger problem is also determinate. 
\qed
\enddemo

\vskip 0.3in
\flushpar{\bf \S 5. Pad\'e Approximants, Continued Fractions, and 
Finite Matrix Approximations}
\vskip 0.1in

In this section, we will primarily consider aspects of the 
Stieltjes moment problem. We will discuss finite matrix 
approximations which converge to the Friedrichs and Krein 
extensions of $A$ and, as a bonus, obtain Theorem~4.19 and 
Theorems~5, 6, 7, and 8. We will see that the convergence of the 
Pad\'e approximants (equivalently, continued fractions of 
Stieltjes) can be reinterpreted as strong resolvent convergence 
of certain finite matrix approximations to $A_F$ and $A_K$.

Before we start, we note that our continued fraction expressions 
are related to but distinct from those of Stieltjes. Our basic 
object (the approximations to $A_F$) are of the form:
$$
\cfrac1 \\
-z+b_0 - \cfrac{a^2_0} \\
-z+b_1 - \cfrac{a^2_1} \\
-z+b_2 + \cdots \endcfrac \tag 5.1a
$$
with two terms (an affine polynomial) at each stage while 
Stieltjes' are of the form
$$
\cfrac1 \\
c_1 w + \cfrac1 \\
c_2 + \cfrac1 \\
c_3 w+\cdots \endcfrac \tag 5.1b
$$
with a single term in each step and alternate constant and linear 
terms. This gives him twice as many continued fractions, so his 
alternate between the two sets of rational approximations we get.
We will say more about this later.

Until Theorem~5.29 below, we will suppose 
$\{\gamma_n\}^\infty_{n=0}$ is a set of Stieltjes moments.

Given the Jacobi matrix (1.16), we will consider two $N\times N$  
approximations. The first is obvious:
$$
A^{[N]}_F = \pmatrix 
b_0 & a_0 & {} & {} & {} \\
a_0 & b_1 & {} & {} & {} \\
{} & {} & \ddots & {} & {} \\
{} & {} & {} & b_{N-2} & a_{N-2} \\
{} & {} & {} & a_{N-2} & b_{N-1} \endpmatrix . \tag 5.2a
$$
The second differs by the value of the $NN$ coefficient:
$$
A^{[N]}_K = \pmatrix 
b_0 & a_0 & {} & {} & {} \\
a_0 & b_1 & {} & {} & {} \\
{} & {} & \ddots & {} & {} \\
{} & {} & {} & b_{N-2} & a_{N-2} \\
{} & {} & {} & a_{N-2} & b_{N-1}-\alpha_{N-1} \endpmatrix , 
\tag 5.2b
$$
where $\alpha_{N-1}$ is chosen so that $A^{[N]}_K$ has a zero 
eigenvalue. That such an $\alpha_{N-1}$ exists is the content of 

\proclaim{Lemma 5.1} There is a unique $\alpha_{N-1}>0$ so that 
{\rom{(5.2b)}} has a zero eigenvalue. $A^{[N]}_K \geq 0$ and 
$\alpha_{N-1}$ obeys
$$
(b_{N-1} - \alpha_{N-1}) P_{N-1}(0) + a_{N-2} P_{N-2}(0)=0 
\tag 5.3
$$
and
$$
\alpha_{N-1} = -a_{N-1}\, \frac{P_N(0)}{P_{N-1}(0)}\, . \tag 5.4
$$
Moreover,
$$
(b_N-\alpha_N)(\alpha_{N-1}) -a^2_{N-1} = 0. \tag 5.5
$$
\endproclaim

\demo{Proof} Any solution of $A^{[N]}_K u=0$ must obey the 
eigenfunction equation (4.1) and so be a multiple of $P_j(0)$, 
that is,
$$
u_j = P_j(0), \qquad j=0,1,\dots,N-1. 
$$
This obeys the condition at the lower right of the matrix if 
and only if (5.3) holds. 

If $P_{N-1}(0)$ were $0$, then $A^{[N-1]}_F$ would have a zero 
eigenvalue, which is impossible since $A^{[N-1]}_F$ is strictly 
positive definite (since the form $S_{N-1}$ is strictly positive 
definite). Thus, $P_{N-1}(0)\neq 0$ and so (5.3) has a unique 
solution. Since $A^{[N]}_F$ is positive definite, this solution 
must have $\alpha_{N-1}>0$.

The eigenfunction equation for $P_N(0)$ then yields (5.4). (5.3) 
for $N\to N+1$ and (5.4) yield two formulas for $\frac{P_N(0)}
{P_{N-1}(0)}$. Setting these to each other yields (5.5). \qed
\enddemo

The first main result of this section, towards which we are 
heading, is 

\proclaim{Theorem 5.2 (includes Theorem 4.19)} Let
$$
f^+_N(z)\equiv (\delta_0, (A^{[N]}_K-z)^{-1}\delta_0) \tag 5.6
$$
and
$$
f^-_N(z) \equiv (\delta_0, (A^{[N]}_F -z)^{-1}\delta_0). \tag 5.7
$$
Then for $x\in (0,\infty)$,
$$
f^+_N(-x) \geq f^+_{N+1}(-x) \text{ converges to } (\delta_0, 
(A_K +x)^{-1}\delta_0)\equiv f^+ (-x)
$$
while
$$
f^-_N (-x) \leq f^-_{N+1} (-x) \text{ converges to } 
(\delta_0, (A_F +x)^{-1}\delta_0) \equiv f^-(-x).
$$
The convergence holds for any $z\in\Bbb C\backslash [0,\infty)$ 
and is uniform on compacts. Moreover, for any solution $\rho$ of 
the moment problem and $x>0$,
$$
f^- (-x) \leq \int\frac{d\rho(y)}{x+y} \leq f^+ (-x). \tag 5.8
$$
\endproclaim

\remark{Remarks} 1. Notice that (5.8) is (4.42).

2. Let $A^{[N]}(\beta)$ be (5.2b) with $\alpha_{N-1}$ 
replaced by $-\beta$. Then $A^{[N]}_K$ is quite a natural 
approximation: It is that $A^{[N]} (\beta)$ which is minimal 
among the non-negative matrices --- reasonable to capture $A_K$, 
the minimal non-negative self-adjoint extension. From this point 
of view, $A^{[N]}_F$ seems less natural. One would instead want 
to consider $\lim_{\beta\to\infty} A^{[N]}(\beta)$. In fact, one 
can show this limit is ``essentially" $A^{[N-1]}_F$ in the sense 
that
$$
\lim_{\beta\to\infty} (\delta_0, (A^{[N]}(\beta)-z)^{-1}\delta_0) 
=(\delta_0, (A^{[N-1]}_F -z)^{-1} \delta_0).
$$
Thus, the $A^{[N]}(\beta)$ interpolate between $A^{[N]}_K$ 
and $A^{[N-1]}_F$. 
\endremark

\proclaim{Proposition 5.3} Let $\tilde A^{[N]}_K$ be the operator 
on $\ell^2$ which is $A^{[N]}_K \oplus 0$. Then $\tilde A^{[N]}_K$ 
converges to $A_K$ in strong resolvent sense.
\endproclaim

\remark{Remark} For self-adjoint operators $\{A_n\}^\infty_{n=1}$ 
and $A_\infty$, we say $A_n$ converges to $A_\infty$ in strong 
resolvent sense if and only if for all $z\in\Bbb C\backslash
\Bbb R$, $(A_n -z)^{-1}\varphi \to (A_\infty - z)^{-1}\varphi$ 
for all vectors $\varphi$. It is a basic theorem (see Reed-Simon 
[\rsI]) that it suffices to prove this for a single $z\in 
\Bbb C\backslash\Bbb R$. The same proof shows that if all 
$A_n$ and $A_\infty$ are non-negative, one then also has 
convergence for $z\in(-\infty, 0)$. Moreover, one has 
convergence uniformly for $z$ in compact subsets of $\Bbb C
\backslash [0,\infty)$.
\endremark

\demo{Proof} Suppose first that the Hamburger problem is 
indeterminate. Thus, $\pi(0)=\mathbreak (P_0(0),\dots) \in 
\ell_2$. Let $P^{[n]}(0)$ be the vector $(P_0(0), \dots, 
P_{n-1}(0), 0,0,\dots )$. Then $A^{[n]}_K P^{[n]} \mathbreak 
=0$ so $(A^{[n]}_K-z)^{-1} P^{[n]}=-z^{-1} P^{[n]}$ for any 
$z\in\Bbb C\backslash\Bbb R$. Since $\| (A^{[n]}_K -z)^{-1}\| 
\leq |\text{Im}\,z|^{-1}$ for any such $z$, and $\|P^{[n]} - 
\pi (0)\| \to 0$, we conclude that
$$
(A^{[n]}_K -z)^{-1} \pi (0) \to -z^{-1} \pi (0) = 
(A_K -z)^{-1} \pi (0) \tag 5.9
$$
since $A_K \pi (0) =0$.

Suppose $\varphi = (A-z)\eta$ for $\eta\in D(A)$, that is, a 
finite sequence. Then $(A_K -z)\eta = \varphi = (A^{[n]}_K -z)
\eta$ for $n$ large, and thus $\lim_{n\to\infty} (A^{[n]}_K 
-z)^{-1} \varphi = \eta= (A_K -z)^{-1} \varphi$. Thus, $(A^{[n]}_K 
-z)^{-1}\varphi \to (A-z)^{-1}\varphi$ for $\varphi\in 
\overline{\text{Ran}(A-z)}$. In the determinate case, that is 
all $\varphi$'s; while in the indeterminate case, we claim that 
$\overline{\text{Ran}(A-z)} + [\pi(0)]$ is all of $\ell^2$ 
so (5.9) completes the proof. 

That leaves the proof of our claim that $\overline{\text{Ran}
(A-z)} + [\pi(0)]$ is $\ell^2$. We first note that $\pi(0)\notin 
D(\bar A)$ by Proposition~4.4. $(A_K -z)^{-1} 
(\,\overline{\text{Ran}(A-z)} + [\pi(0)]) = D(\bar A) + [\pi(0)]$ 
since $(A_K - z)^{-1})\pi(0) =-z^{-1}\pi(0)$. But 
$D(A_K)/D(\bar A)$ has $\dim 1$ and $\pi(0)\notin D(\bar A)$ so 
$(A_K-z)^{-1} (\, \overline{\text{Ran}(A-z)} + [\pi(0)])=
D(A_K)$ and thus, since $\text{Im}\, z\neq 0$, 
$\overline{\text{Ran}(A-z)} + [\pi(0)]=\ell_2$, as claimed. \qed
\enddemo

\proclaim{Proposition 5.4} Let $\Tilde{\Tilde A}^{[N]}_K$ be the 
$(N+1)\times (N+1)$ matrix which is $A^{[N]}_K \oplus 0$, 
that is, it has zeros in its last row and column. Then
$$
\Tilde{\Tilde A}^{[N]}_K \leq A^{[N+1]}_K. \tag 5.10
$$
In particular, for $x>0$,
$$
(\delta_0, (A^{[N+1]}_K + x)^{-1}\delta_0) \leq 
(\delta_0, (\Tilde{\Tilde A}^{[N]}_K + x)^{-1} \delta_0) = 
(\delta_0, (A^{[N]}_K + x)^{-1}\delta_0).
$$
\endproclaim

\remark{Remark} Propositions~5.3 and 5.4 prove the part of 
Theorem~5.2 involving monotonicity of $f^+_N$ and its 
convergence. 
\endremark

\demo{Proof} $B\equiv A^{[N+1]}_K - \Tilde{\Tilde A}^{[N]}_K$ is 
an $(N+1)\times (N+1)$ matrix which has all zeros except for a 
$2\times 2$ block in its lower right corner. This block is
$$
\pmatrix \alpha_{N-1} & a_{N-1} \\
a_{N-1} & b_N - \alpha_N \endpmatrix ,
$$
which is symmetric with a positive trace and determinant zero 
(by (5.5)). Thus, the block is a positive rank one operator so 
$B$ is positive, proving (5.10). \qed
\enddemo

The monotonicity of the $A^{[N]}_F$'s and their convergence is 
a special case of monotone convergence theorems for forms. 
Consider non-negative closed quadratic forms whose domain may not 
be dense. Each such form $q_B$ with $D(q_B)$ is associated to a 
non-negative self-adjoint operator $B$ on $\overline{D(q_B)}$. 
We define $(B -x)^{-1}$ to be $0$ on $D(q_B)^\bot$. This 
essentially sets $B=\infty$ on $D(q_B)^\bot$ consistent with the 
intention that $D(q_B)$ is the set of $\varphi$ for which 
$q_B(\varphi) <\infty$.

Given two forms $q_B$ and $q_C$, we say $q_B \leq q_C$ if 
and only if $D(q_B)\supseteq D(q_C)$ (intuitively $q_C(\varphi) 
< \infty$ implies $q_B(\varphi) <\infty$) and for all $\varphi
\in q_C$, $q_C(\varphi)\geq q_B(\varphi)$. It is a fundamental 
result [\kato, \simjfa] that if $q_B \leq q_C$, then for all 
$x\geq 0$ and all $\varphi$, 
$$
(\varphi, (C+x)^{-1}\varphi) \leq (\varphi, (B+x)^{-1}\varphi). 
\tag 5.11
$$

One monotone convergence theorem for forms [\kato, \simjfa] 
says the following: Let $\{q_{A_n}\}^\infty_{n=1}$ be a sequence 
of quadratic forms with $q_{A_n} \geq q_{A_{n+1}}$. Then $(A_n 
+x)^{-1}$ converges to the resolvent, $(A_\infty +x)^{-1}$, of an 
operator $A_\infty$. It is the operator associated to the form 
$q_{A_\infty}$ defined to be the closure of $\tilde q_{A_\infty}$ 
where $D(\tilde q_{A_\infty}) = \cup D(q_{A_n})$ with 
$\tilde q_{A_\infty}(\varphi) = \lim_{n\to\infty} q_{A_n} 
(\varphi)$ provided that the form $\tilde q_{A_\infty}$ is 
closable.

We can apply this to the $A^{[N]}_F$. Let $q^{[N]}_F$ be defined 
to be the form with $D(q^{[N]}_F)=\{(\varphi_0, \dots, 
\varphi_{N-1}, 0, \dots, 0, \dots)\}$ and $q^{[N]}_F (\varphi) 
= (\varphi, A\varphi)$. Then $D(q^{[N]}_F)\subset D(q^{[N+1]}_F)$ 
and $q^{[N]}_F (\varphi) = q^{[N+1]}_F (\varphi)$ if $\varphi\in 
D(q^{[N]}_F)$. Thus, $q^{[N]}_F \geq q^{[N+1]}_F$ (in essence, 
$\tilde A^{[N]}_F = A^{[N]}_F \oplus \infty$ on $\Bbb C^{N+1}$). 
The monotone convergence theorem applies. $A_\infty$ is just the 
closure of $\varphi\mapsto (\varphi, A\varphi)$ on finite sequences, 
that is, the Friedrichs extension $A_F$, so we have the 
convergence and monotonicity part of Theorem~5.2 for $f^-_N$ 
(using (5.11)):

\proclaim{Proposition 5.5} Let $x>0$. Then $(\delta_0, 
(A^{[N+1]}_F +x)^{-1} \delta_0) \geq (\delta_0, (A^{[N]}_F + x)^{-1}
\delta_0)$ and this converges to $(\delta_0, (A_F+x)^{-1}\delta_0)$ 
as $N\to\infty$. 
\endproclaim

To head towards the remaining part of Theorem~5.2, viz.~(5.8), 
we need expressions for $(\varphi, (A^{[N]}_F -z)^{-1}\varphi)$ 
and $(\varphi, (A^{[N]}_K -z)^{-1}\varphi)$, which are of 
independent interest.

\proclaim{Proposition 5.6}
$$
(\delta_0, (A^{[N]}_F -z)^{-1}\delta_0)= 
-\frac{Q_N(z)}{P_N(z)}\, \tag 5.12
$$
\endproclaim

\demo{Proof} The only possible eigenfunctions of $A^{[N]}_F$ are 
the vectors $\pi^{[N]} (z) \equiv (P_0(z), \dots, \mathbreak 
P_{N-1}(z))$. By the eigenfunction equation (4.1), this is an 
eigenfunction of $A^{(N)}_F$ if and only if $P_N(z)=0$. Thus, the 
$N$ poles of $(\delta_0, (A^{[N]}_F -z)^{-1}\delta_0)$ are 
precisely the $N$-zeros of $P_N(z)$. The zeros are distinct 
because the eigenvalues are simple (since $\pi^{[N]}(z)$ is the 
only potential eigenfunction, since eigenfunctions must have 
$u_0\neq 0$).

Now let the $B^{[N]}_F$ be the $(N-1)\times (N-1)$ matrix obtained 
by removing the top row and left column of $A^{[N]}_F$. By 
Cramer's rule, $(\delta_0, (A^{[N]}_F -z)^{-1}\delta_0) = 
\frac{\det (B^{[N]}_F -z)}{\det (A^{[N]}_F-z)}$, so the $N-1$ 
zeros of $(\delta_0, (A^{[N]}_F -z)^{-1}\delta_0)$ are precisely 
the $N-1$ eigenvalues of $B^{[N]}_F$. The only possible 
eigenfunctions of $B^{[N]}_F$ are $\xi(z)^{[N-1]} = (Q_1(z), 
\dots, Q_{N-1}(z))$ (since $Q_0(z)=0$, $Q_1(z)=\frac{1}{a_0}$). 
Thus, the eigenvalues of $B^{[N]}_F$ are precisely $z$'s with 
$Q_N(z)=0$. It follows that
$$
(\delta_0, (A^{[N]}_F -z)^{-1}\delta_0) = \frac{d_N Q_N(z)}
{P_N(z)}\, , 
$$
and we need only show that $d_N \equiv -1$. 

Since $(\delta_0, (A^{[N=1]}_F -z)^{-1}\delta_0) = (b_0 -z)^{-1}$ 
and $Q_1(z) = \frac{1}{a_0}$, $P_1(z) = \frac{(z-b_0)}{a_0}$, 
we see that $d_1 =-1$. On the other hand, Proposition~4.1 implies 
that
$$
\frac{Q_k (z)}{P_k (z)} - \frac{Q_{k-1}(z)}{P_{k-1}(z)} = 
\frac{1}{a_{k-1} P_k (z) P_{k-1}(z)} = o\biggl(\frac{1}{z}\biggr).
$$
Moreover, $(\delta_0, (A-z)^{-1})\delta_0) = -z^{-1} +O(z^{-2})$, 
so $d_k = d_{k-1}$, that is, $d_N =-1$ for all $N$. \qed
\enddemo

\remark{Remarks} 1. The proof shows that for suitable $c_N$ 
(indeed, an induction shows that $c_N = (-1)^N (a_0 \dots 
a_{N-1})^{-1}$), we have
$$
P_N(z) = c_N \det(A^{[N]}_F -z) \tag 5.13
$$
and
$$
Q_N(z) = -c_N \det(B^{[N]}_F -z). \tag 5.14
$$

2. Since $A^{[N]}_F$ is strictly positive definite, (5.13) shows 
once more that $P_N(0)\neq 0$ in the Stieltjes case. Since 
$B^{[N]}_F$ is also strictly positive definite (as a submatrix 
of $A^{[N]}_F$), (5.14) shows that $Q_N(0)\neq 0$ also in the 
Stieltjes case.
\endremark

(5.13)--(5.14) imply some facts about the zeros of $P_N$ and 
$Q_N$:

\proclaim{Proposition 5.7} All the zeros of each $P_N(z)$ and 
each $Q_N(z)$ are real. Moreover, there is exactly one zero of 
$Q_N(z)$ and one zero of $P_{N-1}(z)$ between any pair of 
successive zeros of $P_N(z)$.
\endproclaim

\demo{Proof} The first assertion follows from the fact that the 
eigenvalues of a real-symmetric matrix are all real. The second 
assertion follows from the fact that if $X$ is an $N\times N$ 
real-symmetric matrix and $Y$ is an $N-1 \times N-1$ submatrix, 
then by the min-max principle, there is exactly one eigenvalue 
of $Y$ between any pair of eigenvalues of $X$. \qed
\enddemo

\remark{Remark} Since the $Q_N$'s are orthogonal polynomials for 
another moment problem (see Proposition~5.16), between any two 
successive zeros of $Q_N(z)$, there is a single zero of $Q_{N-1} 
(z)$. 
\endremark

Now define
$$\align
M_N(z) &= P_N(z) - P_N(0)\, \frac{P_{N-1}(z)}{P_{N-1}(0)} \tag 5.15 \\
N_N(z) &= Q_N(z) - P_N(0)\, \frac{Q_{N-1}(z)}{P_{N-1}(0)}\tag 5.16
\endalign
$$
since $A^{[j]}_F$ is strictly positive, $0$ is never an eigenvalue 
and $P_j(0)\neq 0$ for all $j$, so $M$ is well-defined. Notice 
$M_N(0)=0$.

\proclaim{Proposition 5.8}
$$
(\delta_0, (A^{[N]}_K -z)^{-1}\delta_0) = -\frac{N_N(z)} 
{M_N(z)} \tag 5.17
$$
\endproclaim

\demo{Proof} Let $c_N$ be the constant in (5.13) and let 
$B^{[N]}_K$ be $A^{[N]}_K$ with the top row and left-most 
column removed. We will prove that
$$
M_N(z) = c_N\det(A^{[N]}_K-z); \qquad N_N(z)=-c_N \det 
(B^{[N]}_K -z), \tag 5.18
$$
from which (5.17) holds by Cramer's rule.

Clearly,
$$
\det (A^{[N]}_K -z) = \det(A^{[N]}_F -z) - \alpha_{N-1} 
\det(A^{[N-1]}_F -z)
$$
so
$$
c_N \det(A^{[N]}_K -z) = P_N(z) - \beta_N P_{N-1}(z),
$$
where
$$
\beta_N = \alpha_{N-1} \, \frac{c_N}{c_{N-1}}\, .
$$
But $\det(A^{[N]}_K) =0$ by construction. So $P_N(0) - \beta_N 
P_{N-1}(0)=0$ and thus, $c_N \det (A^{[N]}_K -z) = M_N (z)$.

In the same way, 
$$
c_N \det(B^{[N]}_K -z) = -Q_N(z) + \beta_N Q_{N-1}(z) = -N_N (z). 
\qed
$$
\enddemo

Because of our convergence theorem, we have 

\proclaim{Corollary 5.9} Let $\{\lambda^{[N]}_i\}^N_{i=1}$ 
be the zeros of $P_N(z)$ and let $\nu^{[N]}_i = -
\frac{Q_N (\lambda^{[N]}_i)}{P'_N (\lambda^{[N]}_i)}$. Then 
$\sum_{i=1}^N \nu^{[N]}_i \delta (\lambda - \lambda^{[N]}_i)$ 
converges to $d\mu_F (\lambda)$, the Friedrichs solution 
of the moment problem. A similar formula holds for the Krein 
solution, $d\mu_K(\lambda)$, with $P,Q$ replaced by $M,N$.
\endproclaim

The following concludes the proof of Theorem~5.2. It extends 
the calculation behind (4.4):

\proclaim{Theorem 5.10} Let $d\rho$ solve the Stieltjes moment 
problem. Then for $x> 0$ and $N\geq 1$, 
$$
-\frac{Q_N(-x)}{P_N(-x)} \leq \int \frac{d\rho(y)}{x+y} 
\leq -\frac{N_N(-x)}{M_N (-x)}\, . \tag 5.19
$$
\endproclaim

\demo{Proof} $\frac{P_N (y) - P_N(z)}{y-z}$ is a polynomial in 
$y$ of degree $N-1$ so
$$
\int d\rho(y) P_N(y)\biggl[ \frac{P_N(y)-P_N(z)}{y-z}\biggr] 
=0. \tag 5.20
$$
On the other hand, (4.4) says that
$$
\int d\rho(y)\,  \frac{P_N(y)}{y-z} = P_N(z) \int\frac{d\rho(y)}
{y-z} + Q_N(z). \tag 5.21
$$
Thus for $z=-x$ with $x>0$, 
$$
0\leq \int\frac{d\rho(y) P^2_N(y)}{y+x} = P_N(-x) \int 
\frac{d\rho(y)P_N(y)}{y+x} = P_N (-x)^2 \int \frac{d\rho(y)}
{x+y} + Q_N (-x) P_N (-x). 
$$
Dividing by $P_N(-x)^2$, we obtain the left-most inequality in 
(5.19).

Similarly, since $M_N(0)=0$, $\frac{M_N(z)}{z}$ is a polynomial 
of degree $N-1$ and so $[y^{-1}M_N(y) -z^{-1} M_N(z)]/y-z$ 
is a polynomial of degree $N-2$ in $y$, which is orthogonal to 
$P_N (y)$ and $P_{N-1}(y)$ and so to $M_N(y)$. Thus
$$
\int d\rho(y)\, \frac{M_N(y)}{y-z} \biggl[ \frac{M_N(y)}{y} 
-\frac{M_N(z)}{z}\biggr] =0. \tag 5.22
$$
Since for $N\geq 0$,
$$
Q_N(z) = \int d\rho(y)\, \frac{P_N(y)-P_N(z)}{y-z} \tag 5.23
$$
for each $z$, we see that for $N\geq 1$,
$$
N_N(z) = \int d\rho(y) \, \frac{M_N (y)- M_N(z)}{y-z}\, . 
\tag 5.24
$$
Therefore, for $x>0$,
$$\align
0\leq \int \frac{d\rho(y)}{y+x}\,\, \frac{M_N(y)^2}{y} &= 
\frac{M_N(-x)}{(-x)} \int d\rho(y)\, \frac{M_N(y)}{y+x} 
\qquad \text{by (5.22))} \\ 
&=\frac{M_N(-x)^2}{(-x)} \int \frac{d\rho(y)}{x+y} + 
\frac{M_N(-x)N_N(-x)}{(-x)}\, . 
\endalign
$$
Dividing by $\frac{M_N(-x)^2}{x}$, we obtain the second half 
of (5.19). \qed
\enddemo

Next, we turn to Theorem~6 and the connection of Pad\'e 
approximants to these finite matrix approximations. Given a formal 
power series $\sum_{n=0}^\infty \kappa_n z^n$, we define the 
Pad\'e approximants $f^{[N,M]}(z)$ as follows (see Baker and 
Graves-Morris [\bgm] for background on Pad\'e approximants): We 
seek a function $f^{[N,M]}(z)$ so that
$$
f^{[N,M]}(z) = \frac{A^{[N,M]}(z)}{B^{[N,M]}(z)}\, , \tag 5.25
$$
where $A^{[N,M]}(z)$ is a polynomial of degree at most $N$, 
$B^{[N,M]}(z)$ is a polynomial of degree at most $M$, as $z\to 0$,
$$
f^{[N,M]}(z) - \sum_{j=0}^{N+M} \kappa_j z^j = O(z^{N+M+1}) 
\tag 5.26
$$
and 
$$
B^{[N,M]}(0)=1. \tag 5.27
$$

There is at most one solution, $f$, for these equations since if 
$\tilde A, \tilde B$ are another pair of such polynomials, by 
(5.26), $\tilde A B-A\tilde B = O(z^{N+M+1})$ must be zero since 
$\tilde A B-A\tilde B$ is a polynomial of degree $N+M$. Thus,
$$
A^{[N,M]}(z) \tilde B^{[N,M]}(z) - \tilde A^{[N,M]}(z) 
B^{[N,M]}(z) = 0. \tag 5.28
$$
This implies $A/B = \tilde A/\tilde B$, showing $f$ is uniquely 
determined as an analytic function. If
$$
\deg A=N, \qquad \deg B=M, \qquad A,B \text{ relatively prime}, 
\tag 5.29
$$
then (5.28) shows $\tilde A = A$ and $\tilde B=B$, so $A$ and 
$B$ are uniquely determined. It can be shown ([\bgm]) that if 
$$
\det ((\kappa_{N-M+i+j-1})_{1\leq i,j\leq M})\neq 0 \tag 5.30
$$
then $A,B$ exist and obey (5.29).

There are degenerate cases where a Pad\'e approximant exists, 
but $A,B$ are not unique (e.g., for $\sum_{n=0}^\infty \kappa_n 
z^n = 1+z$, $f^{[2,1]}(z) = 1+z$ can be written as $A(z) = 
(1+z)(1+\alpha z)$, $B(z) = (1+\alpha z)$ for any $\alpha$). 
In any event, if a solution of (5.25)--(5.27) exists, we say the 
$[N,M]$ Pad\'e approximant exists and denote it by $f^{[N,M]}(z)$. 

In the context of Theorem~6, we are interested in the Pad\'e 
approximants for the Taylor series of
$$
f(z) = \int_0^\infty \frac{d\rho(x)}{1+zx}\, , \tag 5.31
$$
where $\rho$ is a measure with finite moments. Without loss, 
normalize $\rho$ so $\int d\rho =1$. Recall the context of 
Theorem~6. We have
$$
\kappa_n = (-1)^n \int_0^\infty x^n\, d\rho(x)
$$
and want to formally sum near $z=0$:
$$
f(z) \sim \sum_{n=0}^\infty \kappa_n z^n.
$$
If we define
$$\gamma_n = \int_0^\infty x^n\, d\rho(x),
$$
then near $w=\infty$, 
$$
G_\rho (w) \equiv \int_0^\infty \frac{d\rho(x)}{x-w} \sim 
-\sum_{n=0}^\infty \gamma_n w^{-n-1}. \tag 5.32
$$

Thus formally,
$$
f(z) = \frac{1}{z}\, G_\rho \biggl(-\frac{1}{z}\biggr). \tag 5.33
$$

We begin by noting:

\proclaim{Proposition 5.11} As $|w|\to\infty$, 
$$\align
-\frac{Q_N(w)}{P_N(w)} &= -\sum_{j=0}^{2N-1} \gamma_j w^{-j-1} 
+O(w^{-2N-1}) \tag 5.34 \\
-\frac{N_N(w)}{M_N(w)} &= -\sum_{j=0}^{2N-2} \gamma_j 
w^{-j-1} + O(w^{-2N}) \tag 5.35
\endalign
$$
\endproclaim

\demo{Proof} Let $A$ be the Jacobi matrix for the moment problem 
$\{\gamma_n\}^\infty_{n=0}$. Then
$$
\gamma_n = \langle \delta_0, A^n \delta_0 \rangle. \tag 5.36
$$
On the other hand, we have an expansion converging near infinity 
for
$$
\langle \delta_0, (A^{[N]}_F -w)^{-1}\delta_0\rangle  = 
-\sum_{j=0}^\infty (\delta_0, (A^{[N]}_F)^j \delta_0)w^{-j-1},
$$
so (5.34) follows from
$$
\langle \delta_0, (A^{[N]}_F)^j \delta_0\rangle = 
\langle \delta_0, A^j \delta_0\rangle , \qquad j=0,1,\dots, 2N-1. \tag 5.37
$$
To prove (5.37), note that
$$
(A^{[N]}_F)^j \delta_0 = A^j_F \delta_0 \tag 5.38
$$
if $j=0,\dots, N-1$ and for $B$ symmetric,
$$
\langle \delta_0, B^{2j}\delta_0\rangle = \|B^j\delta_0\|
$$
and that
$$
\langle \delta_0, B^{2j+1}\delta_0\rangle  = \langle B^j\delta_0, B(B^j 
\delta_0)\rangle. \tag 5.39
$$

To obtain (5.35), note that if $A^{[N]}_F$ is replaced by 
$A^{[N]}_K$, (5.38) still holds for $j=0,\dots, N-1$, but 
(5.39) fails for $j=N-1$ and so the analog of (5.37) only holds 
for $j=0,\dots, 2N-2$. \qed
\enddemo

\remark{Remark} While (5.37) holds for $j=0,1,2,\dots, 2N-1$, 
it never holds for $j=2N$. For if it did, we would have for any 
polynomial, $R$, of degree $2N$ that
$$
\langle \delta_0, R(A^{[N]}_F)\delta_0\rangle = 
\langle \delta_0, R(A)\delta_0\rangle .
$$
But this is false for $R(x) = x^2 P_{N-1}(x)^2$ since
$$
\langle \delta_0, R(A)\delta_0\rangle = 
\|A\delta_{N-1}\|^2 = a^2_{N-1} + a^2_{N-2} + b^2_{N-1},
$$
while
$$
\langle \delta_0, R(A^{[N]}_F)\delta_0\rangle = 
\|A^{[F]}_N \delta_{N-1}\|^2 =a^2_{N-2} + b^2_{N-1}.
$$
\endremark

\demo{Proof of Theorem 6} By (5.33) and (5.34) as $z\to 0$,
$$
-\frac{Q_N(-\frac{1}{z})}{zP_N(-\frac{1}{z})} = 
\sum_{j=0}^{2N-1} \kappa_j z^j +O(z^{2N}). \tag 5.40
$$
Now $z^{N-1} Q_N(-\frac{1}{z})$ is a polynomial of degree $N-1$ 
since $Q_N(0)\neq 0$ and $z^N P_N(-\frac{1}{z})$ is a polynomial 
of degree $N$ since $P_N(0)\neq 0$. Moreover, $\lim_{z\to 0} 
z^N P_N(-\frac{1}{z})\neq 0$ since $P_N$ has degree $N$. Thus 
(5.40) identifies $f^{[N-1,N]}$ and $-z^{N-1} Q_N(-\frac{1}{z})
/ z^N P_N (-\frac{1}{z})$. Similarly, noting that since $M_N(0)
=0$ (but $M'_N(0)\neq 0$ since $A^N_K$ has simple eigenvalues), 
$z^N M_N(-\frac{1}{z})$ is a polynomial of degree $N-1$, and we 
identify $f^{[N-1, N-1]}(z)$ and $-z^{N-1}N_N(-\frac{1}{z}) / 
z^N M_N(-\frac{1}{z})$. Theorem~5.2 thus implies Theorem~6. \qed
\enddemo

\remark{Remarks} 1. Since $P_N$ and $Q_N$ are relatively prime 
(by Proposition~5.7) and similarly for $M_N$, $N_N$, we are in 
the situation where the numerator and denominator are uniquely 
determined up to a constant.

2. Suppose $\gamma_n \leq Ca^n$ so $\rho$ is unique and supported 
on some interval $[0, R]$. Then if $R$ is chosen as small as 
possible, $\sum_{n=0}^\infty \kappa_n z^n$ has radius of 
convergence $R^{-1}$, but the Pad\'e approximants converge in 
the entire region $\Bbb C\backslash (-\infty, -R^{-1}]$!
\endremark

Before leaving our discussion of Pad\'e approximants for series 
of Stieltjes, we will consider the sequences $f^{[N+j, N]}(z)$ 
for $j$ fixed. We will show they each converge as $N\to\infty$ 
with limits that are all distinct (as $j$ varies) if the 
Stieltjes moment problem is indeterminate and all the same if 
a certain set of Stieltjes moment problems are all determinate.

Given a set of moments $\{\gamma_j\}^\infty_{j=1}$ and $\ell=1,2,
\dots$, define for $j=0,1,\dots$
$$
\gamma^{(\ell)}_j \equiv \frac{\gamma_{j+\ell}}{\gamma_\ell}. 
\tag 5.41
$$

\proclaim{Proposition 5.12} Suppose that $\{\gamma_j
\}^\infty_{j=0}$ is a set of Stieltjes \rom(resp.~Hamburger\rom) 
moments. Then $\{\gamma^{(\ell)}_j\}^\infty_{j=0}$ is the set of 
moments for a Stieltjes \rom(resp.~Hamburger\rom) problem for 
$\ell = 1,2,\dots$ \rom(resp.~$\ell=2,4,6,\dots$\rom). This 
problem is indeterminate if the original problem is.
\endproclaim

\demo{Proof} We will consider the Stieltjes case. Let $\rho\in 
\Cal M^S(\gamma)$. Then $\gamma^{-1}_\ell x^\ell\, d\rho(x) 
\equiv d\rho^{(\ell)}(x)$ solves the moment problem for 
$\gamma^{(\ell)}$. If $\rho_1, \rho_2 \in\Cal M^S(\gamma)$, 
then $\rho^{(\ell)}_1, \rho^{(\ell)}_2$ are both in $\Cal M^S 
(\gamma^{(\ell)})$. If $\rho^{(\ell)}_1 = \rho^{(\ell)}_2$, then 
$\rho_1 - \rho_2$ is supported at zero, which means it is zero 
since $\int d\rho_1(x)=\gamma_0=\int d\rho_2 (x)$. Thus, if the 
$\gamma$ problem is indeterminate, so is the $\gamma^{(\ell)}$ 
problem. \qed
\enddemo

\example{Example} As Corollary~4.21 shows, there can be determinate 
moment problems so that $\gamma^{(1)}$ is indeterminate (in the 
language of that corollary, $\tilde\gamma^{(1)}=\gamma$). Thus, 
the converse of the last statement in Proposition~5.12 fails.
\endexample

As an aside, we present a proof of a criterion for indeterminacy 
of Hamburger [\ham], especially interesting because of a way of 
rewriting it in terms of a single ratio of determinants (see 
Theorem~A.7). Note this aside deals with the Hamburger problem.

\proclaim{Proposition 5.13} A necessary and sufficient condition 
for a set of Hamburger moments to be indeterminate is that
\roster
\item"\rom{(i)}" $\sum_{j=0}^\infty |P_j (0)|^2 < \infty$ and
\item"\rom{(ii)}" $\sum_{j=0}^\infty |P_j^{(2)}(0)|^2 < \infty$
\endroster
where $P^{(2)}_j(x)$ are the orthogonal polynomials for the 
$\gamma^{(2)}$ moment problem.
\endproclaim

\demo{Proof} If $\{\gamma_j\}^\infty_{j=0}$ is indeterminate by 
Proposition~5.12, so is $\{\gamma_j^{(2)}\}^\infty_{j=0}$, and 
then (i), (ii) follow by Theorem~3. The converse is more 
involved. The intuition is the following: If $d\rho$ is an even 
measure, then $P^{(2)}_{2j}(x)=\sqrt{\gamma_2} \frac{P_{2j+1}(x)}
{x}$, so $P^{(2)}_m(0)=\sqrt{\gamma_2}\frac{\partial P_{m+1}}
{\partial x} (0)$ (both sides are zero if $m$ is odd). So (ii) 
is equivalent to $\sum_{j=0}^\infty |\frac{\partial P_j(0)}
{\partial x}|^2 <\infty$ and condition (v) of Theorem~3 says that 
(i), (ii) imply indeterminacy. Our goal is to prove in general
that when (i) holds, $P^{(2)}_j(0)$ and $\frac{\partial P_{j+1}}
{\partial x}(0)$ are essentially equivalent.

Let $S_n(x) = P^{(2)}_n (x)$, $\eta_n = \frac{\partial P_n}
{\partial x}(0)$, and define
$$\align
\alpha_n &= \int x S_n(x)\, d\rho(x) \tag 5.42 \\
\beta_n &= \int [x S_n (x)][P_{n+1}(x)]\, d\rho(x), \tag 5.43
\endalign
$$
where $\rho$ is any solution of the $\gamma$ moment problem. 
Since $S_n$ is an orthogonal polynomial for $d\rho^{(2)}$, we 
have
$$
\int [x S_n(x)]x^j\, xd\rho(x)=0, \qquad j=0,1,\dots, n-1
$$
and thus
$$
\int x S_n(x) P_j(x)\, d\rho(x) = \alpha_n P_j (0), 
\qquad j=0,1,\dots, n
$$
so the orthogonal expansion for $xS_n(x)$ is
$$
xS_n (x) = \beta_n P_{n+1}(x) + \alpha_n \sum_{j=0}^n 
P_j (0) P_j(x). \tag 5.44
$$
Since $xS_n(x)$ vanishes at zero,
$$
\beta_n P_{n+1}(0) + \alpha_n \sum_{j=0}^n |P_j (0)|^2 = 0. 
\tag 5.45
$$
Since
$$
\int |xS_n (x)|^2 \, d\rho(x) = \gamma_2 \int S_n(x)^2\, 
d\rho^{(2)}(x) = \gamma_2, 
$$
we have by (5.44) that
$$
\gamma_2 = \beta^2_n + \alpha^2_n \sum_{j=0}^n |P_j(0)|^2. 
\tag 5.46
$$

Moreover, taking derivatives of (5.44) at $x=0$,
$$
S_n(0) = \beta_n \eta_{n+1} + \alpha_n \sum_{j=0}^n 
P_j(0)\eta_j. \tag 5.47
$$
By (5.46), $\beta_n$ is bounded, so by (5.45),
$$
|\alpha_n| \leq C_1 |P_{n+1}(0)|. \tag 5.48
$$
By hypothesis, $\sum_{j=0}^\infty |P_j (0)|^2 <\infty$, so 
$P_{n+1}(0)\to 0$ and thus by (5.46), $\beta_n \to \sqrt{\gamma_2}$ 
as $n\to\infty$. Since $\beta_n >0$ for all $n$ (this follows from 
the fact that both $xS_n(x)$ and $P_{n+1}(x)$ have positive 
leading coefficient multiplying $x^{n+1}$), we see that  
$\beta^{-1}_n$ is bounded. Using (5.48), the Schwarz inequality 
on the sum in (5.47), and $\sum_{j=0}^\infty |P_j(0)|^2 < 
\infty$ again, we see that
$$\align
|\eta_{n+1}|^2 &\leq C\biggl(|S_n(0)|^2 + |P_{n+1}(0)|^2 
\sum_{j=0}^n |\eta_j|^2\biggr) \\
&\leq C (|S_n(0)|^2 + |P_{n+1}(0)|^2) \biggl( 1 + \sum_{j=0}^n 
|\eta_j|^2\biggr).
\endalign
$$
It follows that
$$
\biggl[ 1 + \sum_{j=0}^{n+1} |\eta_j|^2\biggr] \leq [1+ C
(|S_n(0)|^2 + |P_{n+1}(0)|^2)]\biggl[ 1 + \sum_{j=0}^n 
|\eta_j|^2\biggr]
$$
and so by induction that
$$
\sup_{1 \leq n < \infty} \sum_{j=0}^{n+1} |\eta_j|^2 \leq 
\sup_{1 \leq n < \infty} \prod_{j=1}^{n+1} [1+ C(|S_{j-1}(0)|^2 
+ |P_j (0)|^2)] <\infty
$$
since $\sum_{j=0}^\infty |P_j(0)|^2 + \sum_{j=0}^\infty |S_j 
(0)|^2 < \infty$ by hypothesis. Thus, if (i) and (ii) hold, 
$\eta_n \in\ell_2$ and thus, the problem is indeterminate by 
Theorem~3. \qed
\enddemo

\proclaim{Theorem 5.14} Let $\{\gamma_n\}^\infty_{n=0}$ be a 
set of Stieltjes moments. Fix $\ell \geq 1$ and let $P^{(\ell)}_N 
(z)$, $Q^{(\ell)}_N(z)$, $M^{(\ell)}_N(z)$, and $N^{(\ell)}_N 
(z)$ be the orthogonal polynomials and other associated 
polynomials for the $\gamma^{(\ell)}$ moment problem. Let 
$f^{[N,M]}(z)$ be the Pad\'e approximants for the series of 
Stieltjes $\sum_{j=0}^\infty \kappa_j z^j$ where $\kappa_j = 
(-1)^j \gamma_j$. Then
$$
f^{[N+\ell-1, N]}(z) = \sum_{j=0}^{\ell-1} (-1)^j \gamma_j 
z^j + (-1)^\ell \gamma_\ell z^\ell 
\biggl[\frac{Q^{(\ell)}_N (-\frac{1}{z})}
{zP^{(\ell)}_N (-\frac{1}{z})}\biggr] \tag 5.49
$$
and
$$
f^{[N+\ell-1, N-1]}(z) = \sum_{j=0}^{\ell-1} (-1)^j \gamma_j 
z^j + (-1)^\ell \gamma_\ell z^\ell 
\biggl[\frac{N^{(\ell)}_N (-\frac{1}{z})}
{zM^{(\ell)}_N (-\frac{1}{z})}\biggr]. \tag 5.50
$$
In particular:
\roster
\item"\rom{(1)}" For each $\ell$, $(-1)^\ell f^{[N+\ell-1, N]}
(x)$ is monotone increasing to a finite limit for all $x\in
(0,\infty)$.
\item"\rom{(2)}" For all $z\in\Bbb C\backslash (-\infty, 0]$, 
$\lim_{N\to\infty} f^{[N+\ell -1, N]}(z) \equiv f_\ell(z)$ 
exists.
\item"\rom{(3)}" The $\gamma^{(\ell)}$ moment problem is 
\rom(Stieltjes\rom) determinate if and only if $f_\ell (z) = 
f_{\ell+1}(z)$.
\item"\rom{(4)}" If the $\gamma^{(\ell)}$ moment problem is 
determinate, then $f_0(z) = f_1(z) = \cdots = f_{\ell+1}(z)$.
\item"\rom{(5)}" If the $\gamma^{(\ell)}$ moment problem is 
indeterminate, then as $x\to\infty$,
$$
f_{\ell+1}(x) = (-1)^\ell \gamma_\ell \mu^{(\ell)}_K 
(\{0\}) x^\ell + O(x^{\ell-1}), \tag 5.51
$$
where $\mu^{(\ell)}_K$ is the Krein solution of the 
$\gamma^{(\ell)}$ problem. In particular if $\ell >1$, 
$-f_{\ell +1}(x)$ is not a Herglotz function \rom(as it is 
if the problem is determinate\rom).
\endroster
\endproclaim

\remark{Remarks} 1. Thus, it can happen (e.g., if $\gamma$ is 
determinate but $\gamma^{(1)}$ is not (cf.~Corollary~4.21) 
that $f^{[N-1, N]}$ and $f^{[N,N]}$ have the same limit, but 
that $f^{[N+1, N]}$ has a different limit.

2. If the $\{\gamma_j\}^\infty_{j=0}$ obey a condition that 
implies uniqueness and which is invariant under $\gamma_j \to 
\gamma_{j+1}$ (e.g., the condition of (1.12b) of Proposition~1.5), 
then all $\gamma^{(\ell)}$ are determinate; see Theorem~5.19 
below.

3. Even for $\ell=1$ and consideration of $f^{[N-1,N]}$ and 
$f^{[N,N]}$, we have something new --- the remarkable fact that 
$xd\mu_K(x)/\gamma^{-1}_1$ is $d\mu^{(1)}_F(x)$. Multiplication 
by $x$ kills the point measure at $x=0$ and produces a measure 
supported on some set $[R,\infty)$ with $R>0$. We also see that 
monotonicity for $f^-_N$ implies monotonicity of $f^+_N$. The 
direction of monotonicity flips because of the minus sign 
($(-1)^\ell = -1$) in (5.49).

4. In particular, we have $P^{(1)}_N(x) = c_N x^{-1} M_{N+1}
(x)$, that is, for any $\ell\neq m$, \linebreak $\int x^{-1} 
M_m(x) M_\ell(x)\, d\rho(x) = 0$, something that can be checked 
directly.

5. This theorem implies that all the $f^{[N,M]}$ with $N\geq 
M-1$ exist and their denominators obey a three-term recursion 
relation.

6. The connection in Remark~3 extended to $xd\mu^{(\ell)}_K
(x) = c_\ell d\mu^{(\ell+1)}_F (x)$ implies that if any 
$\gamma^{(\ell)}$ is indeterminate, then $d\mu_K(x)$ is a 
point measure. Thus if $\kappa_n = (-1)^n \int_0^\infty 
x^n \, d\rho(x)$ where $\rho$ is associated to a determinate 
Stieltjes problem and $d\rho$ is not a point measure, then all 
the $f_\ell (z)$'s, $\ell=0,1,2,\dots$, are equal.
\endremark

\demo{Proof} It is easy to check that the right side of (5.49)
and (5.50) are ratios of polynomials whose degrees are at most 
of the right size. (Because $P^{(\ell)}_N(0)\neq 0 \neq 
M^{(\ell)'}_N(0)$, it is easy that the denominators are always 
precisely of the right size.) Since $M^{(\ell)}_N (z)$ and 
$P^{(\ell)}_N(z)$ are of degree $n$, the rationalized 
denominators do not vanish at $\frac1{z}=0$. By Proposition~5.11 
for the $\gamma^{(\ell)}$ problem, they have the proper 
asymptotic series to be $f^{[N+\ell-1, N]}(z)$ and 
$f^{[N+\ell-1, N-1]}(z)$, respectively. That proves (5.49) 
and (5.50).

Assertions (1), (2), (3) are then just Theorem~6 for the 
$\gamma^{(\ell)}$ moments. To prove assertion (4), note that 
if $\gamma^{(\ell)}$ is determined by Proposition~5.12, so 
are $\gamma^{(\ell-1)}, \gamma^{(\ell-2)}, \dots, \gamma^{(1)}, 
\mathbreak \gamma$, and so by (3), we have $f_{\ell+1} = f_\ell 
= \cdots = f_0$.

To prove assertion (5), note that if the $\gamma^{(\ell)}$ 
moment problem is indeterminate, $d\mu^{(\ell)}_K(\{0\}) 
\mathbreak >0$. Thus,
$$
\lim_{x\to\infty} \int \frac{d\mu^{(\ell)}_K(y)}{1+xy} 
= \mu^{(\ell)}_K \{0\} > 0. \tag 5.52
$$
By (5.50),
$$
f_{\ell+1}(x) = \sum_{j=0}^{\ell-1} (-1)^j \gamma_j x^j + 
(-1)^\ell \gamma_\ell x^\ell \int_0^\infty 
\frac{d\mu^{(\ell)}_K(y)}{1+xy}\, ,
$$
so (5.52) implies (5.51). \qed
\enddemo

To deal with the sequences $f^{[N-1+\ell, N]}(z)$ with $\ell <0$, 
we need to introduce yet another modified moment problem 
associated to a set of Hamburger (or Stieltjes) moments. To 
motivate what we are looking for, let $\rho\in\Cal M^H(\gamma)$ 
and let $G_\rho(z)$ be the associated Stieltjes transform. Then 
$-G_\rho(z)^{-1}$ is a Herglotz function which has an asymptotic 
series to all orders as $z\to i\infty$. Since $G_\rho(z) \sim 
-\frac1{z} (1+\gamma_1 z^{-1} + \gamma_2 z^{-2} + O(z^{-3}))$,
$$
-G_\rho(z)^{-1} \sim z-\gamma_1 - (\gamma_2 - \gamma^2_1) 
z^{-1} +O(z^{-2})
$$
so (by the proof of Proposition~4.13) the Herglotz representation 
of $-G_\rho(z)^{-1}$ has the form
$$
-G_\rho(z)^{-1} = z - \gamma_1 + (\gamma_2 - \gamma^2_1) 
\int \frac{d\tilde\rho(x)}{x-z}\, , \tag 5.53
$$
where $d\tilde\rho$ is a normalized measure. Its moments, 
which we will call $\gamma^{(0)}_j$, only depend on the 
asymptotic series for $G_\rho(z)$, and so only on the original 
moments. We could do everything abstractly using (5.53), but 
we will be able to explicitly describe the relation between the 
moment problems.

The key is the following formula, a Ricatti-type equation well 
known to practitioners of the inverse problem [\gs] (which we will, 
in essence, prove below):
$$
-m_0 (z)^{-1} = z-b_0 + a^2_0 m_1 (z), \tag 5.54
$$
where $m_0(z) = \langle\delta_0, (A-z)^{-1}\delta_0\rangle$ is 
just $G_\rho(z)$ for a spectral measure, and $m_1(z) = 
\langle \delta_0, (A^{[1]} -z)^{-1}\delta_0\rangle$, where $A^{[1]}$ 
is obtained from $A$ by removing the top row and left-most column. (
5.54) is just (5.53) if we note that $b_0 = \gamma_1$ and $a^2_0 = 
\gamma_2 - \gamma^2_1$. Thus, we are led to define $\gamma^{(0)}_j$ 
as the moments associated to the Jacobi matrix $A^{[1]}$ obtained 
by removing the top row and left-most column, that is,
$$
A^{[1]} = \pmatrix
b_1 & a_1 & 0 & \dots & \dots \\ 
a_1 & b_2 & a_2 & \dots & \dots \\
0 & a_2 & b_3 & a_3 & \dots \\
\vdots & \vdots & \vdots & \vdots & \vdots 
\endpmatrix . 
$$

\proclaim{Proposition 5.15} Let $\{\gamma_j\}^\infty_{j=0}$ 
be a set of Hamburger moments. Then the $\{\gamma^{(0)}_j
\}^\infty_{j=0}$ Hamburger problem is determinate if and only 
if $\{\gamma_j\}^\infty_{j=0}$ is. 
\endproclaim

\demo{Proof} Let $\tilde A^{[1]}$ be $A^{[1]}$ with a row of 
zeros added at the top and column of zeros on the left. Since 
$\tilde A^{[1]}=0 \oplus A^{[1]}$, $\tilde A^{[1]}$ is 
essentially self-adjoint if and only if $A^{[1]}$ is. $A - 
\tilde A^{[1]}$ is a matrix with three non-zero elements and 
essential self-adjointness is provided by bounded perturbations. 
Thus, $A$ is essentially self-adjoint if and only if $A^{[1]}$ 
is. By Theorem~2, we have the equivalence of determinacy for the 
Hamburger problem. \qed
\enddemo

\remark{Remark} As we will see shortly, this result is not true 
in the Stieltjes case.
\endremark

Let $P^{(0)}_N(x)$, $Q^{(0)}_N(x)$, and $f^{(0)}_N(z) = -
\frac{Q^{(0)}_N(z)}{P^{(0)}_N(z)}$ be the polynomials and 
finite matrix approximation for the $\gamma^{(0)}$ problem. Then

\proclaim{Proposition 5.16}
$$\alignat2
&\text{\rom{(i)}} \qquad && P^{(0)}_N(x) = a_0 Q_{N+1}(x) 
\tag 5.56 \\
&\text{\rom{(ii)}} \qquad && Q^{(0)}_N(x) = -\frac1{a_0} \, 
P_{N+1}(x) + \frac{x-b_0}{a_0}\, Q_{N+1}(x) \tag 5.57 \\
&\text{\rom{(iii)}} \qquad && -f^-_{N+1} (z)^{-1} = z -b_0 + 
a^2_0 f^{(0)}_N(z) \tag 5.58
\endalignat
$$
\endproclaim

\remark{Remarks} 1. (i) implies that the $Q_N$'s are orthogonal 
polynomials for some measure.

2. (5.58) is (5.54).
\endremark

\demo{Proof} For each fixed $x$, $P^{(0)}_N(x), Q^{(0)}_N(x)$ 
obey the same difference equation as $P_{N+1}(x)$, $Q_{N+1}(x)$ 
so we need (5.56), (5.57) at the initial points $N=-1,0$ where 
it is required that
$$
P^{(0)}_{-1}(x) = 0, \quad P^{(0)}_0(x)=1; \qquad 
Q^{(0)}_{-1}(x)=-\frac{1}{a_0}, \quad Q^{(0)}_0(x) =0
$$
($Q^{(0)}_{-1}(x) = -\frac1{a_0}$ since we don't make the 
convention $a^{(0)}_{-1} = 1$, but rather $a^{(0)}_{-1} = a_0$ 
consistent with $a^{(0)}_n=a_{n+1}$). This is consistent with 
(5.56) if we note that
$$
Q_0 (x) = 0, \quad Q_1(x) = \frac1{a_0}; \qquad 
P_0 (x) = 1,  \quad P_1 (x) = \frac{(x-b_0)}{a_0}\, .
$$
This proves (i), (ii).

They in turn imply
$$
a^2_0\, \frac{Q^{(0)}_N(x)}{P^{(0)}_N(x)} = x-b_0 - 
\frac{P_{N+1}(x)}{Q_{N+1}(x)}\, ,
$$
which, by (5.12), proves (iii). \qed
\enddemo

\proclaim{Proposition 5.17} Let $\{\gamma_j\}^\infty_{j=0}$ be 
a set of Stieltjes moments. If the Hamburger problem is 
indeterminate, then the $\gamma^{(0)}$ Stieltjes problem is 
indeterminate \rom(even if the $\gamma$ Stieltjes problem is 
determinate\rom).
\endproclaim

\demo{Proof} As we will see below (Proposition 5.22 $\equiv$ 
Theorem~8), a set of Stieltjes moments which is Hamburger 
indeterminate is Stieltjes determinate if and only if $L\equiv 
\lim_{N\to\infty} [-\frac{Q_N(0)}{P_N (0)}]$ is infinite. By 
(5.58), 
$$
a^2_0 L^{(0)} = b_0 - L^{-1}.
$$
Since $L>0$, $L^{(0)}$ is never infinite. \qed
\enddemo

\remark{Remarks} 1. Thus, if $\gamma$ is Hamburger indeterminate 
but Stieltjes determinate, $\gamma$ and $\gamma^{(0)}$ have 
opposite Stieltjes determinacy.

2. The spectral way to understand this result is to note that if 
$\gamma$ is Hamburger indeterminate, the Friedrichs extensions of 
$A$ and $A^{[1]}$ have interlacing eigenvalues. Given that $A \geq 
0$, $A^{[1]}$ must be strictly positive.
\endremark

For $\ell <0$ and integral, we let
$$
\gamma^{(\ell)}_j = [\gamma^{(0)}]^{(-\ell)}_j = 
\frac{\gamma^{(0)}_{j-\ell}}{\gamma^{(0)}_{-\ell}}\, .
$$
Then:

\proclaim{Theorem 5.18} Let $\{\gamma_n\}^\infty_{n=0}$ be a set 
of Stieltjes moments. Fix $\ell \leq 0$  and let $P^{(\ell)}_N(z)$, 
$Q^{(\ell)}_N(z)$, $M^{(\ell)}_N(z)$, and $N^{(\ell)}_N(z)$ be 
the orthogonal polynomials and other associated polynomials for 
the $\gamma^{(\ell)}$ moment problem. Let $f^{[N,M]}(z)$ be the 
Pad\'e aproximants for the series of Stieltjes $\sum_{j=0}^\infty 
\kappa_j z^j$ where $\kappa_j = (-1)^j \gamma_j$. Then
$$
f^{[N+\ell-1,N]}(z) = \biggl\{ 1-\gamma_1 z - 
\sum_{j=0}^{-\ell-1} (-1)^j \gamma^{(0)}_j z^{j+2} + (-1)^{\ell+1} 
\gamma^{(0)}_{-\ell} z^{-\ell +2} \biggl[\frac{Q^{(\ell)}_{N+\ell-1} 
(-\frac{1}{z})}{zP^{(\ell)}_{N+\ell-1} (-\frac1{z})}\biggr] 
\biggr\}^{-1} \tag 5.59
$$
and
$$
f^{[N+\ell-1,N+1]}(z) = \biggl\{ 1-\gamma_1 z - 
\sum_{j=0}^{-\ell-1} (-1)^j \gamma^{(0)}_j z^{j+2} + (-1)^{\ell+1} 
\gamma^{(0)}_{-\ell} z^{-\ell +2} \biggl[\frac{N^{(\ell)}_{N+\ell-1} 
(-\frac{1}{z})}{zM^{(\ell)}_{N+\ell} (-\frac1{z})}\biggr] 
\biggr\}^{-1}. \tag 5.60
$$
In particular,
\roster
\item"\rom{(1)}" For each $\ell$, $(-1)^\ell f^{[N+\ell-1, N]} 
(x)$ is monotone increasing to a finite limit for all $x\in 
(0,\infty)$.
\item"\rom{(2)}" For all $z\in\Bbb C\backslash (-\infty, 0]$, 
$\lim_{N\to\infty} f^{[N+\ell -1, N]}(z) \equiv f_\ell(z)$ 
exists.
\item"\rom{(3)}" The $\gamma^{(\ell)}$ moment problem is 
Stieltjes determinate if and only if $f_\ell(z) = f_{\ell-1}(z)$.
\item"\rom{(4)}" If the $\gamma^{(\ell)}$ moment problem is 
determinate, then $f_0(z) = f_{-1}(z) = \cdots = f_\ell(z) 
= f_{\ell-1}(z)$.
\item"\rom{(5)}" If the $\gamma^{(\ell)}$ moment problem is 
indeterminate, then as $x\to\infty$,
$$
f_{\ell-1}(x) = (-1)^{\ell+1} (\gamma^{(0)}_{-\ell})^{-1} 
x^{\ell-2} \mu^{(\ell)}_K (\{0\})^{-1} + O(x^{\ell-3}), 
$$
where $\mu^{(\ell)}_K$ is the Krein solution of the 
$\gamma^{(\ell)}$ problem. In particular, $-f_{\ell-1}(x)$ is 
not a Herglotz function \rom(as it is if the problem is 
determinate\rom).
\endroster
\endproclaim

\demo{Proof} By (5.58), we have a relation between formal power 
series:
$$
\biggl( \sum_{j=0}^\infty (-1)^j \gamma_j z^j\biggr)^{-1} = 
1 + \gamma_1 z - (\gamma_2 - \gamma^2_1) z^2 \biggl[ 
\sum_{j=0}^\infty (-1)^j \gamma^{(0)}_j z^j\biggr], \tag 5.61
$$
from which one obtains
$$
f^{[M,N+2]}(z) = (1+\gamma_1 z - (\gamma_2 - \gamma^2_1) z^2 
f^{(0)[N,M]}(z))^{-1}.
$$
Thus, (5.59) is just (5.49) and (5.60) is (5.50). The 
consequence (1)--(5) follows as in the proof of Theorem~5.14. 
\qed
\enddemo

We summarize and extend in the following:

\proclaim{Theorem 5.19} Let $\sum_{j=0}^\infty \kappa_j z^j$ 
be a series of Stieltjes. Then for each $\ell = 0, \pm 1, \pm 2, 
\dots$, $f_\ell(z) = \lim_{N\to\infty} f^{[N+\ell, N]}(z)$ 
exists uniformly for $z$ in compact subsets of $\Bbb C 
\backslash (-\infty, 0]$. Moreover,
\roster
\item"\rom{(1)}" If any two $f_\ell$'s are different, then all of 
them are meromorphic.
\item"\rom{(2)}" If the $|\kappa_j|=\gamma_j$ moment problem 
is determinate and the measure solving the moment problem is not 
a discrete point measure, then all $f_\ell$'s are equal.
\item"\rom{(3)}" If $|\kappa_j|\leq C^j (2j)!$, then all 
$f_\ell$'s are equal.
\endroster
\endproclaim

\demo{Proof} The first assertion is Theorem~5.14 for $\ell \geq 
1$ and Theorem~5.18 for $\ell\leq 0$. If some $f_\ell \neq 
f_{\ell+1}$, then some $\gamma^{(\ell)}$ are indeterminate, so 
the corresponding $d\rho_F$'s are pure point and $f_\ell$'s 
meromorphic. This proves (1). Under the hypothesis of (2), 
$f_0$ is not meromorphic, so by (1), all $f$'s are equal. To 
prove (3), an induction using (5.61) proves $|\gamma^{(0)}_j| 
\leq \tilde C^j (2j)!$, so by Proposition~1.5, all the 
$\gamma^{(\ell)}$ moment problems are determinate. \qed
\enddemo

This completes our discusion of Pad\'e approximants for 
series of Stieltjes. We return to consequences of Theorem~5.2 
for the study of the Stieltjes moment problem.

\vskip 0.1in

As we have seen for any $x>0$, $-\frac{Q_N(-x)}{P_N(-x)}$ is 
monotone increasing and $-\frac{xN_N (-x)}{M_N(-x)}$ is monotone 
decreasing. By taking limits (recall $M_N(0)=0$), we see that 
$-\frac{Q_N(0)}{P_N(0)}$ is monotone increasing and 
$\frac{N_N(0)}{M'_N(0)}$ is decreasing.

\proclaim{Proposition 5.20} 
$$\alignat2
&\text{\rom{(i)}} \qquad && \lim_{N\to\infty} - \frac{Q_N(0)}
{P_N(0)} = \int y^{-1}\, d\mu_F (y) \\
&\text{\rom{(ii)}} \qquad && \lim_{N\to\infty} \frac{N_N(0)}
{M'_N(0)} = \mu_K (\{0\}),
\endalignat
$$
where $\mu_F$ \rom(resp.~$\mu_K$\rom) is the Friedrichs 
\rom(resp.~Krein\rom) solution.
\endproclaim

\demo{Proof} By Theorem~5.2 and Proposition~5.6, for $x>0$,
$$
-\frac{Q_N(-x)}{P_N(-x)} \leq \int (y+x)^{-1}\, d\mu_F (y),
$$
so taking $x$ to zero and  $N$ to infinity, we see that
$$
\lim_{N\to\infty} -\frac{Q_N(0)}{P_N(0)} \leq \int y^{-1} \, 
d\mu_F (y).
$$
On the other hand, since $-\frac{Q_N(-x)}{P_N(-x)}=(\delta_0, 
(A^{[N]}_F + x)^{-1}\delta_0)$ is monotone increasing as $x$ 
decreases, we have for each $N$ and $x>0$,
$$
-\frac{Q_N(-x)}{P_N(-x)} \leq \lim_{N\to\infty} 
-\frac{Q_N(0)}{P_N(0)}\, ,
$$
so taking $N$ to infinity for fixed $x>0$ and using Theorem~5.2,
$$
\int (x+y)^{-1} \, d\mu_F (y) \leq \lim_{N\to\infty} 
-\frac{Q_N(0)}{P_N(0)}\, .
$$
Taking $x$ to zero, we see that
$$
\int y^{-1} \, d\mu_F (y) \leq \lim_{N\to\infty} 
-\frac{Q_N(0)}{P_N(0)}\, ,
$$
so (i) is proven.

The proof of (ii) is similar. By Theorem~5.2 and Proposition~5.8, 
$$
-\frac{xN_N(-x)}{M_N (-x)} \geq \int \frac{x}{y+x}\, 
d\mu_K (y) \geq \mu_K (\{0\}),
$$
so taking $x$ to zero and $N$ to infinity,
$$
\lim_{N\to\infty} \frac{N_N(0)}{M'_N(0)} \geq \mu_K (\{0\}).
$$
On the other hand, since $-\frac{xN_N(-x)}{M_N(-x)} = (\delta_0, 
x(A^{[N]}_K + x)^{-1} \delta_0)$ is monotone decreasing as $x$ 
decreases, we have for each $N$ and $x>0$,
$$
-\frac{xN_N(-x)}{M_N(-x)} \geq \lim_{N\to\infty} 
\frac{N_N(0)}{M'_N(0)}\, ,
$$
so taking $N$ to infinity for fixed $x>0$ and using Theorem~5.2,
$$
\int \frac{x}{x+y}\, d\mu_K(y) \geq \lim_{N\to\infty} 
\frac{N_N(0)}{M'_N(0)}\, .
$$
Taking $x$ to zero, we see that
$$
\mu_K (\{0\}) \geq \lim_{N\to\infty} \frac{N_N(0)}
{M'_N(0)}\, ,
$$
so (ii) is proven. \qed
\enddemo

This leads us to define (note that we use $\frac{M'_N(0)}{N_N(0)}$,  
not $\frac{N_N(0)}{M'_N(0)}$ so $M,L\in (0,\infty)\cup \{\infty\}$):
$$\align
L &= \lim_{N\to\infty} -\frac{Q_N(0)}{P_N(0)} \tag 5.62 \\
M &= \lim_{N\to\infty} \frac{M'_N(0)}{N_N(0)}, \tag 5.63
\endalign
$$
so Proposition~5.20 says that
$$\align
L &= \int y^{-1}\, d\mu_F (y) \\
M^{-1} &= \mu_K (\{0\}).
\endalign
$$
By (4.25), $\mu_K (\{0\}) = 1/\sum_{n=0}^\infty |P_N(0)|^2$, so
$$
M=\sum_{n=0}^\infty |P_n(0)|^2 . \tag 5.64
$$

\demo{Note} (4.25) was only proven in the indeterminate case, 
but the argument applies in the determinate case also. If $x$ is 
an eigenvalue of a solution $\mu$ of the moment problem associated 
to a self-adjoint extension, then $\mu (\{x\}) = 1/
\sum_{n=0}^\infty |P_n(x)|^2$.
\enddemo

\proclaim{Theorem 5.21 ($\equiv$ Theorem 7)} Let $\{\gamma_n
\}^\infty_{n=0}$ be the moments of a Stieltjes problem. Then the 
problem is indeterminate if and only if
$$
L < \infty \qquad \text{\rom{and}} \qquad M<\infty.
$$
Equivalently, the problem is determinate if and only if
$$
L=\infty \qquad \text{\rom{or}} \qquad M=\infty.
$$
\endproclaim

\demo{Proof} If $L<\infty$ and  $M<\infty$, then $\int y^{-1} 
d\mu_F (y) <\infty$, while $\mu_K(\{0\}) >0$, so clearly, 
$\mu_F \neq \mu_K$ and the problem is indeterminate. Conversely, 
if the problem is indeterminate, then by Proposition~3.1, $\alpha 
=\inf \, \text{spec}(A_F)>0$, so $\int y^{-1} \, d\mu_F(y) \leq 
\alpha^{-1} < \infty$ and $L<\infty$. Moreover, since the 
Stieltjes problem is indeterminate, so is the Hamburger problem;   
and thus by (5.64), $M<\infty$. \qed
\enddemo

\proclaim{Theorem 5.22 ($\equiv$ Theorem 8)} Let $\{\gamma_n 
\}^\infty_{n=0}$ be a set of Stieltjes moments. Then the 
Stieltjes problem is determinate while the Hamburger problem is 
indeterminate if and only if
$$
\sum_{n=0}^\infty |Q_n(0)|^2 < \infty \tag 5.65
$$
and $L=\infty$.
\endproclaim

\demo{Proof} For a set of Stieltjes moments $-\frac{Q_n(0)} 
{P_n(0)}$ is positive and monotone increasing, so $|Q_n(0)| \geq 
\frac{|Q_1(0) P_n(0)|}{|P_1(0)|}$. Since $Q_1(0)\neq 0$, we see 
that (5.65) implies also that $M<\infty$. Thus, (5.65) is 
equivalent to the Hamburger problem being indeterminate. Given 
that $M<\infty$, Theorem~5.21 says that determinacy of the 
Stieltjes problem is equivalent to $L=\infty$. \qed
\enddemo

As our next topic, we will further examine the conditions $M< 
\infty$ and $L<\infty$ to see they are conditions of Stieltjes 
and Krein in a different form. Define
$$
\ell_n = -\frac{Q_n(0)}{P_n(0)} + 
\frac{Q_{n-1}(0)}{P_{n-1}(0)}\, , \qquad n\geq 1 \tag 5.66
$$
and
$$\align
m_n &= \frac{M'_n(0)}{N_n(0)} - 
\frac{M'_{n-1}(0)}{N_{n-1}(0)}\, , \qquad n\geq 2 \tag 5.67a \\
m_1 &= \frac{M'_1 (0)}{N_1 (0)}\, . \tag 5.67b
\endalign
$$
By the monotonicity properties of this section, $\ell_n >0$, 
$m_n >0$. By definition and $Q_0(0) =0$, 
$$\alignat2
-\frac{Q_N(0)}{P_N(0)} &= \sum_{n=1}^N \ell_n, \qquad && 
L=\sum_{n=1}^\infty \ell_n \\
\frac{M'_N(0)}{N_N(0)} &= \sum_{n=1}^N m_n, \qquad &&
M=\sum_{n=1}^\infty m_n.
\endalignat
$$

\proclaim{Proposition~5.23} For $N\geq 1$,
$$\align
m_N &= |P_{N-1}(0)|^2 \tag 5.68 \\
\ell_N &= -[a_{N-1} P_N(0) P_{N-1}(0)]^{-1}. \tag 5.69
\endalign
$$
\endproclaim

\remark{Remark} Since $A^{[N]}_F >0$, $P_N(z)$ has no zeros on 
$(-\infty, 0]$ and thus, $P_N(z)$ has the same sign near 
$-\infty$ and $0$. But $P_N(z)=c_N z^N + \text{ lower order}$ 
with $c_N >0$, so $(-1)^N P_N(0)>0$ (cf.~(5.13)). Thus, $P_N(0)
P_{N-1}(0)<0$ and the minus sign in (5.69) is just what is needed 
to ensure that $\ell_N >0$.
\endremark

\demo{Proof} $A^{[N]}_K$ has $\{ P_0(0), \dots, P_{N-1}(0)\}$ as 
its eigenfunction with eigenvalue zero. Thus, $\lim_{x\to 0} 
x(\delta_0, (A^{[N]}_K + x)^{-1} \delta_0) = |P_0(0)|^2 / 
\sum_{j=0}^{N-1} |P_j (0)|^2$. Since $P_0(0)=1$, we see that
$$
\frac{M'_N(0)}{N_N(0)} = \sum_{j=0}^{N-1} |P_j (0)|^2, \tag 5.70
$$
which implies (5.68). (5.69) follows immediately from 
Proposition~4.1. \qed
\enddemo

\proclaim{Corollary 5.24} If $A$ given by {\rom{(1.16)}} is the 
Jacobi matrix associated to an indeterminate Stieltjes moment 
problem, then
$$
\sum_{n=0}^\infty a^{-1/2}_n < \infty. 
$$
In particular, if $\sum_{n=0}^\infty a^{-1/2}_n = \infty$ for the 
Jacobi matrix associated to a Stieltjes problem, then the problem  
is determinate. 
\endproclaim

\demo{Proof} If the problem is indeterminate, by Proposition~5.23 
and Theorem~5.21, \linebreak $a^{-1/2}_n |P_n(0) 
P_{n-1}(0)|^{-1/2}$ and $|P_n(0) P_{n-1}(0)|^{1/2}$ both lie in 
$\ell^2$, so their product lies in $\ell^1$. \qed
\enddemo

Now we will define, following Stieltjes and Krein, functions
$$\alignat2
U_n(x) &= \frac{P_n(-x)}{P_n(0)} \, , \qquad && n \geq 0 
\tag 5.71 \\
V_n(x) &= -\frac{Q_n(-x)}{P_n(0)}\, , \qquad && n\geq 0
\tag 5.72 \\
G_n(x) &= -a_{n-1} M_n(-x) P_{n-1}(0), \qquad && n\geq 1 
\tag 5.73 \\
H_n(x) &= a_{n-1} N_n (-x) P_{n-1}(0), \qquad && n\geq 1. 
\tag 5.74
\endalignat
$$
We claim that

\proclaim{Proposition 5.25}
\roster
\item"\rom{(i)}"  $U_n(0)=1$
\item"\rom{(ii)}" $G_n(0) = 0$
\item"\rom{(iii)}" $H_n(0) = 1$
\item"\rom{(iv)}" $U_n(x) H_n(x) - V_n(x) G_n(x) =1 
\text{\rom{ for }} n\geq 1$
\item"\rom{(v)}" $ f^+_n(-x) - f^-_n (-x) = 
\frac{1}{U_n(x) G_n(x)} \text{\rom{ for }} n\geq 1$
\endroster
\endproclaim

\demo{Proof} (i) is immediate from the definition (5.71), (ii) from 
$M_n(0)=0$, and (iii) follows from Proposition~4.1 and the 
definition (5.16) of $N_N$ (and explains why we multiply $M$ and 
$N$ by $a_{n-1}P_{n-1}(0)$). To prove (iv), we note that for some 
constants $\alpha_n, \beta_n, \gamma_n$,
$$\alignat3
U_n(x) &= \alpha_n P_n(-x), & \qquad  V_n(x) 
&= -\alpha_n Q_n (-x) \\
G_n(x) &= \beta_n P_n(-x) + \gamma_n P_{n-1}(-x), & \qquad 
 H_n(x) &= -\beta_n Q_n (-x) - \gamma_n Q_{n-1}(-x), 
\endalignat
$$
so
$$
U_n(x) H_n(x) - V_n(x) G_n(x) = -\alpha_n \gamma_n 
[P_n (-x) Q_{n-1}(-x) - Q_n (-x) P_{n-1} (-x)]
$$
is constant as $x$ is varied for fixed $n$. But at $x=0$, by 
(i)--(iii) this combination is $1$. (v) follows from (iv) and 
the definitions. \qed
\enddemo

The following says that for $x>0$, $U,G$ can be associated with 
the equation of motion of a string of beads (see [\akh]). This 
is the starting point of deep work of Krein [\krstu].

\proclaim{Theorem 5.26}
\roster
\item"\rom{(i)}" $U_n(x) - U_{n-1}(x) = \ell_n G_n (x), \quad 
n\geq 1$
\item"\rom{(ii)}" $G_{n+1}(x) - G_n(x) = m_{n+1}x U_n(x), 
\quad n\geq 1$
\item"\rom{(iii)}" $G_1 (x) = m_1 x \quad U_0(x)=1$
\endroster
\endproclaim

\demo{Proof} (i) By definition of $U_n$, 
$$
U_n (x) - U_{n-1}(x) = \frac{M_n (-x)}{P_n(0)} = \ell_n G_n(x)
$$
by (5.69) and (5.73).

(ii) We have for $n\geq 1$,
$$
a_n P_{n+1} (-x) + b_n P_n (-x) + a_{n-1} P_{n-1} (-x) = 
-x P_n (-x). \tag 5.75
$$
(5.75) for $x=0$ implies that
$$
b_n = -a_n P_{n+1}(0) P_n(0)^{-1} -a_{n-1} P_{n-1}(0) 
P_n(0)^{-1}, \tag 5.76
$$
which we can substitute into (5.75) to obtain
$$
a_n M_{n+1} (-x) - a_{n-1} P_{n-1}(0) P_n(0)^{-1} M_n(-x) = 
-x P_n(x).
$$
Multiplying by $-P_n(0)$ and using the definitions (5.71)/(5.73), 
we see
$$
G_{n+1}(x) - G_n (x) = xP_n(0)^2 U_n(x)
$$
so (5.68) implies the result.

(iii) $U_0(x)$ is a constant so $U_0(0)=1$ implies $U_0(x)\equiv 
1$. $G_1$ is a linear polynomial with $G_1 (0)=0$, so $G_1(x) = 
G'_1 (0)x$.  For any $n$, $G'_n (0)=\lim_{x\downarrow 0} 
\frac{x^{-1}G_n(x)}{H_n(x)} = \sum_{j=1}^n m_j$. \qed
\enddemo

We can use these equations to get further insight into 
Theorem~5.2. First, we obtain explicit bounds on the rate of 
convergence of $f^+_N(-x) - f^-_N(-x)$ to zero in the 
determinate case, where either $\sum m_j$ or $\sum \ell_j$ or 
both diverge.

\proclaim{Theorem 5.27} For $x>0$,
$$
f^+_N (-x) - f^-_N (-x) \leq \biggl[m_1 x^2 \biggl( 
\sum_{j=1}^n m_j \biggr) \biggl(\sum_{j=1}^n \ell_j\biggr) 
\biggl]^{-1}.
$$
\endproclaim

\demo{Proof} It follows from Theorem~5.26 and induction that 
$U_n(x)$ and $G_n(x)$ are non-negative for $x>0$ and then that 
they are monotone in $n$. In particular, $U_n(x)\geq 1$, $G_n(x) 
\geq m_1 x$ from which, by (i), (ii), 
$$\align
U_n(x) &\geq m_1 \biggl( \sum_{j=1}^n \ell_j \biggr) x \\
G_n(x) & \geq \biggl(\sum_{j=1}^n m_j\biggr) x.
\endalign
$$
(v) of Proposition~5.25 then completes the proof. \qed
\enddemo

And we obtain directly that in the indeterminate case, $f^\pm$ 
are meromorphic functions.

\proclaim{Theorem 5.28} {\rom{(i)}} For any $z\in\Bbb C$, we 
have that for $n\geq 1$,
$$\align
|U_n(z)| &\leq \prod_{j=1}^n (1+\ell_j) \prod_{j=1}^n 
(1+m_j |z|) \\
|G_n(z)| &\leq \prod_{j=1}^{n-1} (1+\ell_j) \prod_{j=1}^n 
(1+m_j |z|). 
\endalign
$$

{\rom{(ii)}} If $M<\infty$ and $L<\infty$, then for all $z\in
\Bbb C$, $U_n(z), V_n(z), G_n(z), H_n(z)$ converge to functions 
$U_\infty(z), V_\infty(z), G_\infty(z), H_\infty(z)$, which 
are entire functions obeying
$$
|f(z)| \leq C_\varepsilon \exp(\varepsilon(z))
$$
for each $\varepsilon >0$.

{\rom{(iii)}} $f^-(z) = \frac{V_\infty (-z)}{U_\infty (-z)}$, 
$f^+(z) =\frac{H_\infty (-z)}{G_\infty(-z)}$\, .

{\rom{(iv)}} $f^+(z) - f^- (x) = \frac{1}{U_\infty (-z) 
G_\infty(-z)}$\, .
\endproclaim

\remark{Remark} In terms of the Nevanlinna functions $A,B,C,D$, 
one can see (using the fact that the Friedrichs solution is 
associated to $B_t$ with $t=L$) that $G_\infty (-z) = -D(z)$, 
$H_\infty (-z) = C(z)$, $U_\infty (-z) = -B(z) - LD(z)$, and 
$V_\infty(-z) = A(z) +LC(z)$, where $L=\sum_{j=1}^\infty \ell_j$.
\endremark

\demo{Proof} (i) follows by an elementary induction from 
Theorem~5.26. Similarly, one follows the proof of that theorem to 
show that $V_n, H_n$ obey
$$\alignat2
&V_n(x) - V_{n-1}(x) = \ell_n H_n(x), \qquad && n\geq 1 \\
&H_{n+1}(x) - H_n (x) = m_{n+1}x V_n(x), \qquad && n\geq 1 \\
&H_1(x) =1 \qquad V_0(x)=0
\endalignat
$$
and obtains inductively that
$$\align
|V_n(z)| &\leq \prod_{j=1}^n(1+\ell_j) \prod_{j=2}^n 
(1+m_j |z|) \\
|H_n(z)| &\leq \prod_{j=1}^{n-1} (1+\ell_j) \prod_{j=2}^n 
(1+m_j |z|).
\endalign
$$

Thus, if $L,M <\infty$, we first see that $U,V,H,G$ are bounded 
and then, since $\sum_{j\geq n} \ell_j \to 0$, $\sum_{j\geq n} 
m_j \to 0$, that each sequence is Cauchy as $n\to\infty$. The 
$C_\varepsilon$ bound is easy for products $\prod_{j=1}^\infty 
(1+m_j |z|)$ with $\sum_1^\infty m_j <\infty$. (iii), (iv) are 
then immediate from the definitions. \qed
\enddemo

To make the link to Stieltjes' continued fractions, we need 
to note the relation between $a_n$, $b_n$, $\ell_n$, and $m_n$. 
Immediately from Proposition~5.23, we have
$$
a_n = \bigl( \ell_{n+1} \sqrt{m_{n+1}m_{n+2}}\,\bigr)^{-1}, 
\qquad n\geq 0 \tag 5.77
$$
and by (5.76),
$$\align
b_n &= m^{-1}_{n+1} (\ell^{-1}_n + \ell^{-1}_{n+1}),
\qquad n \geq 1 \tag 5.78 \\
b_0 &= m^{-1}_1 \ell^{-1}_1. \tag 5.79
\endalign
$$

(Parenthetically, we note that these equations can be used 
inductively to define $m_j, \ell_j$ given $a_j, b_j$. We have 
$m_1 =1$, so (5.79) gives $\ell_1$. Given $\ell_1, \dots, \ell_j$ 
and $m_1, \dots, m_j$, we can use $a_{j-1}$ and (5.77) to find 
$m_{j+1}$ and then $b_j$ and (5.78) to get $\ell_{j+1}$.)

Stieltjes' continued fractions are of the form (5.1b). Let 
$$
d_0 = c^{-1}_1, \qquad d_n = (c_n c_{n+1})^{-1} \tag 5.80
$$
for $n\geq 1$ so (5.1b) becomes
$$
\cfrac d_0 \\
w + \cfrac   d_1 \\
1 + \cfrac d_2 \\
w+\dots \endcfrac 
$$
Now use the identity
$$
w + 
\cfrac \beta_1 \\
1 + \cfrac \beta_2 \\
f(w) \endcfrac = w+\beta_1 - \frac{\beta_1\beta_2}{\beta_2 + 
f(w)} 
$$
to see that (5.1b) has the form (5.1a) if $w=-z$, $d_0=1$, and
$$\align
b_0 &= d_1  \tag 5.81a \\
b_n &= d_{2n+1} + d_{2n} \tag 5.81b \\
a^2_n &= d_{2n+1} d_{2n+2}. \tag 5.81c
\endalign
$$
Thus, (5.80) and (5.81) are consistent with (5.77)--(5.79) if 
and only if
$$
c_{2j-1} = m_j, \qquad c_{2j} = \ell_j. \tag 5.82
$$
Thus, we have seen that (5.1a) is (5.1b) if $w=-z$ and the $c$'s 
are given by (5.82). Stieltjes' criterion that depends on whether 
or not $\sum_{j=1}^\infty c_j < \infty$ is equivalent to $L<\infty$ 
and $M<\infty$.

The connection (5.82) is not surprising. In (5.1b) if $w=0$, 
the continued fraction formally reduces to $c_2 + c_4 + \cdots$ 
consistent with $\sum_{j=1}^\infty \ell_j = -\lim_{N\to\infty} 
\frac{Q_N(0)}{P_N(0)}$. On the other hand, if we multiply by $w$,  
the continued fraction formally becomes
$$
\cfrac 1 \\
c_1 + \cfrac 1 \\
c_2 w + \cfrac 1 \\
c_3 + \cdots \endcfrac 
$$
which, formally at $w=0$, is $(c_1 + c_3 + \cdots)^{-1}$, consistent 
with $\sum_{j=1}^\infty m_j = \lim_{N\to\infty} 
\lim_{x\downarrow 0} \mathbreak [-\frac{xN_N(-x)}{M_N(-x)}]$.

We conclude this section by discussing Pad\'e approximants 
for the Hamburger analog of series of Stieltjes. A series of 
Hamburger is a formal power series $\sum_{j=0}^\infty \kappa_j 
z^j$ with
$$
\kappa_j = (-1)^j \int_{-\infty}^\infty x^j \, d\rho(x)
$$
for some measure $\rho$. We will begin with the ``principal" 
Pad\'e approximants $f^{[N-1,N]}(z)$:

\proclaim{Theorem 5.29} Let $\sum_{j=0}^\infty \kappa_j x^j$ 
be a series of Hamburger. Then:
\roster
\item"\rom{(i)}" The $f^{[N-1, N]}(z)$ Pad\'e approximants 
always exist and are given by
$$
f^{[N-1, N]}(z) = (\delta_0, (1+zA^{[N]}_F)^{-1} \delta_0)
= -\frac{z^{N-1} Q_N(-\frac1{z})}{z^N P_N (\frac1{z})}\, . 
\tag 5.83
$$
\item"\rom{(ii)}" If the associated Hamburger moment problem is 
determinate, then for any $z$ with $\text{\rom{Im}}\, z\neq 0$,
$$
\lim_{N\to\infty} f^{[N-1, N]}(z) = 
-\frac{z^{N-1} Q_N(-\frac1{z})}{z^N P_N (\frac1{z})} 
\tag 5.84
$$
exists and equals $\int \frac{d\rho(x)}{1+zx}$ for the unique 
solution, $\rho$, of the moment problem.
\item"\rom{(iii)}" The sequence $f^{[N-1, N]}(z)$ is pre-compact 
in the family of functions analytic in $\Bbb C_+$ \rom(in the 
topology of uniform convergence on compact sets\rom).
\item"\rom{(iv)}" Any limit of $f^{[N-1, N]}(z)$ is of the form 
$\int \frac{d\rho(x)}{1+zx}$ where $\rho$ is a von~Neumann 
solution of the Hamburger problem.
\endroster
\endproclaim

\demo{Proof} (i) The proof that (5.12) holds (Proposition~5.6) 
is the same as in the Stieltjes case, and from there we get 
(5.34) as in the Stieltjes case. Since $\lim_{z\to 0} z^N P_N 
(\frac1{z})\neq 0$, we see $f^{[N-1,N]} (z)$ exists and is given 
by (5.83).

(iii) Since
$$
|zf^{[N-1, N]}(z)| = |(\delta_0, (A^{[N]}_F + z^{-1})^{-1}
\delta_0)| \leq |\text{Im}(z^{-1})|^{-1},
$$
we see that
$$
|f^{[N-1,N]}(z)| \leq |z| \, |\text{Im}(z)|^{-1}
$$
is bounded on compacts. By the Weierstrass-Vitali theorem, such 
functions are pre-compact.

(ii), (iv) By the proof of Proposition~4.13, if $d\rho$ is any 
measure whose first $M+2$ moments are $\gamma_0, \gamma_1, 
\dots, \gamma_{M+2}$, then
$$
\biggl| y^{M+1} \biggl[G_\rho (iy) + \sum_{n=0}^M (-i)^{n+1} 
y^{-n-1} \gamma_n \biggr]\biggr| \leq ay^{-1} \tag 5.85
$$
with $a$ only depending on $\{\gamma\}$ (either $a=\gamma_{M+1}$ 
or $a=\frac12 [\gamma_M + \gamma_{M+2}]$). Since $\langle\delta_0, 
(A^{[N]}_F)^j \delta_0\rangle \mathbreak =\gamma_j$ for $j\leq 2N-1$, 
we see that for $M$ fixed, (5.85) holds uniformly for $N$ large if 
$G_N(z) =-z^{-1} f^{[N-1, N]}(-\frac1{z})$, so any limit point 
of the $G$'s obeys (5.85) and is Herglotz. Thus by Proposition~4.13, 
it is of the form $G_\rho(z)$ with $\rho\in\Cal M^H(\gamma)$. 
Thus the limit for $f^{[N-1,N]}(z)$ as $N\to\infty$, call it 
$f(z)$, has the form
$$
f(z) = \int \frac{d\rho(x)}{1+xz}
$$
with $\rho\in\Cal M^H(\gamma)$. In the determinate case, $\rho$ 
must be the unique solution, so all limit points are equal. Thus  
compactness implies convergence, and we have proven (ii).

In the indeterminate case, we note that $\delta_j = P_j 
(A^{[N]}_F )\delta_0$ for $j=0,\dots, N-1$ so
$$
(A^{[N]}_F -z)^{-1}\delta_0 = \sum_{j=0}^{N-1} \langle P_j 
(A^{[N]}_F)\delta_0, (A^{[N]}_F -z)^{-1}\delta_0\rangle \delta_j.
$$
As in the proof of Theorem~4.3, we conclude that (with $G_N(z) 
=\langle\delta_0, (A^{[N]}_F -z)^{-1}\delta_0\rangle$),
$$
\sum_{j=0}^{N-1} |Q_j (z) + G_N(z) P_j (z)|^2 = 
\frac{\text{Im}\, G_N(z)}{\text{Im}\, z}\, .
$$
Thus, since $Q,P \in \ell_2$ if $G_N(z) \to G(z)$ through a 
subsequence, the limit $\zeta = G(z)$ obeys equality in (4.5). 
By Theorem~4.14, any such solution of the moment problem is a 
von~Neumann solution. \qed
\enddemo

We want to show that in many cases, $\lim f^{[N-1, N]}(z)$ 
will not exist. Consider a set of Stieltjes moments, 
$\{\gamma_n\}^\infty_{n=0}$ and the associated even Hamburger 
problem with moments
$$
\Gamma_{2n} = \gamma_n, \qquad \Gamma_{2n+1} =0. \tag 5.86
$$
Given $\rho\in\Cal M^S(\gamma)$, let $\tilde\rho\in\Cal M^H 
(\Gamma)$ be given by
$$
d\tilde\rho(x) = \tfrac12 (\chi_{[0,\infty)}(x)\, d\rho(x^2) 
+ \chi_{(-\infty, 0]}(x)\, d\rho(x^2)) \tag 5.87
$$
so that Theorem~2.12 says that this sets up a one-one 
correspondence between even integrals in $\Cal M^H(\Gamma)$ 
and $\Cal M^S(\gamma)$. Let $\tilde P_n, \tilde Q_n$ be the 
orthogonal polynomials for $\Gamma$ and $P_n, Q_n, M_n, N_n$ 
for $\gamma$. Define $d_n$ by $d^2_n \int M_n(x)^2 x^{-1}\, 
d\rho(x)=1$. Then we claim that
$$\align
G_{\tilde\rho}(z) &= zG_\rho(z^2) \tag 5.88 \\
\tilde P_{2n}(x) &= P_n(x^2) \tag 5.89 \\
\tilde P_{2n-1}(x) &= \frac{d_n M_n(x^2)}{x} \tag 5.90 \\
\tilde Q_{2n}(x) &= xQ_n (x^2) \tag 5.91 \\
\tilde Q_{2n-1}(x) &= d_n N_n (x^2). \tag 5.92
\endalign
$$

To prove (5.89) and (5.90), we note that the right sides have 
the correct degree, and since $P_n(x^2)$ is even and 
$\frac{M_n(x^2)}{x}$ is odd, they are $d\tilde\rho$ orthogonal 
to each other. Easy calculations prove the $d\tilde\rho$ 
orthogonality of two $P_n(x^2)$'s and two $\frac{M_n(x^2)}{x}$'s. 
The formulae for $\tilde Q$ and $G_{\tilde\rho}$ follow by 
direct calculations using Theorem~4.2 for $Q$. 

\proclaim{Theorem 5.30} Let $\gamma, \Gamma$ be related by 
{\rom{(5.86)}} and let $d\tilde\rho^{(F)}, d\tilde\rho^{(K)}$ 
be the images of $d\rho_F, d\rho_K$ \rom(for $\gamma$\rom) 
under the map $\rho \mapsto \tilde\rho$ given by {\rom{(5.87)}}. 
Let $\tilde f^{[N,M]}(z)$ be the Pad\'e approximants for 
$\sum_{n=0}^\infty (-1)^n \Gamma_n z^n$. Then
$$\align
\lim_{N\to\infty} \tilde f^{[2N-1, 2N]}(z) &= \int 
\frac{d\tilde\rho^{(F)}(x)}{1+xz} \tag 5.93 \\
\lim_{N\to\infty} \tilde f^{[2N, 2N+1]}(z) &= \int
\frac{d\tilde\rho^{(K)}(x)}{1+xz}\, . \tag 5.94
\endalign
$$
In particular, if the $\Gamma$ problem is indeterminate, $\lim 
f^{[N-1, N]}(z)$ does not exist. 
\endproclaim

\demo{Proof} This follows directly from Theorem~6 for the $\gamma$ 
problem and (5.88)--(5.92). \qed
\enddemo

Of course, in the special case, the non-convergence of 
$f^{[N-1, N]}(z)$ is replaced by something almost as nice since 
the even and odd subsequences separately converge. But there is 
no reason to expect this to persist if $\Gamma$ is not even and 
indeed, simple numerical examples ([\kei]) illustrate 
non-convergence is the rule. One can understand why this happens. 
In the indeterminate case, there is a one-parameter family of 
extensions, each with discrete spectrum. There is, in general, 
nothing to tack down $A^{[N]}_F$ so it wanders. In the even case, 
two self-adjoint extensions are even, one with an eigenvalue at 
$0$ and one without, and the $A^{[N]}_F$ for odd $N$ and even 
$N$ track these. This suggests that $A^{[N]}_K$, which is tacked 
down to have a zero eigenvalue, should converge in strong 
resolvent sense and the corresponding $f^{[N,N]}$'s should 
converge. This is true and is not a previously known result:

\proclaim{Theorem 5.31} Let $\{\gamma_n\}^\infty_{n=0}$ be a 
set of Hamburger moments and let $\sum_{n=0}^\infty (-1)^n 
\gamma_n z^n$ be the associated series of Hamburger. Then the 
$f^{[N,N]}(z)$ Pad\'e approximants exist if and only if
$$
P_N(0)\neq 0. \tag 5.95
$$
Moreover, {\rom{(5.95)}} holds if and only if there exists 
$\alpha_N$ so that the matrix $A^{[N+1]}_K$ given by {\rom{(5.2b)}} 
has a zero eigenvalue, and then
$$
f^{[N,N]}(z) = (\delta_0, (1+zA^{[N+1]}_K)^{-1}\delta_0). 
\tag 5.96
$$
If $N\to\infty$ through the sequence of all $N$'s for which 
{\rom{(5.95)}} holds, then
$$
\lim \Sb N\to \infty \\ N:P_N(0)\neq 0 \endSb 
f^{[N,N]}(z) = \int \frac{d\rho(x)}{1+xz}\, , \tag 5.97
$$
where $d\rho$ is the unique solution of the moment problem in 
case it is determinate and the unique von~Neumann solution with 
$\mu (\{0\})>0$ if the moment problem is indeterminate.
\endproclaim

\remark{Remarks} 1. If $P_N(0)=0$, then, by the second-order 
equation $P_N(0)$ obeys, we have that $P_{N+1}(0)\neq 0 \neq 
P_{N-1}(0)$, so (5.95) can fail at most half the time. Typically, 
of course, it always holds (e.g., under a small generic 
perturbation of the Jacobi matrix, one can prove that it will 
hold even if initially it failed).

2. If the $\gamma$'s are a set of even moments, then (5.95) holds 
exactly for $N$ even and
$$
f^{[2M, 2M]}(z) = \frac{z^{2M} Q_{2M+1} (-\frac1{z})}
{z^{2M+1} P_{2M+1} (-\frac1{z})}\, ,
$$
so we get convergence of those measures we called $d\tilde\rho$ 
above.

3. The non-existence of $f^{[N,N]}$ is using the Baker definition 
that we have made. There is a classical definition which would 
define $f^{[N,N]}(z)$ even if $P_N(0)=0$, (but (5.26) and (5.27) 
would both fail!). Under this definition, $f^{[N,N]}(z)$ would 
be $f^{[N-1, N-1]}(z)$ when $P_N(0)=0$ and (5.97) would hold 
using all $N$ rather than just those $N$ with $P_N(0)\neq 0$.
\endremark

\demo{Proof} Define
$$\align
\widetilde M_N(z) &= P_{N-1}(0) P_N(z) - P_N(0) P_{N-1}(z) 
\tag 5.98 \\
\tilde N_N(z) &= P_{N-1}(0) Q_N(z) - P_N(0)P_{N-1}(z).
\tag 5.99
\endalign
$$
Then by the Wronskian calculations (see Proposition~4.1); also 
see (5.70)),
$$\align
\tilde N_N(0) &= \frac{1}{a_{N-1}} \neq 0 \tag 5.100 \\
\widetilde M'_N(0) &= \sum_{j=0}^{N-1} \frac{P_j(0)^2}
{a_{N-1}} \neq 0 \tag 5.101
\endalign
$$
and, of course,
$$
\widetilde M_N(0)=0. \tag 5.102
$$

Define
$$\align
A_N(z) &= z^N \tilde N_{N+1} \bigl(-\tfrac{1}{z}\bigr) 
\tag 5.103 \\
B_N(z) &= z^{N+1} \widetilde M_{N+1} \bigl( -\tfrac{1}{z}\bigr). 
\tag 5.104
\endalign
$$
By (5.100), $\deg A_N=N$ and by (5.101/5.102), $\deg B_N = N$. 
Moreover, by the definition (5.98),
$$
B_N(0) = 0 \Longleftrightarrow P_N(0)=0 \tag 5.105
$$
and
$$
B_N(0) = 0 \Longrightarrow P'_N(0)\neq 0 \tag 5.106
$$
since $P_N(0)=0 \Rightarrow P_{N-1}(0)\neq 0$. 

Moreover, we claim that
$$
A_N(z) - B_N(z) \sum_{j=0}^{2N} \gamma_j z^j = 
O(z^{2N+1}). \tag 5.107
$$
This can be seen either by algebraic manipulation as in the 
remark following the proof of Theorem~A.5 or by noting that 
we have proven it in (5.39) for Stieltjes measures. So if 
$\rho\in\Cal M^H(\gamma)$ and $\mu_0$ is a Stieltjes measure 
and $\rho_t = (1-t)\mu_0 + t\rho$, then $B_N(-z), A_N(-z)$ 
are real analytic functions of $t$ and (5.107) is true for 
all small $t$ in $(0,1)$ since $A^{[N+1]}_F (t) >0$ for small 
$t$ (and fixed $N$) and that is all the proof of (5.106) 
depends on.

If $P_N(0)\neq 0$, $B_N(0)\neq 0$, so $B^{[N,N]}(z) = 
\frac{B_N(z)}{B_N(0)}$ and $A^{[N,N]}(z) = \frac{A_N(z)}
{B_N(0)}$ yield denominator and numerator of $f^{[N,N]}(t)$; 
so $f^{[N,N]}$ exists.

Moreover, the proof of Lemma~5.1 shows that an $\alpha_N$ exists 
giving $A^{[N+1]}_K$ a zero eigenvalue if and only if $P_N(0) 
\neq 0$. In that case, the proof of Proposition~5.8 is valid, 
and we have that (5.17) holds. Noting that $\frac{N_N}{M_N} = 
\frac{\tilde N_N}{\widetilde M_N}$, we see that (5.96) is then 
valid.

On the other hand, suppose that $P_N(0)=0$ and that $f^{[N,N]}
(z)$ exists. Then by (5.107), we have that
$$
A_N(z) B^{[N,N]} - B_N(z) A^{[N,N]}(z) = O(z^{2N+1})
$$ 
and so it is zero since the left side of this is of degree $2N$.

Since $P_N(0) =0$, $\widetilde M_{N+1}(z) = -P_{N+1}(0) P_N(z) 
=c\widetilde M_N(z)$ where $c=-P_{N+1}(0)\mathbreak 
P_{N-1}(0)^{-1}$, and similarly, $\tilde N_{N=1}(z) = 
c\tilde N_N(z)$. It follows that 
$$
A_N(z) = cz A_{N-1}(z), \qquad B_N(z) = czB_{N-1}(z).
$$
Thus, if $P_N(0)=0$ but $f^{[N,N]}(z)$ exists, then
$$
f^{[N,N]}(z) = f^{[N-1, N-1]}(z),
$$
that is, the $[N-1, N-1]$ Pad\'e approximant would need to have an  
error of order $O(z^{2N+1})$.

Since $P_N(0)=0$, $A^{[N]}_K = A^{[N]}_F$ and so we would have 
that $\langle\delta_0, (A^{[N]}_F)^j \delta_0\rangle = \gamma_j$ 
for $j=0,1,\dots, 2N$. But as noted in the remark following 
the proof of Proposition~5.11, this can never happen. We 
conclude that if $P_N(0)=0$, $f^{[N,N]}$ must fail to exist.

To complete the proof, define $A_K$ to be $\bar A$ if the 
problem is determinate and the unique von~Neumann extension 
with eigenvalue zero if the problem is indeterminate (so $A_K 
\pi(0)=0$). Then the proof of Proposition~5.3 goes through without 
change and shows that $A^{[N]}_K$ converges to $A_K$ in strong 
resolvent sense. This implies (5.97). \qed
\enddemo

As for other $f^{[N+\ell -1, N]}(z)$ for series of Hamburger, we 
can only define $\gamma^{(j)}$ moment problems for $j$ an even 
integer (positive or negative). This means that we can use 
(5.49)/(5.59) if $\ell$ is an even integer and (5.50)/(5.60) if 
$\ell$ is an odd integer. The result is the following:

\proclaim{Theorem 5.32} Let $\{\gamma_n\}^\infty_{n=0}$ be a 
set of Hamburger moments and let $\sum_{n=0}^\infty (-1)^n 
\gamma_n z^n$ be the associated series of Hamburger. Then:
\roster
\item"\rom{(i)}" If $\ell$ is any odd integer, the Pad\'e 
approximants $f^{[N+\ell-1, N]}(z)$ exist if and only if 
$P^{(j)}_M(0)\neq 0$ where for $\ell >0$, $j=\ell-1$, and $M=N$ 
and if $\ell <0$, $j=\ell +1$ and $M=N+\ell -1$. The functions 
that exist converge to a finite limit $f_\ell (z)$.
\item"\rom{(ii)}" If $\ell$ is any even integer, the Pad\'e 
approximants $f^{[N+\ell -1, N]}(z)$ all exist and lie in a 
compact subset in the topology of uniform convergence on 
compact subsets of $\Bbb C_+$. Any limit point, $f_\ell (z)$, 
is associated to a von~Neumann solution, $\rho^{(\ell)}$, 
of the $\gamma^{(\ell)}$ moment problem via
$$
f_\ell (z) = \sum_{j=0}^{\ell-1} (-1)^j \gamma_j z^j + 
(-1)^\ell \gamma_\ell z^\ell \int_{-\infty}^\infty 
\frac{d\rho^{(\ell)} (x)}{1+xz} 
$$
if $\ell >0$ and 
$$
f_\ell (z) = \biggl\{ 1 - \gamma_1 z - \sum_{j=0}^{-\ell-1} 
(-1)^j \gamma^{(0)}_j z^{j+2} + (-1)^{\ell +1} 
\gamma^{(0)}_{-\ell} z^{-\ell + 2} \int_{-\infty}^\infty 
\frac{d\rho^{(\ell)}(x)}{1+xz} \biggr\}^{-1}
$$
if $\ell <0$. In particular, if the $\gamma^{(\ell)}$ problem 
is determinate, $f^{[N+\ell -1, N]}(z)$ is convergent.
\item"\rom{(iii)}" Let $\ell$ be an even integer. If the 
$\gamma^{(\ell)}$ problem is determinate \rom(as a Hamburger 
problem\rom), then for $\ell\geq 0$, $f_{\ell + 1}(z) = 
f_\ell (z) = \cdots = f_0(z) = f_{-1}(z)$, and for $\ell \leq 0$, 
$f_{\ell -1}(z) = f_\ell (z) =\cdots = f_0(z) = f_1(z)$. In 
particular, if $|\gamma_j| \leq C^{j+1} j!$, then all 
$f^{[N+\ell -1, N]}(z)$ converges to the same 
$\ell$-independent limits as $N\to\infty$. 
\endroster
\endproclaim

\vskip 0.3in
\flushpar{\bf \S 6. Solutions of Finite Order}
\vskip 0.1in

In this section, we will continue the analysis of the 
indeterminate Hamburger moment problem. Throughout, 
$\{\gamma_n \}^\infty_{n=0}$ will be a set of indeterminate 
Hamburger moments and $\rho$ will be a solution of (1.1), the 
moment problem for $\gamma$.

We have picked out the von~Neumann solutions as coming from 
self-adjoint extensions of the Jacobi matrix, $A$. But in a 
certain sense, every $\rho$ does come from a self-adjoint 
extension of $A$! For let $\Cal H_\rho = L^2 (\Bbb R, d\rho)$, 
let $D(B)=\{f\in\Cal H_\rho \mid \int x^2 |f(x)|^2\, d\rho(x) 
<\infty\}$, and let
$$
(Bf)(x) = xf(x). 
$$
Let $\Bbb C[X]$ be the set of polynomials in $x$ which lies in 
$\Cal H_\rho$ since all moments are finite, and let $\Cal H_0 = 
\overline{\Bbb C[X]}$. Corollary~4.15 says that $\Cal H_0 = 
\Cal H_\rho$ if and only if $\rho$ is a von~Neumann solution. 
Let $A_\rho = B\restriction \Bbb C[X]$. Then to say $\rho$ is 
a solution of the Hamburger moment problem is precisely to say 
that $A_\rho$ is unitarily equivalent to the Jacobi matrix, $A$, 
associated to $\gamma$ under the natural map (so we will drop 
the subscript $\rho$). In that sense, $B$ is a self-adjoint  
extension of $A$, but not in the von~Neumann sense.

$B$ is minimal in the sense that there is no subspace of 
$\Cal H_\rho$ containing $\Cal H_0$ and left invariant by all 
bounded functions of $B$ (equivalently, left invariant by all 
$(B-z)^{-1}$, $z\in\Bbb C\backslash\Bbb R$ or by all 
$e^{iBs}$, $s\in\Bbb R$). So (if $\rho$ is not a von~Neumann 
extension) we are in the strange situation where $D(B)\cap
\Cal H_0$ is dense in $\Cal H_0$, $B[D(B)\cap\Cal H_0]\subset
\Cal H_0$, but $\Cal H_0$ is not invariant for the resolvents 
of $B$!

It is not hard to see that the set of all solutions of the 
moment problem is precisely the set of all self-adjoint 
``extensions" of $A$ in this extended sense, which are minimal 
($\Cal H_0$ is cyclic for $B$) and modulo a natural unitary 
equivalence. This is a point of view originally exposed by 
Naimark [\naI, \naII, \naIII, \naIV] and developed by Livsic 
[\liv], Gil de Lamadrid [\gdl], and Langer [\lang]. See the 
discussion of Naimark's theory in Appendix~1 of [\ag].

In the language of Cayley transforms, the extensions associated 
to $\rho$'s that we will call of order at most $n$ below are 
parametrized by unitary maps, $U$, from $\Cal K_+\oplus\Bbb C^n$ 
to $\Cal K_- \oplus\Bbb C^n$ with $U_1, U_2$ equivalent if and 
only if there is a unitary map $V:\Bbb C^n \to \Bbb C^n$ so $U_1 
(1\oplus V) = (1\oplus V)U_2$. From this point of view, these 
extensions are then parametrized by the variety of conjugacy 
classes of $\Cal U(n+1)$ modulo a $\Cal U(n)$ subgroup. This has 
dimension $(n+1)^2 - n^2 = 2n +1$. Our parametrization below in 
terms of a ratio of two real polynomials of degree at most $n$ also 
is a variety of dimension $2n+1$, showing the consistency of the 
parametrization. (The $\rho$'s of exact order $n$ will be a 
manifold.)

Given any $\rho$, we define the {\it{order}} of $\rho$,
$$
\text{ord}(\rho) = \dim(\Cal H_\rho/\Cal H_0)
$$
so the von~Neumann solutions are exactly the solutions of 
order $0$. Our main result in this section will describe all 
solutions of finite order which turn out to correspond 
precisely to those $\rho$'s whose Nevanlinna function 
$\Phi_\rho$ are ratios of real polynomials.

They will also be $\rho$'s for which $d\rho / \prod_{i=1}^n 
|x-z_i|^2$ is the measure of a determinate moment problem. We 
will therefore need some preliminaries about such problems. 
Given $z_1, \dots, z_n \in\Bbb C_+$, by using partial fractions 
we can write for $x$ real,
$$
\frac{x^m}{\prod_{i=1}^n |x-z_i|^2} = P_{m-2n} (x; z_1, \dots, 
z_n) + \sum_{i=1}^n \biggl[ \frac{z^m_i}{z_i - \bar z_i} \, 
\frac{1}{x-z_i} + \frac{\bar z^m_i}{\bar z_i - z_i} \, 
\frac{1}{x-\bar z_i}\biggr] \tag 6.1
$$
where $P_{m-2n}$ is a polynomial in $x$ of degree $m-2n$. To 
see (6.1), write $\prod_{i=1}^n |x-z_i|^2 = \prod_{i=1}^n  
(x-z_i)(x-\bar z_i)$ and analytically continue. The sum in 
(6.1) comes from computing the residues of the poles of these 
functions. Define
$$
\Gamma^{(0)}_m (z_1, \dots, z_n) = E^\gamma_x (P_{m-2n} 
(x; z_1, \dots, z_n)), \tag 6.2
$$
where, as usual, $E^\gamma_x$ is the expectation with respect 
to any solution of the moment problem. For $\zeta_1, \dots, 
\zeta_n\in\Bbb C_+$, let
$$
\Gamma_m (z_1, \dots, z_n; \zeta_1, \dots, \zeta_n) = 
\Gamma^{(0)}_m (z_1, \dots, z_n) + \sum_{i=1}^n \biggl[ 
\frac{z^m_i}{z_i - \bar z_i}\, \zeta_i + 
\frac{\bar z^m_i}{\bar z_i - z_i}\, \bar\zeta_i \biggr]. 
\tag 6.3
$$

\proclaim{Theorem 6.1} Fix $z_1, \dots, z_n$ and $\zeta_i, 
\dots, \zeta_n$ in $\Bbb C_+$. There is a one-one correspondence 
between $\nu\in\Cal M^H (\Gamma(z_1, \dots, z_n; \zeta_1, 
\dots, \zeta_n))$ and those $\rho$ in $\Cal M^H(\gamma)$ which 
obey
$$
\int \frac{d\rho(x)}{x-z_i} = \zeta_i, \qquad i=1,\dots, n 
\tag 6.4
$$
under the association
$$
d\rho(x) \leftrightarrow d\nu (x) = \prod_{i=1}^n |x-z_i|^{-2}
\, d\rho(x). \tag 6.5
$$
That is, $\rho$ has moments $\{\gamma_m\}^\infty_{m=0}$ with 
subsidiary conditions {\rom{(6.4)}} if and only if $\nu$ 
given by {\rom{(6.5)}} has the moments $\{\Gamma_m (z_1, 
\dots, z_n; \zeta_1, \dots, \zeta_n)\}^\infty_{m=0}$.
\endproclaim

\demo{Proof} This is pure algebra. The functions 
$\{x_m\}^\infty_{m=0}$, $\{\frac{1}{x-z_i}\}^n_{i=1}$ are 
linearly independent as analytic functions (finite sums only). 
(6.1) says that each function $x^m / \prod_{i=1}^n (x-z_i)
(x-\bar z_i)\equiv Q_m(x)$ is in the span, $\Cal S$, of these 
functions. On the other hand, since $\prod_{i=1}^n (x-z_i)
(x-\bar z_i)\equiv L(x)$ is a polynomial in $x$, and $x^\ell Q_m 
(x) = Q_{\ell+m}(x)$, $L(x) Q_m(x)$ is in the span of the 
$\{Q_m(x)\}$. Similarly, since $\frac{L(x)}{(x-z_i)}$ is a 
polynomial, $L(x) Q_0(x) \frac{1}{(x-z_i)}$ lies in that span.

Thus, $\{Q_m(x)\}$ is also a basis of $\Cal S$. It follows that 
there is a one-one correspondence between assignments of numbers 
$Q_m(x)\mapsto \Gamma_m$ and of assignments $x^m \mapsto \gamma_m$, 
$\frac{1}{x-z_i}\mapsto\zeta_i$, $\frac{1}{x-\bar z_i}\mapsto
\lambda_i$ since each defines a linear functional on $\Cal S$. 
For the linear functional to be real in $\Gamma_m$ representation, 
all $\Gamma_m$ must be real. For reality in $(\gamma,\zeta,\lambda)$ 
representation, we must have $\gamma_m$ real and $\lambda_i = 
\bar\zeta_i$. (6.3) is just an explicit realization of the map 
from $\gamma,\zeta, \bar\zeta$ to $\Gamma$. Thus, a real $d\rho$ 
has moments $\gamma_m$ with $\zeta_i$ conditions if and only if 
$\int x^m \prod_{i=1}^n |x-z_i|^{-2}\, d\rho(x)=\Gamma_m$ for 
all $m$. Since (6.5) shows $\nu$ is positive if and only if 
$\rho$ is, we see that the claimed equivalence holds. \qed
\enddemo

As an aside of the main theme of this section, we can construct 
determinate Hamburger moments quite close to indeterminate ones. 
The result is related to Corollary~4.21.

\proclaim{Theorem 6.2} Let $\{\gamma_n\}^\infty_{n=0}$ be a set 
of Hamburger moments. Then for any $a\in (0,\infty)$, there is 
a $c_a$ so that the moment problem with moments
$$\align
\tilde\gamma_{2n+1} &= 0 \\
\tilde\gamma_{2n} &= \gamma_{2n} -a^2 \gamma_{2n-2} + 
a^4 \gamma_{2n-2} + \cdots + (-1)^n a^{2n} \gamma_0 + 
(-1)^{n+1} c_a a^{2n+2} \tag 6.6
\endalign
$$
is a determinate moment problem.
\endproclaim

\demo{Proof} Let $\rho_1$ be any solution for the Hamburger 
problem. Since $\frac12 [d\rho_1 (x) + d\rho_1 (-x)]=d\tilde\rho$ 
has moments
$$
\int x^m\, d\tilde\rho(x) = \cases 0 & m\text{ odd} \\
\gamma_{2m} & m\text{ even} 
\endcases \quad \equiv \gamma'_{2m},
$$
we can suppose $\gamma_{2m+1}=0$. Let $\rho$ be a von~Neumann 
solution to the $\gamma'_m$ moment problem which is invariant 
under $x\to -x$. If the $\gamma'_m$ problem is determinate, then 
the unique solution is invariant. Otherwise, this follows from the 
proof of Theorem~2.13. Let $d\mu(x) = \frac{d\rho(x)}{1+a^2 x^2}$. 
By Corollary~4.15 and Theorem~6.1, $\rho$ is the unique solution 
of the $\gamma'$ moment problem with $\int \frac{d\rho(x)}
{x-ia^{-1}}\equiv\zeta_{a^{-1}}$. Thus, the $\Gamma_m (ia^{-1}, 
\zeta_{a^{-1}})$ problem is determinate. If $c_a = a^{-1}\,
\text{Im}\, \zeta_{a^{-1}}$, then a calculation using the 
geometric series with remainder shows that $\Gamma_m (-ia^{-1}, 
\zeta_{a^{-1}})=\tilde\gamma_m$ given by (6.6). \qed
\enddemo

\vskip 0.1in

Returning to the main theme of this section, we next examine 
when a Herglotz function is a real rational function. Define for 
$z\in\Bbb C_+$,
$$
\varphi_z (x) \equiv \frac{1+xz}{x-z} \tag 6.7a
$$
viewed as a continuous function on $\Bbb R \cup\{\infty\}$ with 
$$
\varphi_z (\infty) = z. \tag 6.7b
$$
If $d\sigma (x) = [\frac{d\mu(x)}{1+x^2}] + c\delta_\infty$ as 
a finite measure on $\Bbb R\cup\{\infty\}$, then (1.19) can be 
rewritten as
$$
\Phi(z) = d + \int_{\Bbb R\cup\{\infty\}}  \varphi_z (x)\, 
d\sigma(x). \tag 6.8
$$
If $\sigma(\{\infty\}) = 0$ and $\int |x|\, d\sigma(x) < \infty$, 
define $\tilde d = d-\int x\, 
d\sigma(x)$.

\proclaim{Proposition 6.3} A Herglotz function is a ratio of two 
real polynomials if and only if the representation {\rom{(6.8)}} 
has a $\sigma$ with finite support. If $\sigma$ has exactly $N$ 
points in its support, there are three possibilities for the 
degrees of $P,Q$ in $\Phi (z) = \frac{P(z)}{Q(z)}$ with $P,Q$ 
relatively prime polynomials:
\roster
\item"\rom{(i)}" $(\sigma(\{\infty\})=0, \tilde d=0), \quad 
\deg P=N-1, \ \deg Q=N$ 
\item"\rom{(ii)}" $(\sigma(\{\infty\})=0, \tilde d\neq 0), \quad 
\deg P=\deg Q=N$
\item"\rom{(iii)}" $(\sigma(\{\infty\})\neq 0), \quad 
\deg P = N, \ \deg Q=N-1$
\endroster
In all cases, we say $N=\max(\deg(P), \deg(Q))$ is the degree of 
$\Phi$.
\endproclaim

\demo{Proof} Elementary. \qed
\enddemo

\remark{Remark} The set of $\sigma$'s with exactly $N$ pure 
points is a manifold of dimension $2N$ ($N$ points, $N$ weights). 
$d\in\Bbb R$ is another parameter so the set of such $\Phi$'s is 
a manifold of dimension $2N+1$.
\endremark

As a final preliminary, for any $z\in\Bbb C\backslash\sigma(B) 
\equiv \Bbb C \backslash\text{ supp}(d\rho)$ and $n=1,2,\dots$, 
we introduce the functions $e_n(z)$ on $\text{supp}(d\rho)$,
$$
e_n(z)(x) = \frac{1}{(x-z)^n},
$$
thought of as elements of $\Cal H_\rho = L^2 (\Bbb R, d\rho)$.

The main result of this section is:

\proclaim{Theorem 6.4} Let $\gamma$ be a set of indeterminate 
Hamburger moments and let $\rho_0 \in \Cal M^H(\gamma)$. Fix 
$N\in \{0,1,\dots\}$. Then the following are equivalent:
\roster
\item"\rom{(1)}" $\rho_0$ has order at most $N$.
\item"\rom{(2)}" For some set of distinct $z_1, \dots, z_N \in
\Bbb C\backslash\Bbb R$, $\Cal H_0 \cup \{e_1 (z_j)\}^N_{j=1}$ 
span $\Cal H_{\rho_0}$.
\item"\rom{(3)}" For any set of distinct $z_1, \dots, z_N\in
\Bbb C\backslash\Bbb R$, $\Cal H_0\cup\{e_1 (z_j)\}^N_{j=1}$ 
span $\Cal H_{\rho_0}$.
\item"\rom{(4)}" For some $z_0 \in \Bbb C\backslash\Bbb R$, 
$\Cal H_0 \cup \{e_j (z_0)\}^N_{j=1}$ span $\Cal H_{\rho_0}$.
\item"\rom{(5)}" For any $z_0\in\Bbb C\backslash\Bbb R$, 
$\Cal H_0 \cup \{e_j (z_0)\}^N_{j=1}$ span $\Cal H_{\rho_0}$.
\item"\rom{(6)}" For some set of $z_0, z_1, \dots, z_N \in 
\Bbb C_+$ and $\zeta_j = G_{\rho_0}(z_j)$, there is no other  
$\rho\in\Cal M^H(\gamma)$ with $G_\rho (z_j)=\zeta_j$.
\item"\rom{(7)}" For all sets of $z_0, z_1, \dots, z_N \in
\Bbb C_+$ and $\zeta_j = G_{\rho_0}(z_j)$, there is no other 
$\rho\in\Cal M^H(\gamma)$ with $G_\rho(z_j)=\zeta_j$.
\item"\rom{(8)}" For some $z_0, z_1, \dots, z_N \in \Bbb C_+$ 
and $\zeta_j = G_{\rho_0}(z_j)$, the $\Gamma (z_0, \dots, 
z_N; \zeta_0, \dots, \zeta_N)$ moment problem is determinate.
\item"\rom{(9)}" For any $z_0, z_1, \dots, z_N \in \Bbb C_+$ 
and $\zeta_j = G_{\rho_0}(z_j)$, the $\Gamma (z_0, \dots, 
z_N; \zeta_0, \dots, \zeta_N)$ moment problem is determinate.
\item"\rom{(10)}" For some $z_1, \dots, z_N\in\Bbb C_+$ and 
$\zeta_j = G_{\rho_0}(z_j)$, $\rho_0$ is a von~Neumann solution 
of the $\Gamma (z_1, \dots, z_N; \zeta_1, \dots, \zeta_N)$ 
moment problem.
\item"\rom{(11)}" For any $z_1, \dots, z_N \in\Bbb C_+$ and 
$\zeta_j = G_{\rho_0}(z_j)$, $\rho_0$ is a von~Neumann solution 
of the $\Gamma (z_1, \dots, z_N; \zeta_1, \dots, \zeta_N)$ 
moment problem.
\item"\rom{(12)}" The Nevanlinna function $\Phi$ of $\rho_0$ 
is a rational function of degree at most $N$.
\endroster

In particular, the Nevanlinna function $\Phi$ of $\rho_0$ has 
degree $N$ if and only if $\rho_0$ has order $N$. Moreover, 
if one and hence all those conditions hold, $\rho_0$ is an 
extreme point in $\Cal M^H(\gamma)$ and is also a pure point 
measure. 
\endproclaim

\remark{Remark} The measures obeying (1)--(12) are what Akhiezer 
calls canonical solutions of order $N$.
\endremark

\proclaim{Lemma 6.5} If $\varphi\in \Cal H_0$ and $z\notin
\sigma(B)$, then
$$
(B-z)^{-1} \varphi \in \Cal H_0 + [e_1(z)].
$$
\endproclaim

\demo{Proof} Since $(B-z)^{-1}$ is bounded and $\Cal H_0 + 
[e_1(t)]$ is closed, it suffices to prove this for $\varphi(x)
=P(x)$, a polynomial in $x$. But then
$$\align
(B-z)^{-1}\varphi &= (x-z)^{-1} P(x) = P(z)(x-z)^{-1} + 
\frac{(P(x)-P(z))}{x-z} \\
&= P(z) e_1(z) + R(x)
\endalign
$$
for a polynomial $R$, that is, $(B-z)^{-1}\varphi \in \Cal H_0 
+e_1(z)$. \qed
\enddemo

\proclaim{Lemma 6.6} Supose that $z_0, z_1, \dots, z_\ell \in 
\Bbb C\backslash\sigma(B)$ are distinct. If $e_m(z_0)$ is in the 
span of $\Cal H_0 \cup \text{\rom{span}}[\{e_j(z_0)\}^{m-1}_{j=1}]
\cup \text{\rom{span}}[\{e_1 (z_k)\}^\ell_{k=1}]$, then so is 
$e_{m+1}(z_0)$.
\endproclaim

\remark{Remark} $\ell$ can be zero.
\endremark

\demo{Proof} By hypothesis, for some $\varphi\in\Cal H_0$ and 
$\{\alpha_j\}^{m-1}_{j=1}$ and $\{\beta_k\}^\ell_{k=1}$ in 
$\Bbb C$,
$$
e_m (z_0) = \varphi + \sum_{j=1}^{m-1} \alpha_j e_j (z_0) 
+ \sum_{k=1}^\ell \beta_k e_1(z_k).
$$
Apply $(B-z_0)^{-1}$ to this using $(B-z_0)^{-1} e_j (z_0) = 
e_{j+1} (z_0)$, Lemma~6.5, and $(B-z_0)^{-1} e_1 (z_k) = 
(z_0 - z_k)^{-1} [e_1 (z_0) - e_1 (z_k)]$ and we see that 
$e_{m+1}(z_0)$ is in the requisite span. \qed
\enddemo

\proclaim{Proposition 6.7} Let $\ell = \text{\rom{ord}}(\rho) 
<\infty$. Then for any set $\{z_i\}^\ell_{i=1}$ of $\ell$ 
distinct points in $\Bbb C\backslash\sigma(B)$, $\Cal H_0 \cup 
\{e_1(z_i)\}^\ell_{i=1}$ span $\Cal H_\rho$.
\endproclaim

\demo{Proof} By the Stone-Weierstrass theorem, linear combinations 
of $\{(x-z)^{-1} \mid z\in\Bbb C\backslash \sigma(B)\}$ are 
dense in the continuous functions on $\sigma(B)$ vanishing at 
infinity, and so they are dense in $L^2 (\Bbb R, d\rho)$. It 
follows that we can find $\{z_i\}^\ell_{i=1}$ so that $\Cal H_0 
\cup \{e_1 (z_i)\}^\ell_{i=1}$ span $\Cal H_\rho$.

Pick $z_0 \in\Bbb C\backslash\sigma(B)$ and let $W = (B-z_\ell)
(B-z_0)^{-1}$. Then $W$ is bounded with bounded inverse, so it 
maps dense subspaces into dense subspaces. By Lemma~6.5, $W
[\Cal H_0] \subset \Cal H_0 + [e_1 (z_0)]$. For $i=1, \dots, 
\ell-1$, $We_1 (z_i) = e_1 (z_i) + (\frac{z_0 - z_\ell}
{z_0 -z_i}) [e_1(z_0) - e_1(z_i)]$ and $We_1 (z_\ell) = e_1 (z_0)$. 
Thus, $W$ maps the span of $\{e_1 (z_i)\}^\ell_{i=1}$ into 
$\{e_1 (z_i)\}^{\ell-1}_{i=0}$. So $\Cal H_0 \cup 
\{e_1 (z_i)\}^{\ell-1}_{i=0}$ span $\Cal H_\rho$. By successive 
replacement, we can move the $z_i$'s to an arbitrary set of 
distinct points. \qed
\enddemo

\remark{Remark} Proposition~6.7 shows (1)--(3) of Theorem~6.4 
are equivalent. 
\endremark

\proclaim{Proposition 6.8} Suppose that $\ell =\text{\rom{ord}}
(\rho)<\infty$. Then for any $z_0\in\Bbb C\backslash\sigma(B)$, 
$\Cal H_0 \cup \{e_j (z_0)\}^\ell_{j=1}$ span $\Cal H_\rho$.
\endproclaim

\demo{Proof} By hypothesis, there must be some dependency 
relation
$$
\varphi + \sum_{j=1}^{\ell + 1} \alpha_j e_j (z_0)=0 \tag 6.10
$$
for $\varphi\in\Cal H_0$ and some $(\alpha_1, \dots, 
\alpha_{\ell+1})\neq 0$. Let $k+1 = \max\{j\mid \alpha_j \neq 0\}$. 
Then solving {\rom{(6.10)}} for $e_{k+1} (z_0)$, we see that 
$e_{k+1} (z_0)$ lies in the span of $\Cal H_0 \cup \{e_j (z_0) 
\}^k_{j=1}$; and so by induction and Lemma~6.6, all $e_m 
(z_0)$ lie in this span and so in the span of $\Cal H_0 \cup 
\{e_j (z_0)\}^\ell_{j=1}$.  $e_1(z)$ is an analytic function in 
$\Bbb C\backslash\Bbb R$ with Taylor coefficients at $z_0$ 
equal to $e_n (z_0)$, so $e_1(z)$ lies in the span of $\Cal H_0 
\cup \{e_j (z_0)\}^\ell_{j=1}$ for $z$ in the same half plane 
as $z_0$. 
By Proposition~6.7, these $e_1$'s 
together with $\Cal H_0$ span $\Cal H_\rho$. \qed
\enddemo

\remark{Remark} Proposition~6.8 shows that (1), (4), (5) of 
Theorem~6.4 are equivalent.
\endremark

\proclaim{Proposition 6.9} Fix distinct points $z_1, \dots, 
z_\ell \in\Bbb C\backslash\Bbb R$. Let
$$
d\mu (x) = \prod_{i=1}^\ell |x-z_i|^{-2}\, d\rho(x).
$$
Then the polynomials are dense in $L^2(\Bbb R, d\mu)$ if and 
only if $\Cal H_0 \cup \{e_1 (z_j)\}^\ell_{j=1}$ span 
$\Cal H_\rho$.
\endproclaim

\demo{Proof} Let $U$ be the unitary map from $\Cal H_\rho$ to 
$\Cal H_\mu$ given by 
$$
(Uf)(x) = \prod_{i=1}^\ell (x-z_i) f(x).
$$
Then $U$ maps the span of $\Bbb C[X]\cup\{e_1(z_i)\}^\ell_{i=1}$ 
onto $\Bbb C[X]$ for $P(x)$ is a polynomial of degree $m$ if 
and only if $\prod_{i=1}^\ell (x-z)^{-1} P(x)$ is a linear 
combination of a polynomial of degree $\max(0, m-\ell)$ and 
$\{(x-z_i)^{-1}\}^\ell_{i=1}$. Thus, the density assertions 
are equivalent. \qed
\enddemo

\remark{Remark} Proposition~6.9 and Corollary~4.15 show the 
equivalence of (2), (3), (10), and (11) of Theorem~6.4 (given 
that we already know that (2) and (3) are equivalent).
\endremark

\proclaim{Proposition 6.10} If $\text{\rom{ord}}(\rho_0)$ is 
finite, then $\rho_0$ is a pure point measure with discrete 
support.
\endproclaim

\demo{Proof} Let $\ell = \text{ord}(\rho_0)$. Pick distinct 
$\{z_i\}^\ell_{i=1}$ in $\Bbb C_+$. Then by Propositions~6.7 and 
6.9, the polynomials are dense in $L^2 (\Bbb R, d\mu_\ell)$ 
where for $k=0,1,\dots, \ell$, we set
$$
d\mu_k (x) = \prod_{j=1}^k |x-z_i|^{-2} \, d\rho(x).
$$
Let $k_0$ be the smallest $k$ for which the polynomials are 
dense in $L^2 (\Bbb R, d\mu_k)$. If $k_0 =0$, then $d\rho$ 
is a von~Neumann solution of the $\gamma$ problem. If $k_0 >0$, 
then $d\mu_{k_0 -1}$ is not a von~Neumann solution of its 
moment problem. So by Proposition~4.15 and Theorem~6.1, the 
moment problem for $\Gamma^{(0)}_m \equiv \Gamma_m (z_1, \dots, 
z_{k_0 -1}, z_{k_0}; \zeta_1, \dots, \zeta_{k_0-1}, \zeta_k)$ 
is indeterminate. But since the polynomials are dense in $L^2 
(\Bbb R, d\mu_k)$, $d\mu_{k_0}$ is a von~Neumann solution 
of an indeterminate problem.

Either way, $d\mu_{k_0}$ is a von~Neumann solution of an 
indeterminate problem. It follows Theorem~4.11 that 
$d\mu_{k_0}$ and so $d\rho$ is a discrete point measure. \qed
\enddemo

As a final preliminary to the proof of Theorem~6.4, we need a 
known result about interpolation of Herglotz functions. Given 
$z_1, \dots, z_n; w_1, \dots, w_n\in C_+$ with the $z$'s distinct, 
define the $n\times n$ matrix $D$ by
$$
D_{ij}(z_1, \dots, z_n; w_1, \dots, w_n) = \frac{w_i -\bar w_j}
{z_i - \bar z_j}\, . \tag 6.11
$$

\proclaim{Theorem 6.11} Pick $z_1, \dots, z_n \in\Bbb C_+$ with 
$n\geq 1$. There exists a Herglotz function $\Phi$ with 
$$
\Phi (z_i) = w_i, \qquad i=1,\dots, n \tag 6.12
$$
if and only if $D(z_1, \dots, z_n; w_1, \dots, w_n)$ is a \rom(not 
necessarily strictly\rom) positive definite matrix. Moreover, the 
following are equivalent, given such a $\Phi$\rom:
\roster
\item"\rom{(1)}" $\det\,D(z_1, \dots, z_n; w_1, \dots, w_n) 
=0$.
\item"\rom{(2)}" There is a unique Herglotz function $\Phi$ 
obeying {\rom{(6.12)}}.
\item"\rom{(3)}" $\Phi$ is a real rational polynomial of degree 
at most $n-1$.
\endroster
\endproclaim

\remark{Remark} Since (3) is independent of the choice of 
$\{z_i\}^n_{i=1}$, so are (1), (2).
\endremark

We will sketch a proof of this result (fleshing out some 
arguments in [\akh]) below.

\demo{Proof of Theorem 6.4} As already noted, Propositions~6.7, 
6.8, and 6.9 show that (1), (2), (3), (4), (5), (10), and (11) 
are equivalent. Theorem~6.1 and Corollary~4.15 prove the 
equivalence of (10) and (11) with (6), (7), (8), and (9). By 
Theorems~4.14 and 6.11, (6) and (12) are equivalent.

Since $\text{ord}(\rho)=N$ is equivalent to $\text{ord}(\rho) 
\geq N$ and the negation of $\text{ord}(\rho)\leq N-1$, we see 
that $\Phi_{\rho_0}$ has degree precisely equal to $\text{ord}
(\rho_0)$. That $\rho_0$ is then an extreme point is proven in 
Appendix~B. \qed
\enddemo

We conclude this section by proving Theorem~6.11.

\proclaim{Proposition 6.12} Let $\Phi$ be a Herglotz function 
and let $w_i = \Phi (z_i)$, $i=1,\dots, n$ for distinct $z_1, 
\dots, z_n \in\Bbb C_+$. Let $D$ be given by {\rom{(6.11)}}. 
Then $D(z_1, \dots, z_n; w_1, \dots, w_n)$ is \rom(not 
necessarily strictly\rom) positive definite and $\det\, D=0$ 
if and only if $\Phi$ is a real ratio of polynomials of degree 
$n-1$ or less.
\endproclaim

\demo{Proof} Using the Herglotz representation (1.19), we get a 
representation of $D_{ij}$,
$$
\frac{w_i-\bar w_j}{z_i - \bar z_j} = c + \int d\mu(x)\, 
\frac{1}{x-z_i}\, \frac{1}{x-\bar z_j}\, ,
$$
from which we obtain for $\alpha\in\Bbb C^N$:
$$
\sum_{i,j=1}^n \bar\alpha_i \alpha_j D_{ij} = 
c \, \biggl| \sum_{i=1}^n \alpha_j \biggr|^2 + 
\int_{-\infty}^\infty d\mu(x) \biggl| \sum_{i=1}^n 
\frac{\alpha_i}{x_i - z_i} \biggr|^2, \tag 6.13
$$
proving the positivity.

Suppose $\det(D) =0$. Then there is a non-zero $\alpha\in 
\Bbb C^n$ so that the right side of (6.13) is $0$. Note that 
$\sum_{i=1}^n \frac{\alpha_i}{(x - z_i)} = \prod_{i=1}^n 
(x-z_i)^{-1} Q(x)$, where $Q$ is a polynomial of degree $n-1$ if 
$\sum_{i=1}^n \alpha_i \neq 0$ and of degree at most $n-2$ if 
$\sum_{i=1}^n \alpha_i =0$. Thus, for the right side of (6.13) 
to vanish for some non-zero $\alpha$ in $\Bbb C^n$, either 
$c\neq 0$ and $\mu$ is supported at $n-2$ or fewer points, or 
else $c= 0$ and $\mu$ is supported at $n-1$ or fewer points. 
Either way, by Proposition~6.3, $\Phi$ is a real rational 
function of degree at most $n-1$.

Conversely, suppose $\Phi$ is a real rational function of degree 
precisely $n-1$. Then either $\mu$ is supported at $n-1$ points 
(say, $x_1, \dots, x_{n-1}$) or $c\neq 0$ and $\mu$ is supported 
at $n-2$ points (say, $x_1, \dots, x_{n-2}$). The map $\psi : 
\Bbb C^n \to \Bbb C^{n-1}$ given by
$$
\psi_j (\alpha) = \sum_{i=1}^n \frac{\alpha_i}{x_j - z_i}
$$
(with $\psi_{n-1}(\alpha) = \sum_{i=1}^n \alpha_i$ if $c\neq 0$) 
has a non-zero kernel by dimension counting. Thus, the right side 
of (6.13) is zero for some $\alpha\in\Bbb C^N$. If degree $\Phi$ 
is smaller than $n-1$, we can find $(\alpha_1, \dots, \alpha_\ell)
\in\Bbb C^\ell$ so that $\sum_{i,j=1}^\ell \bar\alpha_i \alpha_j 
D_{ij}=0$, which still implies that $\det(D)=0$. \qed
\enddemo

\proclaim{Proposition 6.13} Let $z_1, \dots, z_n; w_1, \dots, 
w_n$ lie in $\Bbb C_+$ with the $z_i$ distinct. Suppose the 
matrix $D_{ij}$ given by {\rom{(6.11)}} is \rom(not necessarily 
strictly\rom) positive definite. Then there exists a Herglotz 
function with $\Phi (z_i) = w_i$, $i=1,\dots, n$.
\endproclaim

\demo{Proof} This is a fairly standard use of the Hahn-Banach 
theorem. We will use the representation (6.8) for $\Phi$ and 
the function $\varphi_z$ of (6.7). Begin by noting that without 
loss of generality, we can suppose $z_n =i$ (by mapping $\Bbb C_+ 
\to \Bbb C_+$ with a linear map that takes the original $z_n$ to 
$i$) and that $\text{Re}\, w_n =0$ (by adjusting the constant 
$d$ in (6.8)). Indeed, since
$$
\varphi_i (x) \equiv i, \tag 6.14
$$
this choice $\text{Re}\, \Phi(i)=0$ is equivalent to $d=0$ in the 
representation (6.8).

Let $V$ be the $2n-1$-dimensional subspace of $C(\Bbb R\cup 
\{\infty\})$, the real-valued continuous functions on $\Bbb R\cup 
\{\infty\}$ spanned by $1$, $\{\text{Re}\, \varphi_{z_j}(\,\cdot\,)
\}^{n-1}_{j=1}$ and $\{\text{Im}\, \varphi_{z_j}(\, \cdot \,) 
\}^{n-1}_{j=1}$. Any $f\in V$ can be written:
$$
f= A_n + \sum_{j=1}^{n-1} A_j \varphi_j(x) + \bar A_j \, 
\overline{\varphi_j(x)} \tag 6.15
$$
with $A_1, \dots, A_{n-1}\in\Bbb C$ and $A_n \in\Bbb R$. Define 
a linear functional $f:V\to\Bbb R$ by
$$
\ell (f) = A_n \, \text{Im}\, w_n + \sum_{j=1}^{n-1} A_j 
w_j + \bar A_j \bar w_j \tag 6.16
$$
if $f$ has the form (6.15).

We will prove shortly that
$$
f(x) > 0 \qquad \text{for all } x \Rightarrow \ell(f)>0. \tag 6.17
$$
Assuming this, we let $X = \{f\in C(\Bbb R\cup\{\infty\} \mid 
f(x) >0 \text{ for all }x\}$ and $Y=\{f\in V \mid \ell(f) =0\}$. 
Since $X$ is open and $Y$ closed, by the separating hyperplane 
version of the Hahn-Banach theorem (see [\rsI]), there is a 
linear functional $L$ on $\Bbb C(\Bbb R\cup\{\infty\})$ so 
$L>0$ on $X$ and $L\leq 0$ on $Y$. If $L$ is normalized so 
$L(1) = \text{Im}\, w_n$, it is easy to see that $L$ extends 
$\ell$, and so defines a measure $\sigma$ on $\Bbb R\cup
\{\infty\}$ with 
$$
\Phi(z_i) \equiv \int \varphi_{z_i}(x)\, d\sigma (x) = w_i, 
\qquad i=1,\dots, n.
$$
Thus, if (6.17) holds, we have the required $\Phi$.

Since $f(x) >0$ on the compact set $\Bbb R\cup\{\infty\}$ implies 
$f\equiv\varepsilon + g$ with $g\geq 0$ and $\varepsilon >0$ 
(by hypothesis, $1\in V$) and $\text{Im}\, w_n >0$, we need 
only show $f\geq 0$ implies $\ell(f)\geq 0$.

If $f$ has the form (6.15), we can write
$$
f(x) = \frac{Q(x)}{\prod_{j=1}^{n-1} |x-z_j|^2}\, ,
$$
where $Q$ is a polynomial. $f\geq 0$ implies $Q\geq 0$. Any real 
non-negative polynomial has roots either in complex conjugate  
pairs or double real roots, so $Q$ must have even degree and 
have the form
$$
Q(x) = c^2 \prod_{k=1}^\ell |x-y_k|^2
$$
for suitable $y_1, \dots, y_\ell \in \Bbb R\cup\Bbb C_+$. Thus,  
with $R(x) = c \prod_{k=1}^\ell (x-y_k)$,
$$
f = |h|^2 \qquad \text{with}\qquad h(x) = \frac{R(x)}
{\prod_{j=1}^{n-1} (x-z_j)}\, .
$$
Since $f$ is bounded on $\Bbb R\cup\{\infty\}$, $\deg(R)\leq n-1$, 
and thus $h$ has the form
$$
h(x) = \beta_n + \sum_{j=1}^{n-1} \beta_j (x-z_j)^{-1}. \tag 6.18
$$
Since $(x-z_j)^{-1} = (z_j -i)^{-1} [\frac{(x-i)}{(x-z_j)}-1]$, 
we can rewrite (6.18) as
$$
h(x) = \sum_{j=1}^n \alpha_j\, \frac{x-i}{x-z_j}
$$
for real $x$. Thus,
$$
f(x) = |h(x)|^2 = \sum_{j=1}^n \bar\alpha_i \alpha_j 
\frac{x^2 +1}{(x-\bar z_j)(x-z_j)} = \sum_{j=1}^n \bar\alpha_i 
\alpha_j \biggl[ \frac{\varphi_{z_j}(x) - 
\overline{\varphi_{z_i}(x)}}{z_j - \bar z_i}\biggr]
$$
so that
$$
\ell (f) = \sum_{i,j=1}^n \bar\alpha_i \alpha_j D_{ij},
$$
which is non-negative by hypothesis. \qed
\enddemo

\demo{Proof of Theorem 6.11} The first assertion is a direct 
consequence of Propositions 6.12 and 6.13. Proposition~6.12 
shows the equivalence of (1) and (3).

To prove (1) $\Rightarrow$ (2), suppose (1) holds and, as in the 
last proof, we can suppose that $z_n=i$ and $\text{Re}\, w_n =0$. 
Then by our proof of Proposition~6.12, the set of points where 
the measure $\sigma$ of (6.8) is supported is determined by the 
$\alpha$ with $\sum_i D_{ij} \alpha_j =0$ as an $n-1$ point set 
$x_1, \dots, x_{n-1}$ (with $x_{n-1}=\infty$ allowed). Thus, any 
such $\Phi$ is of the form $P/Q$ where $\deg(P)\leq n-1$ and 
$Q(z)=\prod_{j=1}^{n-1} (z-x_j)$ (if $x_{n-1}=\infty$, the 
product only goes from $1$ to $n-2$), If $\Phi_1$ and $\Phi_2$ 
are two solutions of (6.12), they have the same $Q$ but they 
could be distinct $P$'s, say $P_1$ and $P_2$. But by (6.12), 
$P_1(z) - P_2(z)$ vanishes at $z_1, \dots, z_n$. Since $P_1 - 
P_2$ is a polynomial of degree $n-1$, this is only possible if 
$P_1 - P_2 =0$, that is, $\Phi_1 = \Phi_2$. Thus, (1) 
$\Rightarrow$ (2).

For the converse, suppose the determinant
$$
D(z_1, \dots, z_n; w_1, \dots, w_n) >0. \tag 6.19
$$
If $n=1$, since $z,w\in\Bbb C_+$, it is easy to see there are 
multiple $\Phi$'s obeying (6.12). (For example, as usual we can 
consider the case $z=w=i$ and then that $\Phi(z) = z$ or $\Phi 
(z) = -z^{-1}$.) So we suppose $n\geq 2$.

Consider the function $g(w) = \det[D(z_1, \dots, z_n; w_1, 
\dots, w_{n-1}, w)]$ where we vary $w_n$. $g(w)$ is of the 
form $c|w|^2 + dw + \bar d\bar w + e$ with $c<0$, $e$ real, 
and $d$ complex. ($c$ is strictly negative since it is $-\det 
(D(z_1, \dots, z_{n-1}; w_1, \dots, w_{n-1})$ and we are supposing 
(6.19) and $n\geq 2$.)  Thus, the set $g(w)\geq 0$ is a disk and 
since $g(w_n) >0$, $w_n$ is in its interior. Let $w_0$ be the 
center of this disk, $R$ its radius, and let $w(\theta) = w_0 
+ Re^{i\theta}$. Since we have proven (1) $\Rightarrow$ (2), there 
is a unique $\Phi$ with
$$
\Phi_\theta (z_i) = w_i, \quad i=1,\dots, n-1; \qquad 
\Phi_\theta (z_n) = w(\theta) \tag 6.20
$$
and it is a rational function of degree at most $n-1$.

As usual, we can suppose $z_1 = w_1 = i$ so each $\Phi_\theta 
(z)$ has the form
$$
\Phi_\theta (z) = \int \frac{1+xz}{x-z}\, d\sigma_\theta (x), 
\tag 6.21
$$
where $d\sigma_\theta$ is a probability measure on $\Bbb R 
\cup \{\infty\}$. If $\theta_k \to \theta_\infty$, and 
$d\sigma_{\theta_k}\to d\rho$, then the Herglotz function 
associated to $d\rho$ obeys (6.20) for $\theta_\infty$ and 
so it must be $\Phi_{\theta_\infty}$. It follows (since the 
probability measures are compact) that $d\rho = 
d\sigma_{\theta_\infty}$ and thus $\{d\sigma_\theta\}$ is closed 
and $d\sigma_\theta \mapsto \Phi_\theta$ is continuous. Since a 
continuous bijection between compact sets has a continuous 
inverse, $\theta \mapsto d\sigma_\theta$ is continuous.

Since $\Phi_\theta$ is unique, $d\sigma_\theta$ is a pure point 
measure with at most $n-1$ pure points. Since the function is 
not determined by $(z_1, \dots, z_{n-1})$, there must be exactly 
$n-1$ points. It follows that the points in the support must vary 
continuously in $\theta$.

Note next that for any $\theta$, we can find $\theta'$ and 
$t_\theta \in (0,1)$ so that
$$
w_n = t_\theta w(\theta) + (1-t_\theta)w(\theta').
$$

Suppose that there is a unique $\Phi$ obeying (6.12). It follows 
that for any $\theta$,
$$
\Phi(z) = t_\theta \Phi_\theta (z) + (1-t_\theta) \Phi_{\theta'} 
(z) \tag 6.22
$$ 
and thus $\Phi(z)$ has a representation of the form (6.21) with 
$d\sigma$ a point measure with at most $2n-2$ pure points. By 
(6.22) again, each $d\sigma_\theta$ must be supported in that 
$2n-2$ point set, and then by continuity, in a fixed $n-1$ 
point set. But if $d\sigma_\theta$ is supported in a fixed 
$\theta$-independent $n-1$ point set, so is $d\sigma$, and 
thus (6.19) fails. We conclude that there must be multiple 
$\Phi$'s obeying (6.12). \qed
\enddemo

\vskip 0.3in

\flushpar {\bf Appendix A: The Theory of Moments and 
Determinantal Formulae}
\vskip 0.1in

The theory of moments has a variety of distinct objects constructed 
in principle from the moments $\{\gamma_n\}^\infty_{n=0}$: 
the orthogonal polynomials $\{ P_n(x)\}^\infty_{n=0}$, the 
associated polynomials $\{Q_n(x)\}^\infty_{n=0}$, the sums 
$\sum_{j=0}^n P_j(0)^2$ and $\sum_{j=0}^n Q_j (0)^2$, the 
Jacobi matrix coefficients, and the approximations 
$-\frac{Q_n (x)}{P_n(x)}$ and $-\frac{N_n(x)}{M_n(x)}$. It turns 
out most of these objects can be expressed as determinants. These 
formulae are compact and elegant, but for some numerical 
applications, they suffer from numerical round-off errors in 
large determinants.

We have already seen two sets of determinants in Theorem~1. 
Namely, let $\Cal H_N$ be the $N\times N$ matrix,
$$
\Cal H_N = \pmatrix 
\gamma_0 & \gamma_1 & \dots & \gamma_{N-1}\\
\gamma_1 & \gamma_2 & \dots & \gamma_N \\
\vdots & \vdots & {} & \vdots \\
\gamma_{N-1} & \gamma_N & \dots & \gamma_{2N-2} 
\endpmatrix \tag A.1a
$$
and $\Cal S_N$, the matrix
$$
\Cal S_N = \pmatrix 
\gamma_1 & \gamma_2 & \dots & \gamma_N\\
\gamma_2 & \gamma_3 & \dots & \gamma_{N+1} \\
\vdots & \vdots & {} & \vdots \\
\gamma_N & \gamma_{N+1} & \dots & \gamma_{2N-1} 
\endpmatrix \tag A.1b
$$
and define
$$
h_N = \det (\Cal H_N), \qquad s_N = \det(\Cal S_n) \tag A.2
$$
(for comparison, $D_N$ from [\akh] is our $h_{N+1}$). We will use 
$h_N(\gamma)$ if we need to emphasize what moments $\{\gamma_n 
\}^\infty_{n=0}$ are involved. Thus, Theorem~1 says $h_N >0$ for 
all $N$ is equivalent to solubility of the Hamburger problem and 
$h_N >0$, $s_N>0$ for all $N$ is equivalent to solubility of the 
Stieltjes problem.

There is an interesting use of determinants to rewrite $h_N$ and 
$s_N$ in terms of the moment problem that makes their positivity 
properties evident. Suppose $d\rho$ obeys (1.1). Then
$$\align
h_N &= \int \det ((x^{a+b}_a)_{0\leq a,b\leq N-1}) 
\prod_{a=0}^{N-1} d\rho(x_a) \\
&= \int \biggl[ \prod_{a=1}^n x^a_a\biggr] \det 
((x^b_a))_{0\leq a,b\leq N-1} \prod_{a=0}^{N-1} d\rho(x_a).
\endalign
$$
Permuting over indices, we see that in $\prod_{a=1}^n (x^a_a) 
\det (x^b_a)$, we can replace $x_a$ by $x_{\pi(a)}$ for any 
permutation, $\pi$. Since $\det (x^b_{\pi(a)})=(-1)^\pi 
\det(x^b_a)$, we see that
$$
h_N = (N!)^{-1} \int [\det((x^b_a)_{0\leq a,b\leq N-1})]^2 
\prod_{a=0}^{N-1} d\rho(x_a).
$$
Recognizing the Vandermonde determinant, we have
$$
h_N = (N!)^{-1} \int \prod_{0\leq a < b\leq N-1} (x_a - x_b)^2 
\prod_{a=0}^{N-1} d\rho(x_a). \tag A.3a
$$
Similarly,
$$
s_N = (N!)^{-1} \int x_0 \dots x_{N-1} 
\prod_{0\leq a < b \leq N-1} (x_a - x_b)^2 
 \prod_{a=0}^{N-1} d\rho(x_a). 
\tag A.3b
$$

The most basic formula is:

\proclaim{Theorem A.1} $P_n(x)$ is given by
$$
P_n(x) = \frac{1}{\sqrt{h_n h_{n+1}}}\, 
\det \pmatrix
\gamma_0 & \gamma_1 & \dots & \gamma_n \\
\gamma_1 & \gamma_2 & \dots & \gamma_{n+1} \\
\vdots & \vdots & {} & \vdots \\
\gamma_{n-1} & \gamma_n & \dots & \gamma_{2n-1} \\
1 & x & \dots & x^n 
\endpmatrix. \tag A.4
$$
\endproclaim

\demo{Proof} Let $S_n(x)$ be the determinant on the right side 
of (A.4). Then for any solution $\rho$ of the moment problem, we 
have that $\int S_n(x) x^j\, d\rho(x)$ is given by the same 
determinant, but with the last row replaced by $\gamma_j \ 
\gamma_{j+1} \ \dots \ \gamma_{j+n}$. It follows that
$$\alignat2
\int x^j S_n (x)\, d\rho(x) &=0,  \qquad && j=0,1,\dots, n-1 
\tag A.5 \\
&= h_{n+1} \qquad && j=n.
\endalignat
$$
In particular, since
$$
S_n (x) = h_n x^n + \text{ lower order}, \tag A.6
$$
we see that
$$
\int S_n (x)^2\, d\rho(x) = \ h_n h_{n+1}. \tag A.7
$$
From (A.5)--(A.7), it follows that $P_n(x) = \frac{S_n (x)}
{\sqrt{h_n h_{n+1}}}$. \qed
\enddemo

From (A.4), we can deduce formulae for the coefficients 
$a_n, b_n$ of a Jacobi matrix associated to the moments 
$\{\gamma_n\}^\infty_{n=0}$. Let $\tilde h_n$ be the $n\times n$ 
determinant obtained by changing the last column in $\Cal H_N$:
$$
\tilde h_n = \det \pmatrix
\gamma_0 & \gamma_1 & \dots & \gamma_{n-2} & \gamma_n \\
\vdots & \vdots &{} & \vdots & \vdots \\
\gamma_n & \gamma_{n+1} & \dots & \gamma_{2n-3} & 
\gamma_{2n-1} 
\endpmatrix.
$$
Thus (A.4) implies that
$$
P_n(x) = \sqrt{\frac{h_n}{h_{n+1}}} \, \biggl[ x^n - 
\frac{\tilde h_n}{h_n}\, x^{n-1} + \text{ lower order} 
\biggr] . \tag A.8
$$

\proclaim{Theorem A.2} For $n\geq 0$,
$$\alignat2
&\text{\rom{(i)}} \qquad && a_n = \biggl(\frac{h_n h_{n+2}}
{h^2_{n+1}}\biggr)^{1/2} \\
&\text{\rom{(ii)}} \qquad && \sum_{j=0}^n b_j = 
\frac{\tilde h_{n+1}}{h_{n+1}}\, .
\endalignat
$$
\endproclaim

\demo{Proof} By the definition of $a_n$ and $b_n$, we have that
$$
xP_n (x) = a_n P_{n+1}(x) + b_n P_n (x) + a_{n-1} 
P_{n-1}(x). \tag A.9
$$
Identifying the $x^{n+1}$ and $x^n$ terms in (A.9) using (A.8), 
we see that
$$\gather
\sqrt{\frac{h_n}{h_{n+1}}} = a_n \sqrt{\frac{h_{n+1}}
{h_{n+2}}} \tag A.10 \\
\sqrt{\frac{h_n}{h_{n+1}}} \biggl( - \frac{\tilde h_n}{h_n} 
\biggr) = a_n \sqrt{\frac{h_{n+1}}{h_{n+2}}} \biggl( - 
\frac{\tilde h_{n+1}}{h_{n+1}} \biggr) + b_n 
\sqrt{\frac{h_n}{h_{n+1}}}\, . \tag A.11
\endgather
$$
(A.10) implies (i) immediately, and given (A.10), (A.11) becomes
$$
b_n = \frac{\tilde h_{n+1}}{h_{n+1}} - \frac{\tilde h_n}
{h_n}\, ,
$$
which implies (ii) by induction if we note the starting point 
comes from looking at the constant term in $x P_0(x) =x = a_0 
P_1 (x) + b_0 P_0 (x)$, which implies that $b_0 = 
\frac{\tilde h_1}{h_1} = \frac{\gamma_1}{\gamma_0}$. \qed
\enddemo

\remark{Remark} (ii) has an alternate interpretation. (A.8) says 
that $\frac{\tilde h_{n+1}}{h_{n+1}}$ is the sum of the $n+1$ 
roots of $P_{n+1}(x)$. But $P_{n+1}(x)$ is a multiple of the 
determinant of the Jacobi matrix $A^{[n+1]}_F$. So the sum of 
the roots is just the trace of $A^{[n+1]}_F$, that is, 
$\sum_{j=0}^n b_j$.
\endremark

From (A.4) and Theorem~4.2, (i.e., $Q_n(x) = E_X 
(\frac{P_n(X) - P_n(Y)}{X-Y})$), we immediately get 

\proclaim{Theorem A.3}
$$
Q_n(x) = \frac{1}{\sqrt{h_n h_{n+1}}}\, 
\det \pmatrix
\gamma_0 & \gamma_1 & \dots & \gamma_n \\
\gamma_1 & \gamma_2 & \dots & \gamma_{n+1} \\
\vdots & \vdots & {} & \vdots \\
\gamma_{n-1} & \gamma_n & \dots & \gamma_{2n-1} \\
R_{n,0}(x) & R_{n,1}(x) & \dots & R_{n,n}(x) 
\endpmatrix \tag A.12
$$
where
$$\align
R_{n,j}(x) &= \sum_{k=0}^{j-1} \gamma_{j-1-k} x^k, \qquad 
j \geq 1 \tag A.13a \\
R_{n,j=0} (x) &= 0. \tag A.13b
\endalign
$$
\endproclaim

\demo{Proof} $\frac{x^j - y^j}{x-y} = \sum_{k=0}^{j-1} 
y^{j-1-k}x^k$ so $E_X (\frac{x^j - y^j}{x-y}) = R_{n,j}(x)$. \qed
\enddemo

\remark{Remark} By (A.4), $P_n(x) \sum_{k=0}^{2n-1} \gamma_k 
x^{-k-1}$ has the same form as (A.4) but with the bottom row 
replaced by $S_{n,0}(x) \dots S_{n,n}(x)$ where
$$\align
S_{n,j}(x) &= \sum_{k=0}^{2n-1} \gamma_k x^{j-k-1} = 
\sum_{\ell = -(2n-j)}^{j-1} \gamma_{j-1-\ell} x^\ell \\
&= R_{n,j}(x) + x^{-1} \gamma_j + x^{-2} \gamma_{j+1} + 
\cdots + x^{-n} \gamma_{j+n-1} + O(x^{-n-1}). 
\endalign
$$
Recognizing $x^{-1} \gamma_j + x^{-2} \gamma_{j+1} + \cdots + 
x^{-n-1} \gamma_{j+n-1}$ as $x^{-1}$ times the first row of 
the matrix in (A.4) plus $x^{-2}$ times the second row plus 
\dots, we conclude that
$$
P_n(x) \biggl(\sum_{k=0}^{2n-1} \gamma_k x^{-k-1}\biggr) = 
Q_n(x) + O(x^{-n-1})
$$
and thus
$$
-\frac{Q_n(x)}{P_n(x)} = -\sum_{k=0}^{2n-1} \gamma_k 
x^{-k-1} + O(x^{-2n-1})
$$
consistent with the $f^{[N-1,N]}$ Pad\'e formula (5.28).
\endremark

We saw the quantity $L$ of (5.29) is important. Here is a 
formula for it.

\proclaim{Theorem A.4} $L=\lim_{n\to\infty} - \frac{Q_n(0)}
{P_n(0)}$ and
$$
-\frac{Q_n(0)}{P_n(0)} = \frac{t_n}{s_n}\, , \tag A.14a
$$
where $s_n = \det(\Cal S_N)$ and
$$
t_n = -\det \pmatrix
0 & \gamma_0 & \gamma_1 & \dots & \gamma_n \\
\gamma_0 & \gamma_1 & \gamma_2 & \dots & \gamma_{2n+1} \\
\vdots & \vdots & \vdots & {} & \vdots \\
\gamma_n & \gamma_{n+1} & \gamma_{n+2} & \dots & \gamma_{2n-1} 
\endpmatrix . \tag A.14b
$$
\endproclaim

\demo{Proof} Follows by putting $x=0$ in our formula for $P_n(x)$ 
and $Q_n(x)$. \qed
\enddemo

Next we have explicit formulae for $M_n(x)$ and $N_n(x)$, the 
polynomials introduced for Section~5 (see (5.15) and (5.16)).

\proclaim{Theorem A.5}
$$\align
M_n(x) &= \frac{1}{s_{n-1}} \sqrt{\frac{h_n}{h_{n+1}}} 
\pmatrix
\gamma_1 & \gamma_2 & \dots & \gamma_n \\
\gamma_2 & \gamma_3 & \dots & \gamma_{n+1} \\
\vdots & \vdots & {} & \vdots \\
\gamma_{n-1} & \gamma_{n-2} & \dots & \gamma_{2n-2} \\
x & x^2 & \dots & x^n 
\endpmatrix \tag A.15 \\
N_n(x) & = \frac{1}{s_{n-1}} \sqrt{\frac{h_n}{h_{n+1}}} 
\pmatrix
\gamma_1 & \gamma_2 & \dots & \gamma_n \\
\gamma_2 & \gamma_3 & \dots & \gamma_{n+1} \\
\vdots & \vdots & {} & \vdots \\
\gamma_{n-1} & \gamma_{n-2} & \dots & \gamma_{2n-2} \\
R_{n,1}(x) & R_{n,2}(x) & \dots & R_{n,n}(x)
\endpmatrix \tag A.16
\endalign
$$
where $R_{i,j}(x)$ is given by {\rom{(A.13)}}.
\endproclaim

\remark{Remarks} 1. $P_n$ is given by a $(n+1)\times (n+1)$ 
matrix. To get $M_n$, which has an $n\times n$ matrix, we drop 
the first column and next to last row.

2. It is interesting that (A.15) does not obviously follow from 
the basic definition (5.15). 
\endremark

\demo{Proof} The right side of (A.15) (call it $\widetilde M_n(x)$) 
has the following properties:
\roster
\item"\rom{(i)}" It is a polynomial of degree $n$.
\item"\rom{(2)}" It obeys $\widetilde M_n(0)=0$.
\item"\rom{(3)}" It obeys $E_x (x^j \widetilde M_n(x)) =0$ for 
$j=0,1,\dots, n-2$ since the corresponding matrix has two equal 
rows.
\item"\rom{(4)}" It obeys $\widetilde M_n(x) = \sqrt{h_n/h_{n+1}}
\, x^n + \text{ lower order}$ so it has the same highest degree 
term as $P_n(x)$.
\endroster

These properties uniquely determine $M_n(x)$ so (A.15) is proven. 
(A.16) then follows from (5.24). \qed
\enddemo

\remark{Remark} By mimicking the argument following Theorem~A.3, 
one finds that
$$
M_n(x) \sum_{k=0}^{2n-2} \gamma_k x^{-k-1} = N_n(x) + 
O(x^{-n})
$$
so that
$$
-\frac{N_n(x)}{M_n(x)} = -\sum_{k=0}^{2n-2} \gamma_k 
x^{-k-1} + O(x^{-2n}),
$$
consistent with the $f^{[N-1, N-1]}$ Pad\'e formula used in the 
proof of Theorem~6 (at the end of Section~5).
\endremark

Remarkably, there are also single determinant formulae for 
$\sum_{j=0}^n P_j(0)^2$ and $\sum_{j=0}^n \mathbreak Q_j(0)^2$.

\proclaim{Theorem A.6} We have that
$$\align
\sum_{j=0}^n P_j (0)^2 &= \frac{v_n}{h_{n+1}} \tag A.17 \\
\sum_{j=0}^n Q_j (0)^2 &= - \frac{w_{n+2}}{h_{n+1}}\, , 
\tag A.18
\endalign
$$
where $v_n$ is given by the $n\times n$ determinant,
$$
v_n = \det \pmatrix
 \gamma_2 & \dots & \gamma_{n+1} \\
 \gamma_3 & \dots & \gamma_{n+2} \\
\vdots & {} & \vdots \\
\gamma_{n+1} & \dots & \gamma_{2n} \\
\endpmatrix \tag A.19
$$
and $w_{n+2}$ by the $(n+2)\times (n+2)$ determinant,
$$
w_n = \det \pmatrix
0 & 0 & \gamma_0 & \gamma_1 & \dots & \gamma_{n-1} \\
0 & \gamma_0 & \gamma_1 & \gamma_2 & \dots & \gamma_n \\
\gamma_0 & \gamma_1 & \gamma_2 & \gamma_3 & \dots & \gamma_{n+1} \\
\vdots & \vdots & \vdots & \vdots & {} & \vdots \\
\gamma_{n-1} & \gamma_n & \gamma_{n+1} & \gamma_{n+2} & \dots & 
\gamma_{2n} 
\endpmatrix \tag A.20
$$
\endproclaim

\remark{Remark} Thus by Theorems~3 and 7, indeterminacy for 
both the Hamburger and Stieltjes problems can be expressed in 
terms of limits of ratios of determinants. In the Hamburger 
case, we need (A.17) and (A.18) to have finite limits in order 
that the problem be indeterminate. In the Stieltjes case, (A.17) 
and (A.14) must have finite limits for indeterminacy to hold.
\endremark

\demo{Proof} We actually prove a stronger formula. Let
$$
B_N(x,y)= \sum_{n=0}^N P_n(x) P_n(y) . \tag A.21
$$
We will show that
$$
B_n (x,y) = -h^{-1}_{n+1} \det \pmatrix
0 & 1 & x & \dots & x^N \\
1 & \gamma_0 & \gamma_1 & \dots & \gamma_N \\
y & \gamma_1 & \gamma_2 & \dots & \gamma_{N+1} \\
\vdots & \vdots & \vdots & {} & \vdots \\
y^N & \gamma_N & \gamma_{N+1} & \dots & \gamma_{2N} 
\endpmatrix \tag A.22
$$
(A.17) then follows by setting $x=y=0$ and noting that if $C = 
\{C_{ij}\}_{1\leq i \leq j \leq N}$ is an $N\times N$ matrix 
with $C_{11}=0$, $C_{1j} = \delta_{2j}$, and $C_{i1} =
\delta_{i2}$, then
$$
\det ((C_{ij})_{1\leq i,j\leq N}) = -\det ((C_{ij})_{3\leq 
i,j\leq N}).
$$
(A.18) then follows from $Q_j (z) = E_x 
\bigl(\frac{[P_j (x) - P_j (z)]}{(x-z)}\bigr)$, since
$$
\sum_{n=0}^N Q_n(0)^2 = E_x E_y ([B_N (x,y) - B_N (x,0) - 
B_N (0,y) + B_N (0,0)] x^{-1} y^{-1}). 
$$

Thus we need only prove (A.22). To do this, let $\tilde B_N(x,y)$ 
be the right side of (A.22). Consider $E_x (\tilde B_N(x,y) x^j)$ 
for $0\leq j \leq N$. This replaces the top row in the 
determinant by $(0 \ \gamma_j \ \gamma_{j+1} \ \dots \ 
\gamma_{N+j})$. The determinant is unchanged if we subtract the 
row $(y^j \ \gamma_j \ \dots \ \gamma_{N+j})$ from this row. 
That is, $E_x (\tilde B_N(x,y)x^j)$ is given by the determinant 
with top row $(-y^j \ 0 \ 0 \ \dots \ 0)$. Thus,
$$
E_x (\tilde B_N (x,y) x^j) = -h^{-1}_{n+1} (-y^j) 
h_{n+1} = y^j.
$$
So, $\tilde B_N(x,y)$ is a reproducing kernel. For any polynomial 
$P(x)$ of degree $N$ or smaller, $E_x (\tilde B_N(x,y) P(x)) = 
P(y)$. But
$$
\tilde B_N (x,y) = \sum_{j=0}^N E_x (\tilde B_N (x,y) 
P_j (x)) P_j(x) = B_N (x,y)
$$
since $\{P_j(x)\}^N_{j=0}$ is an orthonormal basis in the 
polynomials of degree $N$ or less. Thus (A.22) is proven. \qed
\enddemo

In terms of the $\gamma^{(\ell)}$ moment problems (with moments 
$\gamma^{(\ell)}_j = \frac{\gamma_{\ell+j}}{\gamma_\ell}$), we 
recognize $v_n$ as $(\gamma_2)^{n-1} h_{n-1}(\gamma^{(2)})$. 
By (A.17) for the $\gamma^{(2)}$ problem,
$$
\sum_{j=0}^{n-1} P^{(2)}_j (0) = \frac{v_{n-1} (\gamma^{(2)})}
{h_n (\gamma^{(2)})}\, .
$$
But $h_n (\gamma^{(2)}) = v_n (\gamma)(\gamma_2)^{-n}$ and 
$v_{n-1} (\gamma^{(2)}) = \gamma^{1-n}_2 y_{n-1}(\gamma)$ where 
$$
y_{n-1} = \pmatrix 
\gamma_4 &\dots & \gamma_{n+2} \\
\gamma_5 & \dots & \gamma_{n+3} \\
\vdots & {} & \vdots \\
\gamma_{n+1} & \dots & \gamma_{2n} 
\endpmatrix . \tag A.23
$$
Thus, using Proposition~5.13:

\proclaim{Theorem A.7} 
$$
\gamma^{-1}_2 \biggl( \sum_{j=0}^n P_j (0)^2 \biggr) 
\biggl( \sum_{j=0}^{n-1} P^{(2)}_j (0)^2\biggr) = 
\frac{y_{n-1}}{h_{n+1}}\, .
$$
In particular, the Hamburger problem is determinate if and only 
if $\lim_{n\to\infty} \frac{y_{n-1}}{h_{n+1}} = \infty$.
\endproclaim

So we have a simple ratio of determinants to determine 
determinacy.

\vskip 0.3in

\flushpar {\bf Appendix B: The Set of Solutions of the 
Moment Problem as a Compact Convex Set}
\vskip 0.1in

In this appendix, we will prove (following [\akh]) that 
$\Cal M^H (\gamma)$ and $\Cal M^S(\gamma)$ are compact convex 
sets whose extreme points are dense. Each set is a subset of 
$\Cal M_+(\Bbb R\cup\{\infty\})$, the set of measures on the 
compact set $\Bbb R\cup\{\infty\}$. This set, with the condition 
$\int d\rho(x) = \gamma_0$, is a compact space in the topology 
of weak convergence (i.e., $\int f(x)\, d\rho_n(x) \to \int f(x)\, 
d\rho(x)$ for continuous functions $f$ on $\Bbb R\cup\{\infty\}$).

\proclaim{Theorem B.1} $\Cal M^H(\gamma)$ and $\Cal M^S(\gamma)$ 
are closed in the weak topology and so are compact convex sets.
\endproclaim

\remark{Remark} Since the $x^n$'s are unbounded, this does not 
follow from the definition of the topology without some additional 
argument.
\endremark

\demo{Proof} By Propositions~4.4 and 4.13, $\mu\in\Cal M^H
(\gamma)$ if and only if
$$
\int (x-z_0)^{-1}\, d\mu(x) \in D(z_0) \tag B.1
$$
for all $z_0\in\Bbb C$. The set of $\mu$'s that obey (B.1) for 
a fixed $z_0$ is closed since $D(z_0)$ is closed and 
$(x-z_0)^{-1}$ is in $C(\Bbb R\cup\{\infty\})$. Thus, the 
intersection over all $z_0$ is closed. $\Cal M^S(\gamma) = 
\Cal M^H(\gamma) \cap \{\mu\mid \int f(x)\, d\mu(x)=0 \text{ if }
\text{supp}\, f\subset (-\infty, 0)\}$ is an intersection of 
closed sets. \qed
\enddemo

\remark{Remark} To get compact sets of measures, we need to consider 
$\Bbb R\cup\{\infty\}$ rather than $\Bbb R$. Thus a priori, the integral 
in (B.1) could have a point at infinity giving a constant term as 
$z_0 = iy$ with $y\to\infty$. Since $D(z_0)\to\{0\}$ as $z_0 = iy$ 
with $y\to\infty$, this term is absent. 
\endremark

\proclaim{Theorem B.2 (Naimark)} $\mu\in\Cal M^H(\gamma)$ 
\rom(resp.~$\Cal M^S(\gamma)$\rom) is an extreme point if and 
only if the polynomials are dense in $L^1 (\Bbb R, d\rho)$.
\endproclaim

\remark{Remark} Compare with density in $L^2 (\Bbb R, d\rho)$ 
which picks out the von~Neumann solutions.
\endremark

\demo{Proof} This is a simple use of duality theory. The 
polynomials fail to be dense if and only if there exists a 
non-zero $F\in L^\infty (\Bbb R, d\rho)$ so that
$$
\int x^n F(x)\, d\rho(x) =0 \tag B.2
$$
for all $n$. We can suppose that $\| F\|_\infty = 1$, in which 
case $d\rho_\pm = (1\pm F)\, d\rho$ both lie in $\Cal M^H 
(\gamma)$ with $\rho=\frac12 (\rho_+ + \rho_-)$, so $\rho$ is not 
extreme.

Conversely, suppose $\rho=\frac12(\rho_+ + \rho_-)$ with 
$\rho_+ \neq \rho_-$ and both in $\Cal M^H(\gamma)$. Then 
$\rho_\pm \leq 2\rho$ so $\rho_+$ is $\rho$-absolutely 
continuous, and the Radon-Nikodyn derivative, $\frac{d\rho_+}
{d\rho}$, obeys $\| \frac{d\rho_+}{d\rho}\|_\infty \leq 2$. 
Let $F=1-\frac{d\rho_+}{d\rho}$, so $\|F\|_\infty \leq 1$ and 
$F\neq 0$ since $\rho_+ \neq \rho_-$. Then
$$
\int F(x) x^n \, d\rho = \int x^n \, d\rho - \int x^n \, 
d\rho_+ = 0.
$$

Thus, we have proven the result for $\Cal M^H(\gamma)$. Since 
$\Cal M^S(\gamma) = \{\rho\in \Cal M^H(\gamma) \mid \int f(x) 
\, d\rho(x)=0 \text{ for $f$ in } C(\Bbb R\cup\{\infty\}) 
\text{ with support in } (-\infty, 0)\}$, $\Cal M^S (\gamma)$ 
is a face of $\Cal M^H (\gamma)$, so extreme points of $\Cal M^S 
(\gamma)$ are exactly those extreme points of $\Cal M^H(\gamma)$ 
that lie in $\Cal M^S(\gamma)$. \qed
\enddemo

\proclaim{Theorem B.3} Let $\rho\in \Cal M^H(\gamma)$ have 
$\text{\rom{ord}}(\rho)<\infty$. Then $\rho$ is an extreme point. 
\endproclaim

\demo{Proof} By Theorem~6.4, for some $z_1, \dots, z_n \in 
\Bbb C_+$, we have that the polynomials are dense in $L^2 (\Bbb R, 
\prod_{j+1}^n |x-z_j|^{-2}d\rho)$. For any polynomially bounded 
continuous function
$$
\int |f(x)|\, d\rho(x) \leq \biggl( \int |f(x)|^2 \prod_{j=1}^n 
|x-z_j|^{-2}\, d\rho(x) \biggr)^{1/2} 
\biggl(\int \prod_{j=1}^n (x-z_j)^2 \, d\rho(x)\biggr)^{1/2}
$$
by the Schwarz inequality. It follows that the identification 
map is continuous from $L^2 (\Bbb R, \prod_{j=1}^n (x-z)^{-2}
\, d\rho)$ into $L^1 (\Bbb R, d\rho)$, so the polynomials are 
dense in $L^1$-norm in the continuous functions, and so in 
$L^1 (\Bbb R, d\rho)$. By Theorem~B.2, $\rho$ is an extreme 
point. \qed
\enddemo

\proclaim{Theorem B.4} The extreme points are dense in $\Cal M^H 
(\gamma)$ \rom(and in $\Cal M^S(\gamma)$\rom). 
\endproclaim

\demo{Proof} The finite point measures are dense in the finite 
measures on $\Bbb R\cup\{\infty\}$. Thus, by the Herglotz 
representation theorem in form (6.8), if $\Phi$ is a Herglotz 
function, there exist real rational Herglotz functions $\Phi_n$ 
so that $\Phi_n(z) \to \Phi(z)$ for each $z\in\Bbb C_+$.

Now let $\rho\in \Cal M^H(\gamma)$ and let $\Phi_\rho$ be the  
Nevanlinna function of $\rho$. Let $\Phi_n$ be as above and 
$\rho_n = \rho_{\Phi_n}$. Then by (4.35), $G_{\rho_n}(z) \to 
G_\rho (z)$ for each $z\in\Bbb C_+$ and so $\rho_n \to \rho$ 
weakly. By Theorems~6.4 and B.3, each such $\rho_n$ is an 
extreme point.

For the Stieltjes case, we need only note that by Theorem~4.18 
and the remark after it, if $\rho\in \Cal M^S(\gamma)$, then 
the approximating $\Phi_n$'s can be chosen so that $\rho_{\Phi_n} 
\in\Cal M^S (\gamma)$. \qed
\enddemo

\proclaim{Theorem B.5} For any indeterminate set of Hamburger 
moments $\{\gamma_n\}^\infty_{n=0}$, $\Cal M^H(\gamma)$ has 
extreme points $\rho$ with $\text{\rom{ord}}(\rho)=\infty$. 
\endproclaim

\demo{Proof} We first pick positive $\alpha_j$ strictly decreasing 
so that for any $\rho\in\Cal M^H(\gamma)$ we have that
$$
\sup_n \int \prod_{j=1}^n (1+\alpha^2_j x^2)^2\, d\rho (x) 
\leq 2. \tag B.3
$$
We can certainly do this as follows: Since the integral only 
depends on the moments $\gamma_j$, we need only do it for some 
fixed $\rho_0 \in\Cal M^H(\gamma)$. Since
$$
\lim_{\alpha_1 \downarrow 0} \int (1+\alpha^2_1 x^2)^2\, 
d\rho(x)= 1,
$$
we can pick $\alpha_1 >0$ so
$$
\int (1+\alpha^2_1 x^2)^2 \, d\rho(x) <2.
$$
We then pick $\alpha_2, \alpha_3, \dots$ inductively so the 
integral is strictly less than 2. Then the $\sup$ in (B.3) is 
bounded by 2. The product must be finite for a.e.~$x$ w.r.t. 
$d\rho$, so
$$
\sum_{j=1}^\infty \alpha^2_j < \infty. \tag B.4
$$
So if
$$
g_n(x) = \prod_{j=1}^n (1+\alpha^2_j x^2),
$$
then for any real $x$,
$$
g(x) = \lim_{n\to\infty} g_n(x)
$$
exists, and by (B.3),
$$
\int g(x)^2\, d\rho(x) < \infty
$$
for any $\rho\in\Cal M^H(\gamma)$.

Now pick $\zeta_1, \zeta_2, \dots$ in $\Bbb C_+$ and 
$\rho_1, \rho_2, \dots$ in $\Cal M^H(\gamma)$ inductively so 
$\rho_k$ obeys
$$
\int (x-i\alpha^{-1}_j)^{-1} \, d\rho_k (x) = \zeta_j, 
\qquad j=1,\dots, k \tag B.5
$$
and then $\zeta_{k+1}$ is the middle of the disk of allowed values 
for $\int (x-i\alpha^{-1}_{j+1})^{-1}\, d\rho(x)$ for those $\rho$ 
in $\Cal M^H(\gamma)$ which obey (B.5). Pick a subsequence of the 
$\rho_j$'s which converges to some $\rho_\infty \in \Cal M^H 
(\gamma)$. Then $\rho_\infty$ obeys (B.5) for all $k$.

Let
$$
\Cal M_k = \{\rho\in\Cal M^H(\gamma)\mid \rho 
\text{ obeys (B.5)}\}
$$ 
and
$$
\Cal M_\infty = \cap \Cal M_k.
$$
Define for all $n,m$,
$$
\Gamma^{(n)}_m = \int x^m \prod_{j=1}^n (1+\alpha^2_j x^2)^{-1} 
\, d\rho_n (x).
$$
By Theorem~6.1, $\mu\in\Cal M^H(\Gamma^{(n)})$ if and only if 
$d\rho \equiv g_n\, d\mu$ lies in $\Cal M_k$. $\Gamma^{(n)}_m$ 
is decreasing to $\Gamma^{(\infty)}_m$ and
$$
\Gamma^{(\infty)}_m = \int x^m g(x)^{-1}\, d\rho(x)
$$
by a simple use of the monotone convergence theorem.

We claim that $\mu\in\Cal M^H(\Gamma^{(\infty)})$ if and only if 
$d\rho = g\, d\mu$ lies in $\Cal M_\infty$. For
$$
\gamma_n = \lim_{n\to\infty} \int x^m g(x)^{-1} g_n(x)\, 
d\rho(x)
$$
and the right side only depends on the moments 
$\Gamma^{(\infty)}_m$ and similarly for calculation of $\int 
(x-i\alpha_j)^{-1}\, d\rho(x)$.

Now let $\mu$ be a von~Neumann solution of the $\Gamma^{(\infty)}$ 
moment problem and $d\rho=g\,d\mu$. Since $\int g^2\, d\rho < 
\infty$ by construction, the proof of Theorem~B.3 shows that $\rho$ 
is an extreme point. On the other hand, since $\rho$ obeys (B.5) 
and $\zeta_j$ is not in the boundary of its allowed circle, 
$\text{ord}(\rho) = \infty$. \qed
\enddemo

\vskip 0.3in

\flushpar {\bf Appendix C: Summary of Notation and Constructions}
\vskip 0.1in

Since there are so many objects and constructions associated to 
the moment problem, I am providing the reader with this summary of 
them and where they are discussed in this paper.

\definition{C1 \quad Structure of the Set of Moments}
\vskip 0.1in

$\{\gamma_n\}^\infty_{n=0}$ is called a set of {\it{Hamburger 
moments}} if there is a positive measure $\rho$ on $(-\infty, 
\infty)$ with $\gamma_n = \int x^n\, d\rho(x)$ and a set of 
{\it{Stieltjes moments}} if there is a $\rho$ with support in 
$[0,\infty)$. We take $\gamma_0=1$ and demand $\text{supp}(\rho)$ 
is not a finite set. Theorem~1 gives necessary and sufficient 
conditions for existence. If there is a unique $\rho$, the 
problem is called {\it{determinate}}. If there are multiple 
$\rho$'s, the problem is called {\it{indeterminate}}. If a set of 
Stieltjes moments is Hamburger determinate, it is a fortiori 
Stieltjes determinate. But the converse can be false (see the 
end of Section~3).

$\Cal M^H(\gamma)$ (resp.~$\Cal M^S(\gamma)$) denotes the set 
of all solutions of the Hamburger problem (resp.~all $\rho\in
\Cal M^H(\gamma)$ supported on $[0,\infty)$). They are compact 
convex sets (Theorem~B.1) whose extreme points are dense 
(Theorem~B.4). Indeterminate Hamburger problems have a 
distinguished class of solutions we have called {\it{von~Neumann 
solutions}} (called $N$-extremal in [\akh] and extremal in 
[\st]). They are characterized as those $\rho\in\Cal M^H
(\gamma)$ with the polynomials dense in $L^2 (\Bbb R, d\rho)$. 
They are associated to self-adjoint extensions $B_t$ of the 
Jacobi matrix associated to $\gamma$ (see {\bf{C4}} below for 
the definition of this Jacobi matrix). The parameter $t$ lies in 
$\Bbb R\cup \{\infty\}$ and is related to $B_t$ and the solution 
$\mu_t\in \Cal M^H(\gamma)$ by
$$
t = (\delta_0, B^{-1}_t \delta_0) = \int x^{-1} d\mu_t(x). 
\tag C.1
$$

In the indeterminate Stieltjes case, there are two distinguished 
solutions: $\mu_F$, the {\it{Friedrichs solution}}, and $\mu_K$, 
the {\it{Krein solution}}. Both are von~Neumann solutions of the 
associated Hamburger problem and are characterized by $\inf 
[\text{supp}(\mu_F)] > \inf[\text{supp}\,\rho]$ for any other 
$\rho\in\Cal M^H(\gamma)$ and by $\mu_K(\{0\})>0$ and $\mu_K 
(\{0\}) >\rho(\{0\})$ for any other $\rho\in\Cal M^H(\gamma)$ 
(see Theorems~3.2 and 4.11 and 4.17). All solutions of the 
Stieltjes problem lie between $\mu_F$ and $\mu_K$ in the sense 
of (4.42) (see Theorem~5.10).

We define
$$
\Bbb C_+ = \{ z\in\Bbb C \mid\text{Im}\, z>0\}.
$$

For any probability measure $\rho$, we define its {\it{Stieltjes 
transform}} as a function of $\Bbb C\backslash\text{supp}(\rho)$ 
by
$$
G_\rho(z) = \int \frac{d\rho(x)}{x-z}\, .
$$
These functions map $\Bbb C_+$ to $\Bbb C_+$ and have an 
asymptotic series
$$
G_\rho (z) \sim -z^{-1} \biggl( \sum_{j=0}^\infty 
\gamma_j z^{-j}\biggr) \tag C.2
$$
as $|z| \to \infty$ with $\min(|\text{Arg}(z)|, |\text{Arg}(-z)|) 
>\varepsilon$. (C.2) characterizes $\Cal M^H(\gamma)$ and the 
asymptotics are uniform over $\Cal M^H(\gamma)$ (Proposition~4.13).

For an indeterminate moment problem, we defined $\text{ord}(\rho)$, 
the {\it{order}} of $\rho$, to be the codimension of the closure 
of the polynomials in $L^2 (\Bbb R, d\rho)$. The solutions of 
finite order are dense in $\Cal M^H(\gamma)$ and in $\Cal M^S 
(\gamma)$, and each is an extreme point. Solutions of finite order 
(and, in particular, the von~Neumann solutions) are discrete pure 
point measures; equivalently, their Stieltjes transforms are 
meromorphic functions (Theorems~4.11 and 6.4).
\enddefinition

\vskip 0.1in

\definition{C2 \quad Nevanlinna Parametrization}

\vskip 0.1in

We used $\Cal F$ to denote the analytic maps of $\Bbb C_+$ to 
$\bar\Bbb C_+$ where the closure is in the Riemann sphere. The 
open mapping theorem implies that if $\Phi\in\Cal F$, either 
$\Phi(z) \equiv t$ for some $t\in\Bbb R\cup\{\infty\}$ or else  
$\Phi$ is a Herglotz function which has a representation of 
the form (1.19).

Associated to any indeterminate moment problem are four entire 
functions, $A(z)$, $B(z)$, $C(z)$, $D(z)$, obeying growth 
condition of the form $|f(z)|\leq c_\varepsilon \exp(\varepsilon 
|z|)$. We defined them via the transfer matrix relation (4.16), 
but they also have explicit formulae in terms of the orthogonal 
polynomials, $P$ and $Q$ (defined in {\bf{C4}} below)---these 
are given by Theorem~4.9.

We define a fractional linear transformation $F(z) : \Bbb C \cup 
\{\infty\} \to \Bbb C\cup\{\infty\}$ by
$$
F(z)(w) = -\frac{C(z)w + A(z)}{D(z)w + B(z)}\, . \tag C.3
$$

For indeterminate Hamburger problems, there is a one-one 
correspondence between $\rho\in\Cal M^H(\gamma)$ and $\Phi\in 
\Cal F$ given by
$$
G_\rho(z) = F(z)(\Phi(z))= - \frac{C(z)\Phi(z) + A(z)}
{D(z) \Phi(z) + B(z)} \tag C.4
$$
for $z\in\Bbb C_+$ (Theorem~4.14). $\Phi$ is called the Nevanlinna 
function of $\rho$, denoted by $\Phi_\rho$. The von~Neumann 
solutions correspond precisely to $\Phi(z) = t$ where $t$ is 
given by (C.1) (Theorem~4.10). The solutions of finite order 
correspond precisely to the case where $\Phi$ is a ratio of real 
polynomials (Theorem~6.4).

Given a set of Stieltjes moments, the $\rho\in\Cal M^S(\gamma)$ 
are precisely those $\rho\in\Cal M^H(\gamma)$ whose Nevanlinna 
function has the form (Theorem~4.18)
$$
\Phi(z) = d_0 + \int_0^\infty \frac{d\mu(x)}{x-z}
$$
with $\int d\mu(x) < \infty$ and $d_0 \geq t_F = \int d\rho_F(x) 
= (\delta_0, A^{-1}_F\delta_0)$.

For $z\in\Bbb C_+$, the image of $\bar\Bbb C_+ \cup \{\infty\}$ 
under $F(z)$ is a closed disk denoted by $\Cal D(z)$. Von~Neumann 
solutions $\rho$ have $G_\rho(z)\in\partial\Cal D(z)$ for all $z$ 
while other solutions have $G_\rho(z)\in\Cal D(z)^{\text{\rom{int}}}$ 
for all $z$ (Theorem~4.3, Proposition~4.4, and Theorem~4.14).
\enddefinition

\vskip 0.1in

\definition{C3 \quad Derived Moment Problems}
\vskip 0.1in

Any set of moments has many families of associated moments. For 
each real $c$, $\gamma(c)$ is defined by
$$
\gamma_n (c) = \sum_{j=0}^n \binom{n}{j} c^j \gamma_{n-j}. 
\tag C.5
$$
There is a simple map ($\rho\mapsto \rho (\,\cdot\, -c)$) that 
sets up a bijection between $\Cal M^H(\gamma)$ and $\Cal M^H
(\gamma(c))$ and, in particular, $\gamma$ is Hamburger 
determinate if and only if $\gamma(c)$ is Hamburger determinate. 
But the analog is not true for the Stieljes problem. Indeed, 
(end of Section~3), if $\gamma$ is a set of indeterminate 
Stieltjes moments and $c_F = -\inf\,\text{supp}(d\mu_F)$, then 
$\gamma(c)$ is a set of Stieltjes moments if and only if $c 
\geq c_F$, and it is Stieltjes determinate if and only if $c = 
c_F$. The orthogonal polynomials for $\gamma(c)$ and $\gamma$ 
are related via $P^{(\gamma(c))}_N(z) = P_N(z-c)$.

Given a set of Stieltjes moments $\{\gamma_n\}^\infty_{n=0}$, 
one defines Hamburger moments $\{\Gamma_n\}^\infty_{n=0}$ by
$$
\Gamma_{2m} = \gamma_m, \qquad \Gamma_{2m+1} = 0. \tag C.6
$$
There is a simple map ($d\rho\mapsto d\mu(x) = \frac12 
\chi_{[0,\infty)}(x)\, d\rho(x^2) + \frac12 \chi_{(-\infty, 0]}
(x)\, d\rho(x^2)$) that sets up a bijection between $\Cal M^H 
(\gamma)$ and those $\mu\in\Cal M^H(\Gamma)$ with $d\mu (-x) 
=d\mu(x)$. Then $\gamma$ is Stieltjes determinate if and only 
if the $\Gamma$ problem is Hamburger determinate (Theorem~2.13). 
There are simple relations between the orthogonal polynomials 
for $\gamma$ and those for $\Gamma$ ((5.89)--(5.92)).

For $\ell=1,2,\dots$, define
$$
\gamma^{(\ell)}_j = \frac{\gamma_{\ell+j}}{\gamma_\ell}\, . 
\tag C.7
$$
If $\gamma$ is a set of Stieltjes moments, so are $\gamma^{(\ell)}$ 
for all $\ell$, and if $\gamma$ is a set of Hamburger moments, so 
are $\gamma^{(\ell)}$ for $\ell = 2,4,\dots$. The map $\rho \mapsto 
d\rho^{(\ell)} = x^\ell \frac{d\rho(x)}{\gamma_\ell}$ is a map 
from $\Cal M^H(\gamma)$ to $\Cal M^H(\gamma^{(\ell)})$ which is 
injective. But in the indeterminate case, it is {\it{not}}, in 
general, surjective. If $\gamma$ is indeterminate, so is 
$\gamma^{(\ell)}$, but the converse may be false for either the 
Hamburger or Stieltjes problems (Corollary~4.21). For general 
$\ell$, the connection between the orthogonal polynomials for 
$\gamma$ and $\gamma^{(\ell)}$ is complicated (see the proof of 
Proposition~5.13 for $\ell=2$), but (see the remarks following 
Theorem~5.14)
$$
P^{(1)}_N(z) = \frac{d_N [P_N(0) P_{N+1}(z) - P_{N+1}(0) P_N(z)]} 
{z}\, .
$$

Given a set of Hamburger moments $\{\gamma_j\}^\infty_{j=0}$, 
one defines a new set $\{\gamma^{(0)}_j\}^\infty_{j=0}$ by the 
following formal power series relation:
$$
\biggl[ \sum_{n=0}^\infty (-1)^n \gamma_n z^n \biggr]^{-1} 
= 1 - \gamma_1 z + (\gamma_2 - \gamma^2_1) z^2 
\biggl( \sum_{n=0}^\infty (-1)^n \gamma^{(0)}_n z^n \biggr). 
\tag C.8
$$
This complicated formula is simple at the level of Jacobi 
matrices. If $\{a_n\}^\infty_{a=0}$, $\{b_n\}^\infty_{n=0}$ 
(resp.~$\{a^{(0)}_n\}^\infty_{n=0}$, $\{b^{(0)}_n
\}^\infty_{n=0}$) are the Jacobi matrix coefficients associated 
to $\gamma$ (resp.~$\gamma^{(0)}$), then
$$
a^{(0)}_n = a_{n+1}, \qquad b^{(0)}_n = b_{n+1}. 
$$
The map $\rho\mapsto \tilde\rho$ given by (5.53) sets up a 
bijection between $\Cal M^H(\gamma)$ and $\Cal M^H(\gamma^{(0)})$,  
and, in particular, $\gamma$ is a Hamburger determinate if and 
only if $\gamma^{(0)}$ is (Proposition~5.15). This is not, in 
general, true in the Stieltjes case (see Proposition~5.17). There 
is a simple relation between the orthogonal polynomials for 
$\gamma$ and for $\gamma^{(0)}$; see Proposition~5.16. In 
particular, 
$$
P^{(0)}_N(z) = a_0 Q_{N+1}(z). \tag C.9
$$
For $\ell= -1, -2, \dots$, we defined
$$
\gamma^{(\ell)} = (\gamma^{(0)})^{(-\ell)} = 
\frac{\gamma^{(0)}_{j-\ell}}{\gamma^{(0)}_{-\ell}}\, . 
\tag C.10
$$
\enddefinition

\vskip 0.1in

\definition{C4 \quad Associated Polynomials}

Given a set of Hamburger moments, the fundamental orthogonal 
polynomials $P_N(z)$ are defined by requiring for any $\rho\in 
\Cal M^H(\gamma)$ that
$$
\int P_j (x) P_\ell(x)\, d\rho(x) = \delta_{j\ell} \tag C.11
$$
and $P_N(x) = c_{NN} x^N + \text{ lower order terms}$ with 
$c_{NN} >0$. They obey a three-term recursion relation
$$
xP_N(x) = a_N P_{N+1}(x) + b_N P_N(x) + a_{N-1} P_{N-1}(x). 
\tag C.12
$$
The Jacobi matrix $A$ associated to $\gamma$ is the tridiagonal 
matrix given by (1.16).

The associated polynomials $Q_N(x)$ are of degree $N-1$ and 
defined to obey (C.12) but with the starting conditions,
$Q_0(x)=0$, $Q_1(x) = \frac1{a_0}$. They are related to $P_N$ 
by (Theorem~4.2)
$$
Q_N(x) = \int \frac{P_N(x) - P_N(y)}{x-y}\, d\rho(y) \tag C.13
$$
for any $\rho\in\Cal M^H(\gamma)$. As noted in (C.9), they are 
up to constants, orthogonal polynomials for another moment problem, 
namely, $\gamma^{(0)}$.

In the Stieltjes case, we defined polynomials $M_N$ by
$$
M_N(x) = P_N(x) - \frac{P_N(0)P_{N-1}(x)}{P_{N-1}(0)}\, . 
\tag C.14
$$
They vanish at $x=0$ so $\frac{M_N(x)}{x}$ is also a 
polynomial. It is of degree $N-1$ and is up to a constant, the 
principal orthogonal polynomial of the moment problem 
$\gamma^{(1)}$. $N_N(x)$ is defined analogously to (C.13) with 
$P_N$ replaced by $M_N$. In Section~5, it was useful to change 
the normalizations and define multiples of $P_N(-x)$, $Q_N(-x)$, 
$M_N(-x)$, and $N_N(-x)$ as functions $U_N(x)$, $V_N(x)$, 
$G_N(x)$, $H_N(x)$ given by (5.71)--(5.74).
\enddefinition

\vskip 0.1in

\definition{C5 \quad Pad\'e Approximants and Finite Matrix 
Approximations}
\vskip 0.1in

If $\{\gamma_n\}^\infty_{n=0}$ is a set of Hamburger 
(resp.~Stieltjes) moments, the formal power series 
$\sum_{n=0}^\infty (-1)^n \gamma_n z^n$ is called a 
{\it{series of Hamburger}} (resp.~{\it{a series of Stieltjes}}). 
Formally, it sums to
$$
\int \frac{d\rho(x)}{1+xz}
$$
if $\rho\in\Cal M^H(\gamma)$. The Pad\'e approximants 
$f^{[N,M]}(z)$ to these series, if they exist, are defined by 
(5.25)--(5.27) as the rational function which is a ratio of a 
polynomial of degree $N$ to a polynomial of degree $M$, whose 
first $N+M+1$ Taylor coefficients are $\{(-1)^n\gamma_n
\}^{N+M}_{n=0}$.

For a series of Stieltjes, for each fixed $\ell = 0,\pm 1, \dots$,
$$
\lim_{N\to\infty} f^{[N+\ell-1, N]}(z) \equiv f_\ell (z)
$$
exists for all $z\in\Bbb C\backslash (-\infty, 0)$ and defines 
a function analytic there. Indeed, for $x\in [0,\infty)$, 
$(-1)^\ell f^{[N+\ell -1, N]}(x)$ is monotone increasing 
(Theorem~6, Theorem~5.14, and Theorem~5.18). Moreover,
$$
f_0(z) = \int \frac{d\rho_F(x)}{1+xz}\, , \qquad 
f_1(z) = \int \frac{d\rho_K (x)}{1+xz}\, ,
$$
where $\rho_F$, $\rho_K$ are the Friedrichs and Krein solutions 
of the moment problem. In particular, the moment problem is 
Stieltjes determinate if and only if $f_0 = f_1$.

There is a connection between $f^{[N+\ell-1,N]}(z)$ and the 
moment problems $\gamma^{(\ell)}$ and $\gamma^{(\ell-1)}$ (if 
$\ell < 0$, $\gamma^{(\ell)}$ and $\gamma^{(\ell+1)}$) given 
in Theorem~5.14 and Theorem~5.18. In particular, if $\ell >0$ and 
$\gamma^{(\ell)}$ is Stieltjes determinate, then $f_{-1}(z) = 
f_0(z) = \cdots = f_{\ell+1}(z)$, and if $\ell <0$ and 
$\gamma^{(\ell)}$ is Stieltjes determinate, then $f_1(z) = 
f_0(z) = \cdots = f_\ell(z) = f_{\ell-1}(z)$.

The situation for series of Hamburger is more complicated, but, 
in general, \linebreak $f^{[N+\ell-1, N]} (z)$ converges for 
$z\in\Bbb C_+$ and $\ell = \pm 1, \pm 3, \dots$ (Theorem~5.31 and 
Theorem~5.32; see Theorem~5.31 for issues of existence of the 
Pad\'e approximant).

There are connections between the Pad\'e approximants and the 
polynomials $P,Q,M,N$ as well as two finite matrix approximations 
$A^{[N]}_F$ and $A^{[N]}_K$ to the Jacobi matrix, $A$. $A^{[N]}_F$ 
is the upper right $N\times N$ piece of $A$. $A^{[N]}_K$ differs 
by adjusting the $NN$ matrix element of $A^{[N]}_F$ so $\det 
(A^{[N]}_K)=0$. Then
$$\align
f^{[N-1, N]}(z) &= \langle\delta_0, (1+zA^{[N]}_F )^{-1} \delta_0\rangle 
= -\frac{z^{N-1}Q_N (-\frac1{z})}{z^N P_N(-\frac1{z})} \\
f^{[N,N]}(z) &= \langle\delta_0, (1+zA^{[N+1]}_K)^{-1} \delta_0\rangle =
-\frac{z^N N_{N+1} (-\frac1{z})}{z^{N+1}M_{N+1} (-\frac1{z})}\, . 
\endalign
$$

The connection to the continued fractions of Stieltjes is 
discussed after Theorem~5.28.
\enddefinition

\vskip 0.1in

\definition{C6 \quad Criteria for Determinacy}
\vskip 0.1in

Criteria for when a Hamburger problem is determinate can be 
found in Proposition~1.5, Theorem~3, Corollary~4.5, 
Proposition~5.13, Theorem~A.6, and Theorem~A.7.

Criteria for when a Stieltjes problem is determinate can be 
found in Proposition~1.5, Theorem~7, Corollary~4.5, 
Theorem~5.21, Corollary~5.24, Theorem~A.4, and Theorem~A.6. In 
particular, if one defines
$$\align
c_{2j} &= \ell_j = -[a_{j-1} P_j (0) P_{j-1}(0)]^{-1} \\
c_{2j-1} &= m_j = |P_{j-1}(0)|^2
\endalign
$$ 
(the $c$'s are coefficients of Stieltjes continued fractions; 
the $m$'s and $\ell$'s are natural parameters in Krein's theory), 
then the Stieltjes problem is determinate if and only if 
(Theorem~5.21 and Proposition~5.23) $\sum_{j=0}^\infty c_j = 
\infty$.
\enddefinition

\vskip 0.3in
\definition{Acknowledgments} I would like to thank George Baker, 
Percy Deift, Rafael del Rio, Fritz Gesztesy, Mourad Ismail, 
Uri Keich, and Sasha Kiselev for valuable comments.
\enddefinition

\vskip 0.3in

\Refs
\endRefs
\vskip 0.1in

\item{\akh.} N.I.~Akhiezer, {\it The Classical Moment Problem}, 
Hafner Publishing Co., New York, 1965.
\gap
\item{\ag.} N.I.~Akhiezer and I.M.~Glazman, {\it Theory of Linear 
Operators in Hilbert Space, Vol.~2}, Frederick Ungar, New York, 
1961.
\gap
\item{\as.} \ref{A.~Alonso and B.~Simon}{The Birman-Krein-Vishik 
theory of self-adjoint extensions of semibounded operators}
{J.~Oper.~Th.}{4}{1980}{251--270}
\gap
\item{\bgm.} G.~Baker and P.~Graves-Morris, {\it Pad\'e  
Approximants}, 2nd~ed., Cambridge University Press, New York, 1996.
\gap
\item{\ds.} N.~Dunford and J.~Schwartz, {\it Linear Operators, 
II. Spectral Theory}, Interscience Publishers, New York, 1963.
\gap
\item{\gs.} F.~Gesztesy and B.~Simon, {\it $m$-functions and 
inverse spectral analysis for finite and semi-infinite Jacobi 
matrices}, to appear in J.~d'Anal.~Math.
\gap
\item{\gdl.} \ref{J.~Gil de Lamadrid}{Determinacy theory for the 
Liv\v sic moments problem}{J.~Math. Anal.~Appl.}{34}{1971}
{429--444}
\gap
\item{\ham.} H.~Hamburger, {\it \"Uber eine Erweiterung des 
Stieltjesschen Momentenproblems}, Math. Ann. {\bf 81} (1920), 
235--319; {\bf 82} (1921), 120--164, 168--187.
\gap
\item{\im.} \ref{M.E.H.~Ismail and D.~Masson}{$q$-hermite polynomials, 
biorthogonal rational functions, and $q$-beta integrals}
{Trans.~Amer.~Math.~Soc.}{346}{1994}{63--116}
\gap
\item{\ir.} M.E.H.~Ismail and M.~Rahman, {\it The $q$-Laguerre 
polynomials and related moment problems}, to appear in 
J.~Math.~Anal.~Appl.
\gap
\item{\jl.} S.~Jitomirskaya and Y.~Last, {\it Power law 
subordinacy and singular spectra, I. Half-line operators}, in 
preparation.
\gap
\item{\kato.} T.~Kato, {\it Perturbation Theory for Linear 
Operators}, 2nd ed., Springer, New York, 1980. 
\gap
\item{\katz.} Y.~Katznelson, {\it An Introduction to Harmonic 
Analysis}, 2nd corrected ed., Dover Publications, New York, 
1976.
\gap
\item{\kei.} U.~Keich, {\it private communication}.
\gap
\item{\krch.} M.G.~Krein, {\it Chebyshev's and Markov's ideas in 
the theory of limiting values of integrals and their further 
development}, Uspekhi matem.~Nauk {\bf 44} (1951), in Russian.
\gap
\item{\krsti.} M.G.~Krein, {\it On a generalization of an 
investigation by Stieltjes}, Dokl.~Akad.~Nauk SSSR {\bf 87} 
(1952), in Russian.
\gap 
\item{\krstu.} M.G.~Krein, {\it Some new problems in the theory 
of oscillations of Sturm systems}, Priklad.~matem.~i mekh. 
{\bf 16} (1952), 555--568, in Russian.
\gap
\item{\krkol.} \ref{M.G.~Krein}{On an extrapolation problem due 
to Kolmogorov}{Dokl.~Akad.~Nauk SSSR}{46}{1945}{306--309}
\gap
\item{\land.} \ref{H.~Landau}{The classical moment problem: 
Hilbertian proofs}{J.~Funct.~Anal.}{38}{1980}{255-272}
\gap
\item{\lang.} \ref{R.W.~Langer}{More determinacy theory for the 
Liv\v sic moments problem}{J.~Math.~Anal. Appl.}{56}{1976}
{586--616}
\gap
\item{\liv.} \ref{M.S.~Liv\v sic}{On an application of the 
theory of hermitian operators to the generalized problem of 
moments}{Dokl.~Akad.~Nauk SSSR}{26}{1940}{17--22}
\gap
\item{\loe.} \ref{J.J.~Loeffel, A.~Martin, B.~Simon, and 
A.S.~Wightman}{Pad\'e approximants and the anharmonic oscillator}
{Phys.~Lett.}{30B}{1969}{656--658}
\gap
\item{\mmc.} \ref{D.~Masson and W.~McClary}{Classes of 
$C^\infty$-vectors and essential self-adjointness}{J.~Funct.~Anal.}
{10}{1972}{19--32}
\gap
\item{\naI.} \ref{M.A.~Naimark}{Self-adjoint extensions of the 
second kind of a symmetric operator}{Izv.~Akad.~Nauk SSSR}{4}
{1940}{90--104}
\gap
\item{\naII.} \ref{M.A.~Naimark}{Spectral functions of a 
symmetric operator}{Izv.~Akad.~Nauk SSSR}{4}{1940}{309--318}
\gap
\item{\naIII.} \ref{M.A.~Naimark}{On spectral functions of a 
symmetric operator}{Izv.~Akad.~Nauk SSSR}{7}{1943}{285--296}
\gap
\item{\naIV.} \ref{M.A.~Naimark}{On extremal spectral functions 
of a symmetric operator}{Dokl.~Akad. Nauk SSSR}{54}{1946}{7--9}
\gap
\item{\nel.} \ref{E.~Nelson}{Analytic vectors}{Ann.~of Math.}
{70}{1959}{572--615}
\gap
\item{\nev.} R.~Nevanlinna, {\it Asymptotische Entwickelungen 
beschr\"ankter Funktionen und das Stieltjessche Momentenproblem}, 
Ann.~Acad.~Sci.~Fenn. {\bf A 18} (1922).
\gap
\item{\nusI.} \ref{A.~Nussbaum}{Quasi-analytic vectors}
{Ark.~Mat.}{6}{1965}{179--191}
\gap
\item{\nusII.} \ref{A.~Nussbaum}{A note on quasi-analytic 
vectors}{Studia Math.}{33}{1969}{305--310}
\gap
\item{\pea.} D.B.~Pearson, {\it Quantum Scattering and Spectral 
Theory}, Academic Press, London, 1988.
\gap
\item{\rsI.} M.~Reed and B.~Simon, {\it Methods of Modern 
Mathematical Physics, I. Functional Analysis}, Academic Press, 
New York, 1972.
\gap
\item{\rsII.} M.~Reed and B.~Simon, {\it Methods of Modern 
Mathematical Physics, II. Fourier Analysis, Self-Adjointness}, 
Academic Press, New York, 1975.
\gap
\item{\st.} J.A.~Shohat and J.D.~Tamarkin, {\it The problem 
of moments}, Amer.~Math.~Soc.~Surveys, No.~1 (1943).
\gap
\item{\simap.} B.~Simon, {\it Coupling constant analyticity for 
the anharmonic oscillator} (with an appendix by A.~Dicke), 
Ann.~Phys. {\bf 58} (1970), 76--136.
\gap
\item{\simjfa.} \ref{B.~Simon}{A canonical decomposition for 
quadratic forms with applications to monotone convergence 
theorems}{J.~Funct.~Anal.}{28}{1978}{377--385}
\gap
\item{\sti.} T.~Stieltjes, {\it Recherches sur les fractions 
continues}, Anns.~Fac.~Sci.~Univ.~Toulouse (1894--95), 
{\bf 8}, J1--J122; {\bf 9}, A5--A47.
\gap
\item{\sto.} M.H.~Stone, {\it Linear Transformations in Hilbert 
Space}, Amer.~Math.~Soc.~Colloq.~Publ. XV, New York, 1932.
\gap

\enddocument